\documentclass[preprint,12pt]{elsarticle}

\usepackage{bbm}
\usepackage{graphics}
\usepackage{amsfonts,amsbsy}

\usepackage{subfig}
\usepackage{graphicx}
\usepackage{amssymb,mathtools}

\usepackage{bm}
\usepackage{color}
\usepackage{bm,amsmath,amssymb}

\numberwithin{equation}{section}

\long\def\comment#1{ }

\newcommand{\eqn}[1]{Eq.~\eqref{#1}}

\newcommand{\beq}{\begin{eqnarray}}
\newcommand{\eeq}{\end{eqnarray}}
\newcommand{\nn}{\nonumber\\}
\newcommand{\bel}[1]{\begin{eqnarray}\label{#1}}

\def\inel{\text{inel}}
\def\el{\text{el}}

\def\eq{\text{eq}}

\def\mfp{\text{mfp}}
\def\coh{\text{coh}}
\def\BH{\text{BH}}

\def\rel{\text{rel}}
\def\bnab{{\boldsymbol \nabla}}
\def\KZ{\text{KZ}}
\def\st{\text{st}}

\newcommand{\rmd}{{\rm d}}
\newcommand{\rme}{{\rm e}}

\def\q{{\bm q}}
\def\p{{\boldsymbol p}}
\def\l{{\boldsymbol l}}
\def\k{{\boldsymbol k}}

\def\x{{\boldsymbol x}}
\def\y{{\boldsymbol y}}

\def\D{{\boldsymbol D}}

\def\u{{\boldsymbol u}}
\def\v{{\boldsymbol v}}
\def\w{{\boldsymbol w}}

\def\B{{\boldsymbol B}}

\def\min{\text{--}}
\def\max{\text{+}}
\def\tr{\text{Tr}}
\def\for{\text{f}}

\def\qp{{{\bm q}_\perp}}
\def\pp{{{\boldsymbol p}_\perp}}
\def\lp{{{\boldsymbol l}_\perp}}
\def\kp{{{\boldsymbol k}_\perp}}

\def\cK{{\cal K}}
\def\cC{{\cal C}}

\newcommand{\del}{\partial}

\newcommand{\abar}{\bar{\alpha}}

\begin{document}


\begin{frontmatter}

\title{{\bf Nonlocal wave turbulence\\ in non-Abelian plasmas}}

\author{Yacine Mehtar-Tani}

\address{Institute for Nuclear Theory, University of Washington, \\Seattle, WA 98195-1550, USA}

\begin{abstract}
We investigate driven wave turbulence in non-Abelian plasmas, in the framework of kinetic theory where 
both elastic and inelastic processes are considered in the small angle approximation. The gluon spectrum, that forms in the presence of a steady source, is shown to be controlled by nonlocal interactions in momentum space, in contrast to the universal Kolmogorov-Zakharov spectra.  Assuming strongly nonlocal interactions, we show that inelastic processes are dominant in the IR and cause a thermal bath to form below the forcing scale, as a result of a detailed balance between radiation and absorption of soft gluons by the hard ones. Above the forcing scale, the inelastic collision term reduces to an inhomogeneous diffusion-like equation yielding a spectrum that spreads to the UV as $t^{1/2}$, similarly to elastic processes. Due to nonlocal interactions the non-universal turbulent spectrum is not steady and flattens when time goes on toward the thermal distribution. This analysis is complemented by numerical simulations, where we observe that in the explored time interval the spectral exponent of the nonlocal turbulent cascade is close to $-2$ in agreement with simulations of classical Yang-Mills equations. 

\end{abstract}

\end{frontmatter}
 \begin{flushright}
INT-PUB-16-040
 \end{flushright}
\section{Introduction}
\label{sec:intro}
For more than a decade, the approach to thermal equilibrium of the quark gluon plasma in the early time dynamics of ultrarelativistic heavy ion collisions, has been thoroughly studied. While a lot of progress has been made in uncovering the relevant phenomena at play, such as plasma instabilities, turbulence, particle production, izotropization, etc, the exact mechanisms by which the system reaches an early hydrodynamical behavior observed in the data is still unclear and remains the objet of intense research (see Ref.~\cite{Berges:2012ks,Huang:2014iwa,Fukushima:2016xgg,Kurkela:2016vts} for reviews). 

In this context, it was recognized early that plasma instabilities, such as Chromo-Weibel instabilities  \cite{Mrowczynski:1988dz,Arnold:2003rq}, may play a role in spurring the isotropization of the system in weakly coupled plasmas (see e.g. Refs.~\cite{Strickland:2006cv} and references therein). P. Arnold and G. D. Moore have shown in 2005  \cite{Arnold:2005ef}, in particular, by solving numerically the Vlasov equation coupled to the classical Yang-Mills equation, that energy carried by unstable magnetic modes, that develop due to the initial anisotropy of the momentum distribution of the relatively hard particles, grows exponentially in a first stage, similarly to Abelian plasmas, then linearly due to gauge field self-interaction.  During this phase, higher modes are excited generating a direct turbulent energy cascade characterized by a fitted power spectrum $~ \omega^{-2}$  \cite{Arnold:2005ef,Ipp:2010uy}. They suggested, in a follow-up paper \cite{Arnold:2005qs}, that the observed quasi-steady spectrum may be due to diffusion of hard modes on the strong classical field related to the unstable modes. However, this heuristic parametric argument appears to be in contradiction with the fact that diffusion conserves particle number and not energy \cite{Ipp:2010uy}.

Indeed, A. H. Mueller and collaborators argue, in a paper that appeared shortly after, that in non-Abelian plasmas, interactions are nonlocal in momentum space forcing the system in contact with a steady source of energy to be in thermal equilibrium at late times \cite{Mueller:2006up}, that is, the spectrum must scale as $
\omega^{-1}$ which is also a fixed point of the diffusion equation, as we shall see. 

Later, simulations of classical Yang-Mills equations of motion, for a system that undergoes longitudinal expansion, have shown no evidence for plasma instabilities \cite{Berges:2013eia}, confirming the standard bottom-up thermalization scenario \cite{Baier:2000sb}. Nevertheless, the turbulent spectrum obtained by Arnold and Moore in the Hard-Loop approximation is of a fundamental interest and remains to be fully understood. 

The present work is an attempt to clarify this issue in the framework of kinetic theory, where both elastic and inelastic processes are taken into account. A complete analytic treatment proves to be difficult, however, some insights can be gained by analyzing the kinetic equation under certain, well motivated, approximations. We find for instance that strongly nonlocal inelastic interactions obey a diffusion equation with a source term that conserves energy, and hence, is different from standard diffusion in momentum space.

Besides the aforementioned works \cite{Arnold:2005ef,Mueller:2006up}, most studies on turbulence in non-Abelian plasma are numerical and focus on the so-called free (decaying) turbulence \cite{Kurkela:2011ti,Kurkela:2012hp,Berges:2013eia,Berges:2013fga}, that sets the early stages of thermalization in heavy ion collisions. After a transient regime, the system evolves in a self-similar way until the occupation number drops below unity where quantum effects begin to play a role in the final stage of thermalization. On the other hand, in forced (steady) turbulence where the system is in contact with a steady source, a stationary state is expected to be reached.  

Forced turbulence is a chaotic phenomenon that is characterized by the transport of conserved quantities such as energy, in open systems, with constant fluxes through a wide interval of scales, the so-called inertial range, connecting the scale where energy is injected into the system, the forcing scale $\omega_f$, and the scale where it is dissipated. In the seminal work by Kolmogorov and Obukhov, known as KO41 theory, the famous $5/3$ power law was derived for the energy spectrum of a fluid in a turbulent state \cite{Zakharov1992-rev,Nazarenko2011} under the assumption of scale invariance.
This approximation is valid only if the relevant interactions are local, that is, confined to eddies of similar sizes, such that the physics in the inertial range is not sensitive to the injection and dissipation scales. 

The phenomenon described above for classical fluids, occurs in a wide range of physical systems, in particular, in non-linear wave systems, such as surface water waves, atmospheric waves, waves in plasmas, non-linear optics and Bose-Einstein condensates for instance. In the early 1960's, V. E. Zakharov laid the foundation of the theory of (weak) wave turbulence theory by discovering that the kinetic equation that describes weakly non-linear waves admits non-thermal stationary solutions in the form of power spectra associated with a constant flux for each integral of motion of the system \cite{Zakharov1992-rev,Nazarenko2011}. Later, it was shown that the Boltzmann equation also admits similar power law solutions \cite{KKMN}.

In the high occupancy limit, non-abelian plasmas described by classical Yang-Mills equations fall into the class of non-linear wave systems whose statistics can be studied in the framework of weak turbulence theory. The quantity of interest is the {\it wave occupation number}, $n_k\equiv n({\boldsymbol k},t)$ which is related to the equal time two-point correlation function $\langle a_{\boldsymbol k}(t) a^\ast_{{\boldsymbol k}'}(t) \rangle \equiv n(k,t) \, \delta({\boldsymbol k}-{\boldsymbol k}')$\footnote{Here, color and polarization of the gluon field is omitted. }, where $a_k$ is the classical analog of the gluon creation operator and the angle brackets $\langle ...\rangle $ stands for ensemble average with respect to the statistics of the forcing in driven(forced) turbulence or the initial condition for decaying turbulence. In the regime of weak non-linearity, that is when the coupling constant $g\ll 1$ for Yang-Mills theory, the occupation number obeys a kinetic equation of the form \footnote{For a recent derivation of the kinetic equation (with only 2 to 2 elastic processes considered) see Ref.~\cite{Mathieu:2014aba}}, 
\beq\label{kinetic-eq}
\frac{\del n_k }{\del t} =I_k[n],
\eeq 
where the collision integral scales as $I_k[n] \sim n^{N-1}$ for dominant $N$ particle interactions (for instance $N=3$ for elastic 2 to 2 elastic scatterings). In deriving the wave kinetic equation (\ref{kinetic-eq}) one averages over random wave phases which implies a Gaussian statistics and allows to express higher order correlation functions as a product of two-point correlation functions, yielding a closed equation for $n_k$. 

As alluded to above, the purpose of this work is to study wave turbulence in non-Abelian plasmas, with the aim of better understanding the dynamics underlying the turbulent cascade resulting from Chromo-Weibel instabilities \cite{Arnold:2005ef}. Our setup is similar to that of Ref.~\cite{Mueller:2006up}: We consider an open system of gluons in contact with an isotropic and spectrally narrow steady source in order to mimic the growth of the unstable modes. Although instabilities are generated in anisotropic system, we assume that in the inertial range away from the source the system is statistically isotropic. Furthermore, we shall proceed in the framework of the effective kinetic theory proposed in Refs.~\cite{Baier:2000sb,Arnold:2002zm}, that accounts for 2 to 2 elastic scattering and collinear effective 1 to 2 gluon branchings induced by small angle multiple scattering, that are parametrically of the same order (see discussion below). There are a few caveats that are as follows:
\begin{enumerate}
\item First, we restrict our analysis to the purely gluonic case and we assume the medium to be homogeneous and isotropic. It follows that the gluon occupation number is only function of the modulus of the wave number, i.e., $|{\boldsymbol k}|\equiv k$,
\beq
(2\pi)^3\frac{\rmd N}{\rmd^3\x\rmd^3\k } \equiv n(\k,\x)\equiv n(k),
\eeq
where $N$ is the number of gluons of a given spin and color. 
\item In addition, we assume that the gluon momentum is much larger than its thermal mass, i.e., $k \gg m$  in order to use the homogeneous dispersion relation
\beq
\omega(\k)\equiv |\k|, 
\eeq 
and apply the small angle scattering approximation to the kinetic equations. In effect, gluon-gluon Coulomb scattering is dominated by small momentum transfer of the order of the Debye mass $m_D\sim m  \ll k $, 
, where \cite{Blaizot:2001nr,Arnold:2002zm} 
\bel{therm-mass}
m^2_D=2 m^2\equiv  \frac{g^2 N_c}{\pi^2}   \int \rmd k k^2 \, \frac{\del n(k)}{\del k}.
\eeq
for Yang-Mills plasma. 
\item We shall work in the classical limit of weakly non-linear waves where a kinetic description applies, namely, 
\beq
1\, \ll\,  n(k)\,  \ll \, \frac{1}{g^{2}}, 
\eeq
with $g \, \ll 1 $, the Yang-Mills coupling constant in a perturbative regime.
 \item  Finally, we consider the standard setup in the study of developed turbulence consisting in an open system in contact with an external source that injects gluons at a forcing frequency $\omega_f$, and sinks, that  dissipate energy in the ultraviolet and the infrared at $\omega_\max$ and $\omega_\min$, respectively, with a wide scale separation: between the source and the sinks, $\omega_\min\, \ll\,  \omega_f\,  \ll \,\omega_\max$.
\end{enumerate}
In general, at leading order in the weak coupling $g$, one expects the dominant process to be 2 to 2 elastic scattering (four wave interaction) whose matrix element  squaredis of order $g^4$. The corresponding kinetic equation admits two integrals of motion: energy and particle number\footnote{In general, momentum is also conserved but in an isotropic system it averages to zero.}. Therefore, two non-thermal stationary spectra might form in the presence of a steady source. A direct energy cascade, characterized by a constant energy flux density, $P$, and an inverse particle cascade carrying a particle flux density $Q$. 

Owing to the fact that the Yang-Mills coupling is dimensionless and the dispersion relation is linear, there is no other dimensionfull quantity besides the frequency of the measured gluon, $\omega$. Hence, for a scale invariant system the collision integral scales parametrically as
\beq\label{param-n-4}
\dot n(\omega) \equiv  I^{2\to 2}[n] \sim g^4\, \omega \,n^3,
\eeq
with $n\equiv n(\omega)$. On the other hand, the power spectra that carry constant flux of energy and constant flux of particles, respectively, verify 
\beq\label{param-P-4}
 P \equiv \dot E \sim \omega^4 \dot n \equiv \text{const.} \quad  \text{and} \quad  Q \equiv \dot N \sim \omega^3 \dot n \equiv \text{const.}
\eeq
Now, using Eq.~(\ref{param-n-4}) in Eq.~(\ref{param-P-4}) and solving for $n(\omega)$ as a function of constant energy flux $P$, and similarly for constant particle flux $Q$, we readily obtain the Kolmogorov-Zakharov (KZ) spectra associated with the direct energy cascade and inverse particle cascade, respectively, 
\bel{KZ-E-N}
n(\omega) \sim P^{1/3}\omega^{- 5/3} \quad  \text{and} \quad n(\omega) \sim Q^{1/3}\omega^{- 4/3}. 
\eeq
The direction of the fluxes can be determined by the following simple argument: in a steady state, the energy and particle number pumping rate at $\omega_f$ is exactly balanced by the rate at which they are removed at the sinks, located at $\omega_\max$ and $\omega_\min$, therefore, we must have $\dot E_f= \omega_f \dot N_f = \omega_\min \dot N_\min + \omega_\max \dot N_\max$ and $\dot N_f = \dot N_\min + \dot N_\max$. Now, solving for $\dot N_\max$ and $\dot N_\min$ in the limit  $\omega_\min \ll \omega_f \ll \omega_\max$ we find that, $\dot N_\min \simeq \dot N_f\quad  \text{and} \quad \dot E_\max = \omega_\max \dot N_\max \simeq \dot E_f$,
that means that particle number must be dissipated at $\omega_\min$ and energy at  $\omega_\max$. 

The realization of the KZ spectra (\ref{KZ-E-N}) depends upon two conditions: 
\begin{enumerate}
\item Locality of interactions in momentum space, that is, the dynamics in the inertial range does not depend on the physics at the forcing or damping, besides the fluxes of the conserved quantities. This allows to use scale invariance and seek for power law solutions. 

\item Suppression of inelastic processes (or more generally higher order processes) by extra powers of $g$, compared to elastic 2 to 2 scatterings.
\end{enumerate}
However, while these conditions are satisfied  for $\phi^4$ scalar theory (for instance) as we shall see in Section \ref{sec:BN-equation},  neither of them is fulfilled by Yang-Mills equations.  At weak coupling, higher order processes involve collinear divergences that are cutoff by thermal masses $m$, yielding a $m^{-2}$ pre-factor in the collision integral that breaks scale invariance. Moreover, since the thermal mass squared is proportional to $g^2 n $, the supposedly suppressed 2 to 3 process (of order $g^6 n^4$) is enhanced in the region where two of the outgoing (or incoming) gluons are collinear (see \ref{app:inelatis-2to3} for a detailed demonstration). In turn, the collision integral for five wave interactions (2 to 3 scattering) is parametrically of the order of the four wave interaction term, that is, $I^{2\to 3}[n] \sim g^6 \omega^3 m^{-2}  n^4 \sim g^4 \omega n^3$, and thus it is a leading order effect \cite{Baier:2000sb,Arnold:2002zm,Blaizot:2011xf}. In the quasi-collinear splitting approximation, the  2 to 3 inelastic process reduces to an effective 1 to 2 collinear splitting triggered by the exchange of a low momentum gluon (of the order $m_D$) in the $t$-channel   that can be extended to processes that involve multiple scatterings acting coherently during the collinear branching process. 
Analogous approaches have been considered to cope with nonlocal collision integrals, for instance, in describing kelvin waves in superfluids \cite{Naz2010} or interactions of hard modes with a Bose-Einstein condensate \cite{Zakharov1992,Micha:2004bv}, that cause the breakdown of the 4 wave interaction into an effective three wave interaction of acoustic type \cite{Zakharov1970}. 

Let us now summarize our results. In Section \ref{sec:BN-equation}, as a warm up exercise, we discuss $\phi^4$ scalar theory that admits local KZ spectra as non-thermal stationary states, in order to illustrate and highlight the differences with Yang-Mills theories. We compute in particular the non-universal KZ constant factors.  In the following two sections, we turn to classical Yang-Mills theory where elastic and inelastic collision integrals are first discussed separately. In Section \ref{elastic-YM}, 2 to 2 elastic processes are first analyzed. Because the dominant process is the small angle Coulomb scattering, the collision integral can be reduced to a Fokker-Planck (FP) equation which is solved in the stationary case. In the inertial range to the left of the forcing scale (away from the source and sinks), the steady state power spectrum is of the form $Q_f^{1/3}\omega_f^{-1/3}\omega^{-1}$. It is related to an inverse particle cascade. The explicit dependence on the forcing scale $\omega_f$ reflects the non universality of the stationary state. In addition, we find that although energy is conserved, there is no finite steady state associate with the direct energy cascade. It follows that above the forcing scale, where the direct cascade is expected to form, the spectrum vanishes in the limit of infinitely large inertial range, i.e., when energy is dissipated at infinity. Numerical simulations of the FP equation are also performed to investigate quantitatively the approach toward the steady state obtained analytically. 

In Section \ref{inelastic-YM}, we turn to inelastic processes that can be described by effective quasi-collinear 1 to 2 scattering. This process conserves energy only. We recover the two KZ spectra $\omega^{-2}$ and $\omega^{-7/4}$ that correspond to the Bethe-Heitler (BH), single scattering, and Landau-Pomeranchuk-Migdal (LPM) multiple scattering regimes, respectively, separated by the scale $\omega_\BH\sim \hat q \ell^2_\mfp$, where $\hat q $ is the momentum space diffusion coefficient and $\ell_\mfp$ is the mean free path. Following Zakharov's approach, we show that the latter spectrum is nonlocal and hence can not be realized, whereas the former yields a logarithmically divergent energy flux and thus is marginally nonlocal. This means that up to a mild logarithmic factor the KZ spectrum $\omega^{-2}$ is physically relevant. If follows that at least in the LPM regime strongly asymmetric gluon splittings are expected to be the dominant processes at late times. In this limit, the collision integral reduces to a inhomogeneous diffusion-like equation.

As a result, the turbulent spectrum forms in the wake of a wave front that scales as $t^{1/2}$ in the ultraviolet with a diffusion coefficient of the order of the elastic diffusion coefficient $\hat q$. In this diffusion process energy is directly transferred from the source to the hard gluons via soft gluon absorption. On the other hand, radiation is suppressed due to the rapid fall off of the spectrum as long as the spectrum is steeper than $\omega^{-1}$. 

Because of the inhomogeneity of both the elastic and inelastic collision integrals in the diffusion approximation the resulting spectrum is not steady in the wake of the UV wave front and flattens as time goes on. A numerical simulation of the effective kinetic equation in the LPM regime shows the flattening of the spectrum, however the observed power exponents are closer to $-2$ than $-1$. 
 
Below the forcing scale, deep in the infrared regime, the dominant nonlocal processes are radiation and absorption of soft gluons by the source that balance each other due to the singular behavior of the radiation rate, forcing the soft sector to be in thermal equilibrium although the system is constantly driven out of equilibrium. This phenomenon is analyzed in detail in a recent work on thermalization of isotropic plasmas \cite{Blaizot:2016iir}. 

In Section \ref{sec:el-inel}, the interplay of elastic and inelastic processes is discussed. Numerical simulations of the complete collision integral, including elastic and inelastic processes, are performed, and we observe that the late time behavior is similar to that of inelastic alone, namely, the existence of a thermal bath below the forcing scale and a quasi-steady power spectrum close to $\omega^{-2}$ develops. As in the case of the inelastic scatterings alone, we argue that this spectrum corresponds to a long lived transient regime and that asymptotically a flattening of the spectrum toward the thermal one may occur, with a small deviation that would account for finite energy flux. 

Finally, in the last section we summarize and conclude.

\section{KZ steady state spectra in $\phi^4$ theory}
\label{sec:BN-equation}
Before we proceed with the analysis of wave turbulence in Yang-Mills theory, let us first address the somewhat academic example of $\phi^4$ scalar theory, to demonstrate how the power spectra derived from dimensional analysis in the introduction are indeed stationary solutions of the kinetic equation. The latter is 
of the form of the Boltzmann-Nordheim (BN) equation, which describes the transport properties of a gas of bosons that undergo elastic 2 to 2 scattering \cite{Nordheim}. It is widely used in the context of Bose-Einstein condensation (see Ref.~\cite{Pomeau} for instance). 

\begin{figure}[ht]
 \begin{center}
\subfloat[]{ \includegraphics[width=5cm]{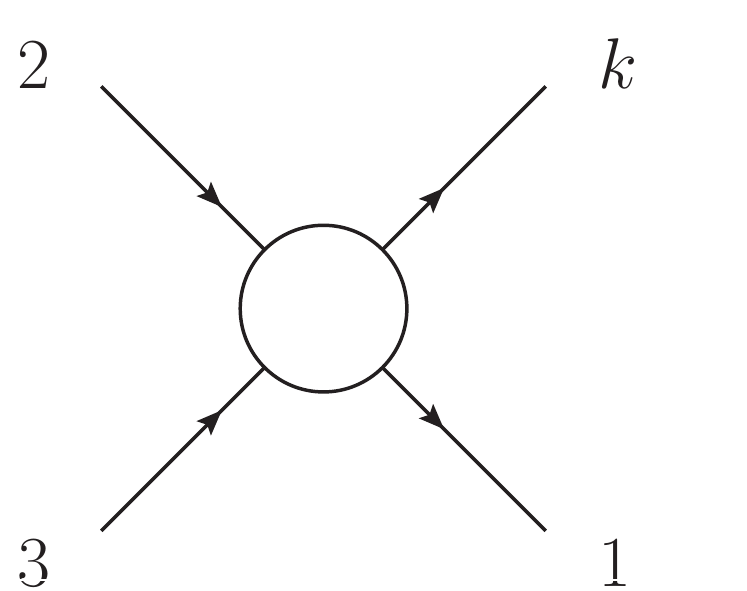} } \qquad \subfloat[]{ \includegraphics[width=5cm]{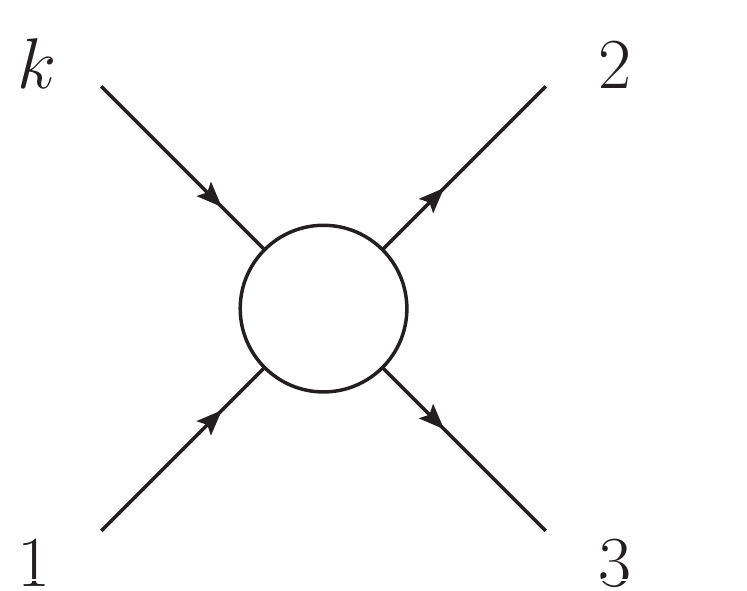} } 
 \end{center}
		\caption{Illustration of the 2 to 2 scattering process. The two contributions of \eqn{F-occup}, corresponding to the gain (left) and loss (right) terms, are depicted. }
		\label{matrix-2to2}
\end{figure}

Given the dispersion relation $\omega_k\equiv \sqrt{{\boldsymbol k}^2+m^2}$, where $m$ is the mass of the bosonic particles, the leading order contribution to the collision integral at small coupling (weak nonlinearity) is the 2 to 2 scattering process, even if there was a cubic interaction term in the Lagrangian.  Indeed, the 3-wave resonance condition, that is, $\omega_1+\omega_2 = \omega_3$ for ${\boldsymbol k}_1+{\boldsymbol k}_2={\boldsymbol k}_3$ for 1 to 2 particle decay (and the inverse process), is only fulfilled when all momenta are collinear and hence, is kinematically suppressed. Likewise, the 1 to 3 and 0 to 4 particles decay vanish and only the 2 to 2 matrix element contributes to the collision integral that takes the standard form 
\beq\label{BN-eq}
I^{2\to2}[n]=&&\frac{1}{2!}\int_{123} \frac{1}{2\omega_k}  |M_{23 \to 1k}|^2 \, F[n]\,(2\pi)^4\delta(k+k_1-k_2-k_3),
\eeq  
where $|M_{23 \to 1 k }|^2 = g^4$ is the (squared) matrix element for the elastic process $k_2+ k_3 \to k + k_1$ as illustrated in Figure~\ref{matrix-2to2} for the interaction Lagrangian density $g^2 \phi^4 /4!$, and 
\beq\label{F-occup}
F[n]=n_2n_3 (1+n_k)(1+n_1)-n_kn_1 (1+n_2)(1+n_3),
\eeq
where $n_i\equiv n({\boldsymbol k}_i)$.
Internal degrees of freedom like spin are averaged over for the measured boson $(k)$ and summed over for the integrated ones $(1,2,3)$. 
The symmetry factor $1/2!$, in \eqn{BN-eq},  accounts for the exchange of the incoming (outgoing) particles 1 and 2. Also, we have used the shorthand notation 
\beq
\int_i\equiv \int  \frac{\rmd^3 k_i}{2(2\pi)^3 \omega_i}
\eeq
for the invariant measure. For isotropic media the kinetic equation admits two integrals of motion, energy and particle number, respectively,
\beq
E=\int \frac{\rmd^3 k}{(2\pi)^3}\, \omega\, n_k\quad\text{and}\quad N=\int \frac{\rmd^3 k}{(2\pi)^3} \, n_k, 
\eeq
where $\omega\equiv \omega_k$. Using the symmetries of the 2 to 2 matrix element under the exchange of bosons in the initial and final state, and under time reversal, $|M_{12 \to 3k}|^2=|M_{21 \to 3k}|^2 =|M_{12 \to k3}|^2=|M_{3k \to 12}|^2$, and assuming that the integrant vanishes at infinite momenta such that there is no `` leakage'' of energy or particle number at infinity, it is easy to verify that $E$ and $N$ are conserved, that is,
\beq 
\dot N=\int \frac{\rmd^3 k}{(2\pi)^3} \frac{\del {n_k}}{\del t}= \int \frac{\rmd^3 k}{(2\pi)^3} I^{2\to2}_k[n] =0.
\eeq
for the particle number, and 
\beq
\dot E=\int \frac{\rmd^3 k}{(2\pi)^3} \,\omega \,\frac{\del {n_k}}{\del t}= \int \frac{\rmd^3 k}{(2\pi)^3} \omega I^{2\to2}_k[n] =0
\eeq 
for energy. Now, owing to the Boltzmann's $H$-theorem, an isolated system must evolve toward a state of maximum entropy, that corresponds to thermal equilibrium characterized by the Bose-Einstein (BE) distribution. At equilibrium, the collision integral vanishes $F[n]=0$ (as well as all fluxes), as a result of the detailed balance between the gain and loss terms in \eqn{F-occup}, independently of the form of the matrix element and in particular its degree of homogeneity, in contrast with the non-thermal fixed point. The occupation number is thus given by the BE distribution, 
\beq\label{BE-dist}
n_\text{BE} (k) =\frac{1}{\exp\left(\frac{\omega-\mu}{T}\right)-1 },
\eeq
where $T$ is the equilibrium temperature and $\mu$ the chemical potential. 
In the regime in which $n\gg 1$ (classical wave regime), one can neglect the quadratic terms in $n$ and Eq.~(\ref{F-occup}) becomes
\bel{F-classical}
F[n]\simeq n_2n_3 n_k+n_2n_3 n_1-n_1n_k n_2-n_1n_k n_3.
\eeq
In this case, the thermal fixed point reduces to the Rayleigh-Jeans distribution, that can be deduced from the BE distribution by taking the limit $\omega-\mu \ll T$,
\bel{RJ-dist-0}
n_\text{RJ} (k)= \frac{T}{\omega-\mu}.
\eeq
Let us turn now to the finite flux KZ solutions of \eqn{BN-eq} with $F[n]$ given by \eqn{F-classical} \footnote{Note that scale invariant KZ spectra exist only for homogenous collision integrals.}.
For scale invariant media, besides the thermal fixed point (\ref{RJ-dist-0}), Eq.~(\ref{BN-eq}) admits non-thermal stationary power solutions, known as the Kolmogorov-Zakharov (KZ) spectra, associated with  constant energy and particle fluxes. In general, KZ spectra are expected to form in open systems in contact with a source, which steadily injects energy into the systems, and sinks for dissipation. 

\begin{figure}[ht]
 \begin{center}
 \includegraphics[width=6cm]{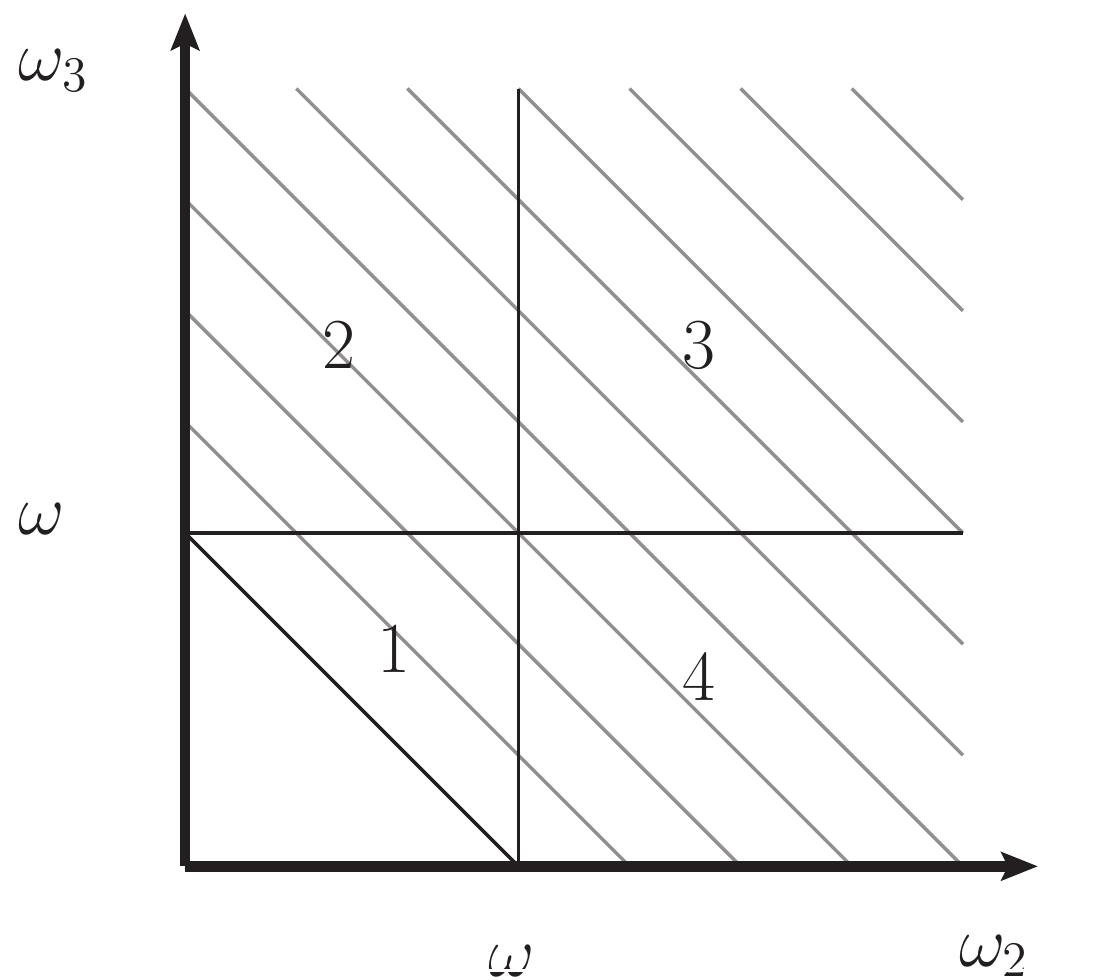} 
 \end{center}
\caption{The integration regions for the integral given in \eqn{coll-int-omega-2} where the lower left triangle is excluded. }
		\label{KZ-int-regions}
\end{figure}

For isotropic media $F[n]$ is independent of angles. Hence, one can integrate over the solid angle (see \ref{angular-integration} for details) and the collision integral, \eqn{BN-eq}, yields 
\beq\label{coll-int-omega}
I^{2\to2}[n]=\frac{1}{\omega^2}\int  \rmd\omega_1 \rmd \omega_2 \rmd \omega_3 \,U(\omega,\omega_1,\omega_2,\omega_3)\, F[n]\, \delta(\omega+\omega_1-\omega_2-\omega_3),\nn
\eeq
with
\bel{U-def}
U(\omega,\omega_1,\omega_2,\omega_3)= \frac{g^4 }{2(4\pi)^3} \text{min}(\omega, \omega_1,\omega_2,\omega_3),
\eeq
where  $\rmd \Omega_i$ ($i=1,2,3$) is the integration measure for the solid angle.
Note that $U(\omega,\omega_1,\omega_2,\omega_3)$ possesses the same symmetries as $|M_{k1\to23}|^2$, namely,
\bel{sym-U-22}
U(\omega,\omega_1,\omega_2,\omega_3)=U(\omega_1,\omega,\omega_2,\omega_3)=U(\omega,\omega_1,\omega_3,\omega_2)=U(\omega_2,\omega_3,\omega,\omega_1).\nn
\eeq
Integrating over $\omega_1$, \eqn{coll-int-omega} can be written as
\beq\label{coll-int-omega-2}
I^{2\to2}[n]=&& \frac{1}{\omega^2}\int _\Delta\rmd \omega_2\,\rmd \omega_3 \,U(\omega,\omega_1,\omega_2,\omega_3)\, F[n],
\eeq
where the integration domaine in the $(\omega_2,\omega_3)$ plane,
\bel{int-regions}
\Delta\equiv \Big\{ \omega_2+\omega_3>\omega, \omega_2>0, \omega_3>0\Big\},
\eeq
can be divided into four subregions (cf. Figure~\ref{KZ-int-regions}). 

Now, let us look for power law solutions of the form
\bel{power-spect}
n(\omega)\equiv A\, \omega^x,
\eeq 
where $A$ and $x$ are to be determined. As already discussed in the introduction, the spectra exponent is universal, and can be derived by dimensional analysis. The constant $A$, known as the KZ constant, depends however on the details of the collision integral. 
Using \eqn{power-spect} in \eqn{coll-int-omega-2}, the KZ spectra can be made manifest after performing the Zakharov conformal transformations (cf. \ref{Z-transformation}), that allow us to map all integration regions onto that where $\omega$ is the largest frequency of the quartet $(\omega_1,\omega_2,\omega_3,\omega)$. Thus, \eqn{coll-int-omega} yields
\beq\label{coll-int-trans}
I^{2\to2}[n]&& =  \int_0^\omega  \rmd \omega_3 \int_{\omega-\omega_2}^\omega\rmd \omega_2 \,U(\omega,\omega_1,\omega_2,\omega_3)\,\nn
&&\left[1+\left(\frac{\omega_1}{\omega}\right)^y-\left(\frac{\omega_2}{\omega}\right)^y-\left(\frac{\omega_3}{\omega}\right)^y\right] \,F(\omega,\omega_1,\omega_2,\omega_3),\nn
\eeq
with  $y=-3x-5-\nu$, where $\nu=1$  is the degree of homogeneity of $U$ and 
\beq
F(\omega,\omega_1,\omega_2,\omega_3)\equiv A^3 \omega^{x} \omega_1^{x}\omega_2^x \omega_3^x \left[ \omega^{-x}+\omega_1^{-x}-\omega_2^{-x}-\omega_3^{-x}\right],
\eeq
We can readily extract the power exponent for which $I^{2\to2}[n]=0$. First, we recover the thermal fixed points that correspond to the two limits of the RJ distribution (\ref{RJ-dist-0}), $
x = 1  \quad \text{and} \quad x = 0$.  For $y=1$  or  $y=0$, we obtain the KZ non-thermal fixed points, respectively, 
\bel{KZ-spect-4}
x = -\frac{5}{3}  \quad \text{and} \quad x = -\frac{4}{3},
\eeq
that were infered by dimensional analysis in the introduction.   
We turn now to the determination of the non-universal constant $A$. Note that since $\text{min}(\omega,\omega_1,\omega_2,\omega_3) = \omega_1$ in \eqn{coll-int-trans}, we have $
U \sim \omega_1 $. 
Now, introducing the scaling variables: $z_1=\omega_1/\omega$, $z_2=\omega_2/\omega$ and $z_3=\omega_3/\omega$, Eq.~(\ref{coll-int-trans}) yields
\beq\label{coll-int-omega-5}
I^{2\to2}[n]= \frac{g^4 A^3 }{2(4\pi)^3}  \, f(x)\,\omega^{3x+1},\,\nn
\eeq
where the dimensionless function $f(x)$ is given by
\bel{dimless-func}
f(x) &=&\int^1_0\rmd z_2\,\int_{1-z_2}^1 \rmd z_3 \,z_1^{x+1}z_2^x z_3^x  \nn 
&&\times \left(1+z_1^y-z_2^y-z_3^y\right)\left( 1+z_1^{-x}-z_2^{-x}-z_3^{-x}\right),\nn
\eeq
with $z_1=z_2+z_3-1$ from energy conservation.  It is straightforward to check that $f(x) $ converges on the KZ spectra (\ref{KZ-spect-4}) ensuring the locality of interactions for relativistic 2 to 2 scattering in $\phi^4$ theory. A more complete analysis would include the study of the stability of the KZ spectra under small perturbations, but this is beyond the scope of the present work. 

Let us turn now to the fluxes associated with the KZ spectra.  The energy and particle fluxes through a sphere of radius $\omega$ read, respectively,
\bel{P-def}
P(\omega) \equiv  \int\limits_{\omega_0<\omega} \frac{\rmd^3 \k_0}{(2\pi)^3} \omega_0 \frac{\del n(\omega_0) }{\del t}= \frac{1}{2\pi^2} \int_0^\omega \rmd\omega_0\, \omega_0^3\, I^{2\to 2}[n],
\eeq
and 
\bel{Q-def}
Q(\omega) \equiv  \int\limits_{\omega_0<\omega} \frac{\rmd^3 \k_0}{(2\pi)^3} \, \frac{\del n(\omega_0) }{\del t}= \frac{1}{2\pi^2} \int_0^\omega \rmd\omega_0\, \omega_0^2\, I^{2\to 2}[n].
\eeq
Inserting  \eqn{coll-int-omega-5} in \eqn {P-def} and integrating over $\omega_0$ assuming $3x+5>0$, one finds
\beq\label{energy-flux}
P(\omega) =- \frac{2 g^4 A^3}{(4\pi)^5} \,  \frac{\omega^{3x+5}}{3 x+5}\, f(x). \nn
\eeq
Evidently, the energy flux (\ref{energy-flux}) diverges in the limit $x \to  -5/3$, but at the same time $f(-5/3)=0$, yielding an undetermined form.  Using the L'Hopital rule, we find that the limit is finite and reads,
\beq\label{f-limit-P}
&&\lim_{x\to -5/3} \,\frac{f(x) }{3 x+5} =\frac{1}{3}\lim_{x\to -5/3}\,f'(x)  = \int^1_0\rmd z_2\,\int_{1-z_2}^1 \rmd z_3 \,z_1^{-2/3}z_2^{-5/3} z_3^{-5/3} \, \nn && \times\left(z_1 \ln z_1-z_2  \ln z_2-z_3  \ln z_3\right) \left( 1+z_1^{5/3}-z_2^{5/3}-z_3^{5/3}\right) \simeq 2.6\nn
\eeq
It follows that the energy flux (\ref{energy-flux}) is negative and independent of $\omega$. Therefore, it is directed toward increasing $\omega$ and thus, energy must dissipate in the ultraviolet. By solving Eq.~(\ref{energy-flux}) for the constant $A$ as a function of a constant flux of energy 
$
P(\omega) \equiv -P_f < 0,
$
we obtain the KZ constant associated with a direct energy cascade, 
\beq
n(\omega) = C_\KZ \, g^{-4/3} P_f^{1/3}\omega^{-5/3},  \quad \text{where }\quad C_\KZ= 39.2
\eeq
Note that on this KZ spectrum, the particle flux vanishes, i.e., $Q(\omega) =0$ (cf. \eqn{particle-flux}). 

Similarly to the energy flux, from \eqn{coll-int-omega-5} together with \eqn{P-def}, one finds
\beq\label{particle-flux}
Q(\omega)=- \frac{2 g^4 A^3}{(4\pi)^5} \,  \frac{\omega^{3x+4}}{3 x+4}\, f(x), 
\eeq
which for $x\to -4/3$ yields a constant flux with
\beq
\lim_{x\to -4/3}\,\,\frac{f(x) }{3 x+4}  =&&\int^1_0\rmd z_2\,\int_{1-z_2}^1 \rmd z_3 \,z_1^{-1/3}z_2^{-4/3} z_3^{-4/3} \nn
 && \times \ln\left(\frac{z_1}{z_2z_3}\right) \left( 1+z_1^{4/3}-z_2^{4/3}-z_3^{4/3}\right)\simeq -0.9\nn
 \eeq
In agreement with the heuristic argument given in the introduction for the direction of the fluxes, we see that the KZ spectrum $x=-4/3$ is associated with a positive particle flux, $Q(\omega)\equiv Q_f>0 $, namely, an inverse particle cascade. The corresponding KZ spectrum then reads
\beq
n(\omega) = C_\KZ\, Q_f ^{1/3}g^{-4/3}\omega^{-4/3},  \quad \text{where}\quad C_\KZ= 55.8.
\eeq
In this case, $P(\omega)=0$.

The convergence of the collision integral on the KZ spectra is a reflection of the local nature of the interactions in $\phi^4$ theory, namely, of the fact that the spectrum is determined by interactions of particles of comparable momenta. For wide enough inertial range, the locations of the source and sinks are then irrelevant and can be sent to 0 and infinity, respectively, for the energy cascade and conversely for the particle cascade.

In the next sections, we shall discuss the kinetic equations that describe a gas of gluons. We shall in particular see that interactions are strongly nonlocal and as a result the steady state spectra differ from the KZ spectra.

\section{Steady state spectra in YM theory: elastic processes}\label{elastic-YM}
Let us turn now to classical Yang-Mills (YM) theory.  In this section, we shall first discuss in detail elastic processes alone. 

At small coupling, $g\ll 1$, the collision integral can be expressed as a powers series in $g^2$. The leading order corresponds to elastic 2 to 2 gluon scattering process, whose (squared) matrix element reads
\beq\label{matrix-22-g}
|M_{23\to1k}|^2 \equiv 8 g^4 N_c^2  \left(3-\frac{tu}{s^2}-\frac{us}{t^2}-\frac{st}{u^2}\right).
\eeq
where
\bel{mandelstam}
s =(k+k_1)^2,\quad u =(k-k_3)^2,\quad  t =(k-k_2)^2,
\eeq
are the Mandelstam variables and  $k_i\equiv (\omega_i,\k_i)$ are 4-vectors. 
Similarly to $\phi^4$ theory, the degree of homogeneity of the matrix element (\ref{matrix-22-g}) is zero. Therefore, the latter is invariant under a uniform rescaling of all momenta. Hence, one could expect the power spectra (\ref{KZ-spect-4}) derived in the previous section to remain valid for elastic processes in Yang-Mills theories. However, as we shall shortly see, this is not the case since the dominant interactions are nonlocal in momentum space. 

 \begin{figure}[ht]
 \begin{center}
 \includegraphics[width=6cm]{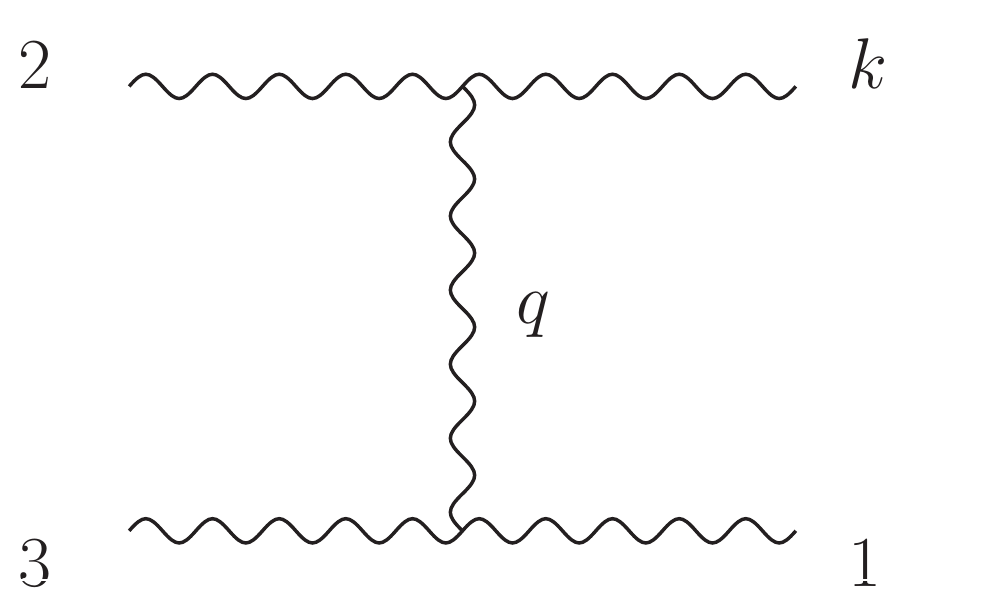} 
 \end{center}
\caption{Diagrammatic representation of the dominant contribution to 2 to 2 gluon scattering in the small angle approximation where the momentum transfer $q \ll k_i $ ($i=1,2,3$). }
		\label{Coulomb}
\end{figure}
In order to derive the actual turbulent spectrum, we first note that the matrix element (\ref{matrix-22-g}) possesses an infrared singularity in the $t$ and $u$ channels inherent to Coulomb scattering. This implies that a typical scattering is characterized by a small momentum exchange, resulting in  a small angle scattering as depicted in Figure~\ref{Coulomb}. Keeping only the singular part and using the symmetry of the kinetic equation under the exchange $u \leftrightarrow t$, \eqn{matrix-22-g} yields
\beq
|M_{23\to1k}|^2 \simeq  16 N_c^2 g^4 \frac{s^2 }{t^2}.
\eeq
This divergence should  in fact be cutoff at the Debye mass, i.e., $t \to t+m^2_D$. Therefore, to apply the small angle approximation, we assume (and will verify it {\it a posteriori}) that the screening mass lies beyond the inertial range where wave turbulence develop, that is, $k \gg m_D$. We can thus perform a gradient expansion around the Coulomb singularity. This power-like divergence present in the gain and loss terms cancels out yielding a milder logarithmic divergence (cutoff at the Debye mass). The details of the calculation is presented in  \ref{gradient-exp}. As a result, the 2 to 2 kinetic equation reduces to a Fokker-Planck (FP) equation (also known as the Landau equation) \cite{Landau1981,Blaizot:2013lga}
\beq\label{I-el-1}
I^{2\to 2}[n] \sim  I_\el [n] \equiv  4\pi^3  \abar^2\, \lambda \, \bnab \cdot \left[  I_ a\,\bnab  n(k)\, + \frac{\k}{|\k|}I_b \,n^2(k)\right],
\eeq
where $ \lambda\simeq \ln k/m_D$ is the Coulomb logarithm, which we assume to be a constant in what follows, $\abar\equiv  g^2N_c/4\pi^2$, and
\bel{Ia-Ib}
I_a \equiv \int \frac{\rmd^3 \p}{(2\pi)^3} n^2 (\p)\quad \text{and} \quad I_b \equiv 2 \int \frac{\rmd^3 \p}{(2\pi)^3} \frac{n(\p)}{|\p|}.
\eeq
The first term in \eqn{I-el-1} describes a process of diffusion in momentum space $k$ and the second term the drag. 

Recalling that we focus on isotropic media, it follows that 
Eq.~(\ref{I-el-1}) reduces to a one dimensional continuity equation, 
\beq\label{I-el-2}
 I_\el [n] \equiv  \frac{2\pi^2}{ \omega^2}\frac{\del }{\del \omega} Q(\omega), 
\eeq
where
\bel{flux-Q-FP}
Q(\omega) \equiv  \frac{1}{4}\hat q \,\omega^2 \left( \, \frac{\del }{\del \omega}  n(\omega)\, + \frac{n^2(\omega)}{T_\ast}\right),
\eeq
stands for the particle flux. Here, we have introduced the diffusion coefficient 
\bel{qhat-def}
\hat q \equiv 16 \pi^3 \abar^2 \lambda \, I_a,
\eeq
also known as the jet-quenching parameter \cite{Baier:1996sk},
and the effective temperature 
\bel{T-def}
T_\ast =\frac{I_a}{I_b}, 
\eeq
that equals the temperature at thermal equilibrium. 

Particle number conservation is straightforward owing to the fact that $\omega^2  I_\el [n]$ is a total derivative. 
Hence, when the flux vanishes at 0 and $\infty$, we have
\beq
\dot N = \int \frac{\rmd^3 \k}{(2\pi)^3} \, I_\el [n] = Q(\infty)-Q(0) = 0.
\eeq
Energy conservation relies, on the other hand, on the specific form of the integrals $I_a$ and $I_b$ that depend globally on the occupation number. We find
\bel{E-conserv-FP}
\dot E &= &\int \frac{\rmd^3 \k}{(2\pi)^3}  \,  \omega I_\el [n]= \int_0^\infty \rmd \omega \omega  \frac{\del }{\del \omega} Q(\omega)= - \int_0^\infty \rmd \omega Q(\omega)\nn
&=&  -\frac{1}{4}\hat q \int_0^\infty \rmd\omega \omega^2 \left( \, \frac{\del }{\del \omega}  n(\omega)\, + \frac{n^2(\omega)}{T_\ast}\right)\nn
&=& \frac{1}{4 I_a} \hat q  \left[ 2 I_a  \int_0^\infty \rmd\omega \omega  n(\omega)\, - I_b \int_0^\infty \rmd\omega \omega^2 n^2(\omega)\right],
\eeq
where we have repeatedly integrated by part. In the last line, one recognizes the integrals $I_b$ and $I_a$ defined in \eqn{Ia-Ib}, in the first and second terms, respectively, that is, 
\bel{Ia-Ib-iso}
 2 \int_0^\infty \rmd\omega \omega  n(\omega) = 2\pi^2 I_b\quad \text{and}    \quad \int_0^\infty \rmd\omega \omega^2  n^2(\omega) = 2\pi^2 I_a. 
\eeq
Therefore, we have $\dot E =0$. 

\eqn{I-el-2} was studied in the context of Bose Einstein condensation in an overpopulated system of gluons \cite{Blaizot:2013lga}. To accommodate the excess of gluons that do not fit the Bose Einstein distribution at equilibrium, a finite flux of particles builds up a $\omega=0$ to fill the condensate. In the next section, we shall see that an analogous phenomenon occurs in the presence of a steady source. 

In the classical regime, for a closed system, the particle flux vanishes identically on the thermal fixed-point (\ref{RJ-dist-0}) with $T=T_\ast$. However, an initial distribution with compact support in $\omega$ space never relaxes to thermal equilibrium with a finite value of the temperature. Instead, it evolves in a self-similar way characterized by a spreading of the distribution toward higher frequencies and asymptotically we must have $T_\ast=0$. This is a general feature of decaying wave systems. This can be seen by solving for the temperature the implicit equation \eqn{T-def} where the occupation number $n$ is given by the thermal distribution  \eqn{RJ-dist-0}. We find $T_\ast=T=0$. 

Of course, when $n\sim 1$ the classical approximation ceases to be valid and the complete quantum statistic must be taken into account. The Bose-Einstein distribution is then the true equilibrium distribution characterized by a finite temperature. 
\subsection{Stationary state solutions}
\label{el-forced-turb}
 
Recall that we are interested in the response of the system when energy is injected at the forcing scale $\omega_f$, at a constant rate $P_f=Q_f \omega_f$. If the interactions were local in momentum space one should expect the KZ spectra characterized by the exponents (\ref{KZ-spect-4}), to be realized in the inertial range. Obviously, those KZ spectra are not fixed points of \eqn{I-el-2}. Moreover, the integrals $I_a$ and $I_b$ diverge for any power spectrum and in particular on the KZ spectra $\omega^{-5/3}$ and $ \omega^{-4/3}$. Hence, these integrals strongly depends on boundary conditions, that is, on either the forcing or the dissipation scales.  This demonstrates nonlocality of interactions in momentum space for the 2 to 2 matrix element (\ref{matrix-22-g}). Note that the small angle approximation has nothing to do with the loss of the KZ fixed-points. As a counter example, Kats {\it et al} have shown that the stationary solutions of the non-relativistic Landau equation in the low occupancy limit are indeed given by the corresponding local KZ spectra $\omega^{-5/4}$ and $\omega^{-3/4}$  \cite{KKMN}.

In order to investigate the steady state spectra associated with a finite flux we add a spectrally narrow source term to the collision term (\ref{I-el-2}), 
\bel{Landau-source}
\frac{\del n(\omega)}{\del t} = I_\el[n]+ S(\omega),
\eeq
where the source term 
\beq\label{source}
S(\omega)= \frac{2\pi^2 Q_f}{\omega^2}\delta(\omega - \omega_f),
\eeq
is normalized to yield the particle flux as follows
\beq
Q_f =\int \frac{\rmd^3\k}{(2\pi)^3}\, S(\omega).
\eeq
In the presence of a source and sinks we expect the system to reach a stationary state after a transient regime. We find that  the stationary solutions of \eqn{Landau-source} obey a Ricatti type differential equation
\bel{Ricatti}
Q = \frac{1}{8\pi^2} \hat q \,\omega^2 \left(  \frac{\del }{\del \omega}  n(\omega)\, + \frac{n^2(\omega)}{T}\right) + Q_f \,\Theta(\omega - \omega_f)=\text{const}, 
\eeq
which necessarily implies a constant flux of particles. Hence, there is no solution explicitly related to energy conservation which is accounted for globally through a implicit dependence of the integrals $I_a$ and $I_b$ on $n$.  Curiously, as we shall shortly show, this fact implies the vanishing of the spectrum above the forcing scale where the energy cascade forms. 
 
The exact solution of \eqn{Ricatti} is derived in \ref{solution-FP} (cf. Eqs.~(\ref{g-sol-min}) and (\ref{g-sol-max})). Here, we only focus on the homogeneous (power like) solution, that can be inferred from the exact solution far from the source and sinks. For $\omega_\min \ll  \omega < \omega_f$, below the forcing scale, we obtain
\beq\label{left-spec-exact}
n(\omega) = \frac{A}{\omega}, \quad\text{with } \quad A = \frac{T_\ast}{2}\left(1+\sqrt{1+ \frac{32\pi^2 Q}{\hat q T_\ast}}\right),
\eeq
and for $ \omega_f < \omega \ll \omega_\max $, above the forcing scale, 
\beq\label{right-spec-exact}
n(\omega)= \frac{B}{\omega}, \quad\text{with }  \quad B = \frac{T_\ast}{2}\left(1-\sqrt{1+\frac{32 \pi^2 (Q-Q_f)}{\hat q T_\ast}}\right).
\eeq
Note that $A>B$ and thus, the solution is discontinuous at the forcing scale, $\omega_f$. In the absence of an extra boundary condition to fixe the value of $Q$, we conclude that it should depend on initial conditions and the location of the sinks. Moreover, since particle number must flow toward the infrared, we expect $Q\simeq Q_f>0$. As a result, the steady state solution for large inertial range simply reads
\beq\label{sol-st-FP}
n(\omega) = \begin{dcases} \,  \frac{1}{2} \left(1+\sqrt{1+ \frac{32 \pi^2 Q_f}{\hat q T_\ast}}\right) \frac{T_\ast }{\omega} \quad \text{for} \quad\omega<\omega_f.\\
\\
\qquad \qquad 0 \qquad \qquad\qquad \qquad\text{for} \quad\omega>\omega_f.
\end{dcases}
\eeq

Of course, energy is conserved and carried by the modes at the ultraviolet boundary, $\omega_\max$. The vanishing of the occupation number to the right of $\omega_f$ reflects the fact that  it must depend on the location of the sink $\omega_\max$ where energy is dissipated. Note also that $T_\ast$ and $\hat q $ depend implicitly on $n$. We shall discuss these points, and in particular the approach to the steady spectrum (\ref{sol-st-FP}), in more detail in what follows.

\subsection{Approach to the steady state}
\begin{figure}[ht] 
\centering
 \includegraphics[width=10cm]{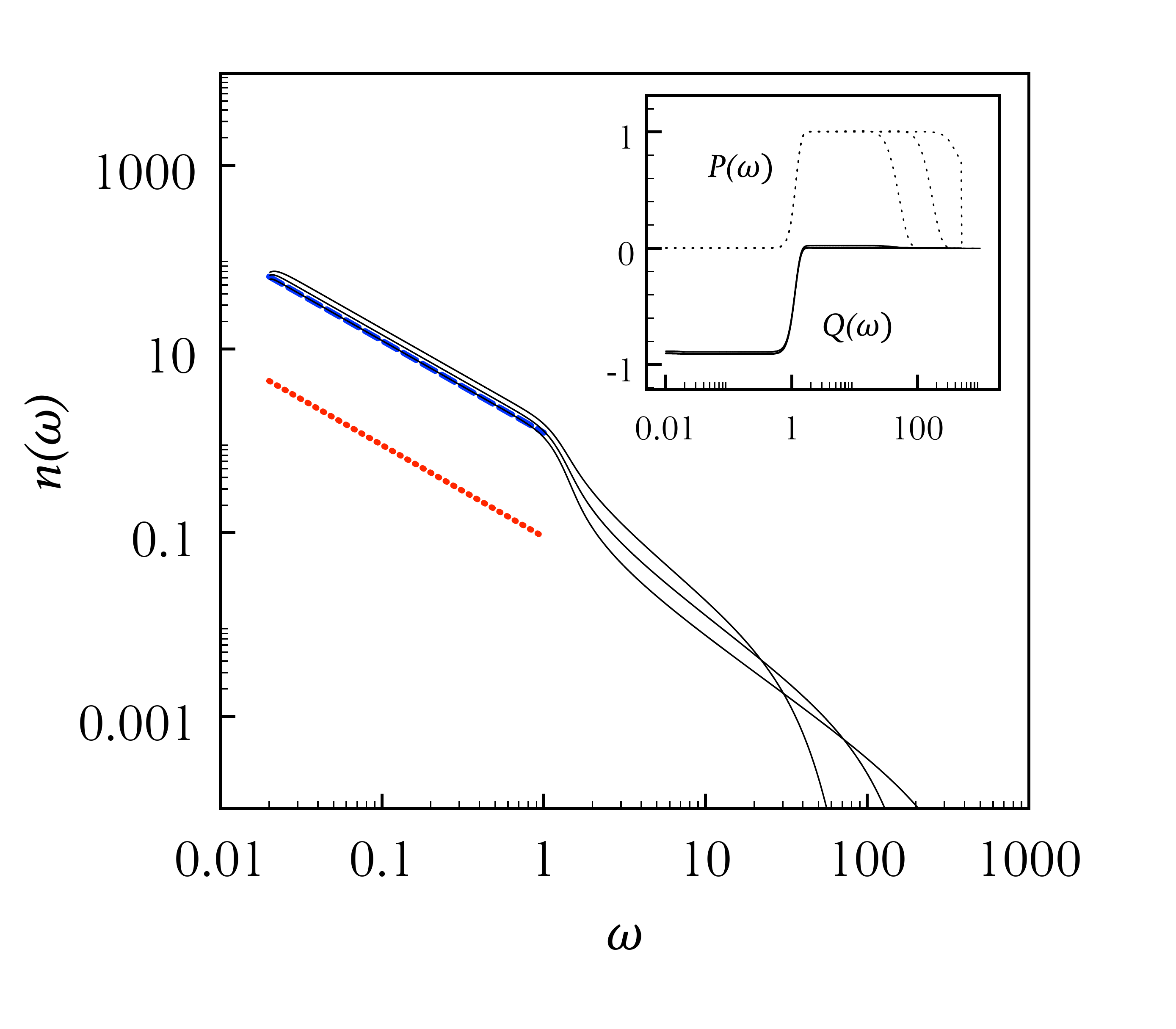}   
\caption{Numerical solution of the Fokker-Planck equation (\ref{Landau-source}) in the presence of a source at $\omega_f=1$, pumping particles at a rate $Q_f\simeq 0.9$. The occupation number at late times is depicted by full (black) curves corresponding to $\tau=$ 1018, 16286 and 256504, from top to bottom. The dashed (blue) curve is the asymptotic spectrum \eqn{sol-st-FP}. The dotted (red) curve corresponds to the thermal spectrum $T_\ast/\omega$ with $T_\ast\simeq 0.09$ (at $\tau=256504$) and shown as a reference. In the inset, the corresponding particle flux (full curves) and energy flux (dotted curves) are plotted. 
} 

\label{el-steady}
\end{figure}

Here, we shall perform numerical simulations (see \ref{numerics} for details) of the Fokker-Planck equation (\ref{Landau-source}) to study the approach toward the steady state (\ref{sol-st-FP}). The results will be plotted as a function of the rescaled time variable
\bel{tau-def}
\tau =4\pi^3  \abar^2\, \lambda \, t. 
\eeq
Furthermore, we set $\omega_f=1$ and $Q_f=1$ and the source is modeled by a narrow Gaussian 
\bel{Num-source}
S(\omega)=\frac{2 \pi^2 Q_f}{\sqrt{2\pi} \sigma \omega_f^2} \exp\left[(\omega-\omega_f)^2/2\sigma^2\right]
\eeq
with $\sigma^2=0.05$.  The sinks are accounted for by removing particles outside the inertial range, that is, letting $n(\omega)=0$, for $\omega > \omega_\max =500$ and $\omega < \omega_\min =0.02$. Because the sources are not exactly vanishing at $\omega=0$, \eqn{Num-source} yields the numerical values $Q_f\simeq 0.9$ and $P_f\simeq 1$ (see inset in Figure~\ref{el-steady}). 

The late time evolution of the full spectrum is shown in Figure~\ref{el-steady} for $\tau=$ 102, 1076  and 100057. 
We have checked that the spectrum to the left of the forcing scale ($\omega<\omega_f =1$) agrees with \eqn{sol-st-FP}  with $A\simeq 1.17$where $\hat q \simeq 4.60$, $T_\ast\simeq 0.09$ and  $Q=Q_f\simeq 0.9$ are computed dynamically, evaluated at $\tau=256504$ using the numerical solution for $n(\omega)$ in Eqs.~(\ref{Ia-Ib}), (\ref{qhat-def}) and (\ref{T-def}). To the right of the forcing scale, we see that the distribution decreases with time and thus, is expected to vanish asymptotically matching the stationary solution (\ref{sol-st-FP}).
 
\begin{figure}[ht] 
\centering

\includegraphics[width=9cm]{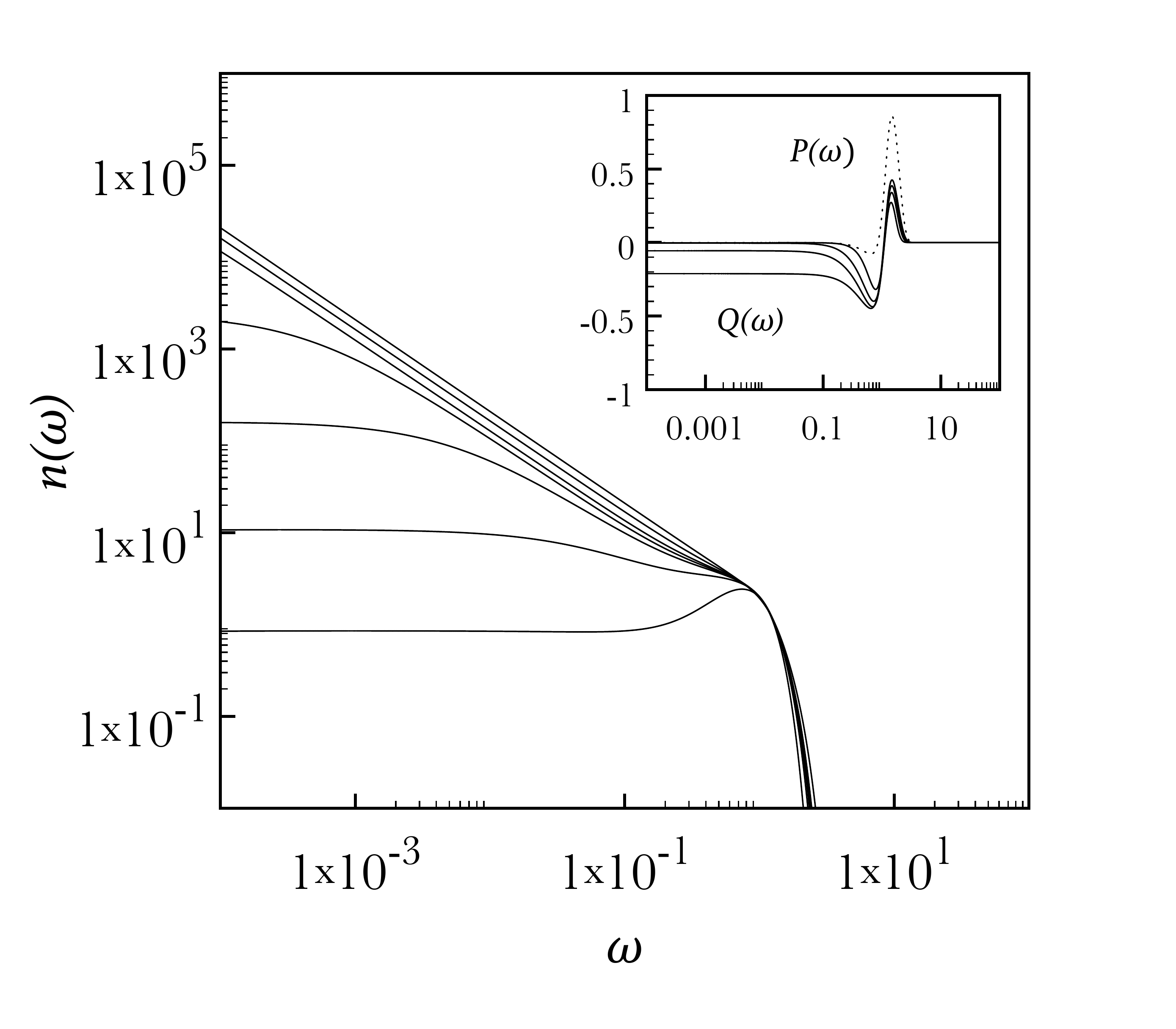} 
\caption{Time evolution of the occupation number before the onset of condensation. From bottom to top $\tau=$ 1.00, 1.20, 1.26, 1.28, 1.30, 1.35 and 1.50.  In the inset, the time evolution of the particle flux (full lines from top to bottom $\tau=$ 1.00, 1.20, 1.28 and 1.35 ) and energy flux (dashed line for $\tau=1.35$) (see also Figure~\ref{el-steady} for late time behavior), is shown. The asymptotic $\omega^{-1}$ spectrum forms after the onset of condensation, at $\tau \gtrsim 1.3$, that is when the particle flux $Q\neq 0$ at $\omega=0$. }  \label{el-BEC}
\end{figure}

Moreover, we confirm that $Q\simeq Q_f - 0.9$ and $P\simeq 0$, to the left of the forcing scale as shown in Figure \ref{el-steady} (inset), which corresponds to the inverse particle cascade. Conversely, to right of the forcing scale, we should have $Q\simeq 0 $ and $P\simeq P_f$.  
To the right of the forcing scale, $Q\simeq 0$ and $P \simeq P_f \approx 1$, reflecting the energy transfer to the ultraviolet. 
Also, we have also analyzed the early times dynamics. In the infrared sector, i.e. $\omega<\omega_f$, the power spectrum $\omega^{-1}$ can contain a finite number of particles, thus, it is of finite capacity. A finite flux of particle fills the entire spectrum all the way down to $\omega=0$ (in the absence of damping) in a finite amount of time $t_c$, which corresponds  to the time it takes for a finite flux of particle to develop at the origin. To see this, we have plotted the evolution of spectrum before the onset of condensation in Figure~\ref{el-BEC}.

This phenomenon is akin to Bose-Einstein condensation and has been discussed in Ref.~\cite{Blaizot:2011xf,Blaizot:2013lga} (see also \cite{Pomeau} for a detailed discussion of BEC in the framework of the Boltzmann-Nordheim equation).
In this case the steady state is not sensitive to position of the infrared sink $\omega_\min \ll \omega_f$. 

Because the occupation number is expected to drop to the  right  of the forcing scale at late times, it is reasonable to assume and verify it afterwards that the section $\omega>\omega_f$ does not contribute asymptotically to the integrals $I_a$ and $I_b$.  Under this assumption we have 
\beq\label{Ia-left}
I_a \sim \int_0^{\omega_f} \rmd \omega \omega^2 n^2 \sim A^2\omega_f \quad\text{and}\quad
I_b \sim \int_0^{\omega_f} \rmd \omega \omega n \sim A\omega_f,
\eeq
where we have used \eqn{left-spec-exact} in \eqn{Ia-Ib-iso}.
Therefore, we have, parametrically, $A\sim T_\ast$.
However, because there is a finite flux of particles, we must have $A>T$. Putting all the pieces together, Eqs.~(\ref{left-spec-exact}) and (\ref{Ia-left}) yield
\beq\label{factor-A-res}
n(\omega)=\frac{A}{\omega}  \quad\text{with } \quad A\sim \frac{Q_f^{1/3}}{ \bar\alpha^{2/3} \omega_f^{1/3} }.
\eeq
This spectrum corresponds to the nonlocal turbulent spectrum associated with the inverse particle cascade. The nonlocality is reflected in the explicit dependence of the occupation number on the forcing scale. 

To right of the forcing scale the time evolution of the spectrum is controlled by a diffusion process toward to ultraviolet. As shown in Figure~\ref{el-steady}, a power spectrum forms in the wake of a wave front that increases with time as 
\beq\label{front-steady}
\omega_\ast(t) \sim \sqrt{\hat q  t},
\eeq
with
\beq
\hat q \sim \bar\alpha^2 A^2\omega_f \sim  \bar\alpha^{2/3} Q_f^{2/3} \omega_f^{1/3}.
\eeq
It remains to analyze the evolution of the constant $B$ and verify that $B=0$ is indeed an attractor.  
Note that because the slope of the spectrum is larger that $-4$, the total energy contained in the spectrum is dominated by the ultraviolet cutoff $\omega_\ast$. Hence, the total energy injected in the system at the time $t$ reads
\bel{energy-param}
E \sim P_f t \sim \int_0^{\omega_\ast(t)} \rmd \omega \omega^3 n^2 \simeq \omega^3_\ast(t) B.
\eeq
Since $\omega_\ast\sim t^{1/2}$, it follows that the spectrum to the right of the forcing scale decreases with $t$ as $B\sim t^{-1/2}$. The wave front reaches the sink, $\omega_\ast \sim \omega_\max$, at the time $t_\max\sim \omega^2_\max /\hat q$. So, letting $t=t_\max$ in \eqn{energy-param} and solving for $B$, we obtain
\bel{B-factor}
n(\omega)=\frac{B}{\omega}  \quad\text{with } \quad B\sim \frac{Q_f^{1/3} \omega_f^{2/3}}{ \bar\alpha^{2/3} \omega_\max}.
\eeq
Therefore, the turbulent spectrum is suppressed to the right of the forcing scale in the limit $\omega_\max \to \infty$. 
As a consistency check, it can be verified that the contribution of the section $\omega> \omega_f$ to the integrals $I_a$ and $I_a$ is suppressed in for the former and  independent of  $\omega_+$ for the latter, as assumed in \eqn{Ia-left}.

\section{Steady state spectrum in Yang-Mills: inelastic processes}
\label{inelastic-YM}
Higher order corrections to the collision integral involve inelastic processes in general, and hence do not conserve particle number.  
Naively, one expects these correction to be suppressed owing to the fact that an extra gluon radiation (absorption) costs a factor $g^2 n \ll 1$. However, as alluded to in the introduction, inelastic processes are not suppressed by higher powers of the coupling constant and must be taken into account for a complete treatment. In effect, the 2 to 3 gluons matrix element possesses a collinear divergences associated with Coulomb scattering and gluon branching, which is be regulated by the Debye mass and the gluon thermal, respectively. This effect is non-perturbative, as it results from the resummation of hard loops in the gluon propagator.  Contrary to the elastic scattering this power divergence does not reduce to a logarithmic one thanks to cancellations between the gain and loss terms. The angular (or transverse momentum) integration generates a factor $1/m^2$ in the collision integral and because $m^2\sim g^2 n$ (cf. \eqn{therm-mass}), inelastic 2 to 3 processes turn out to be parametrically as important as elastic 2 to 2 processes (cf. \ref{app:inelatis-2to3}). 

\begin{figure}[ht]
\centering
\subfloat[]{ 
\includegraphics[width=5.5cm]{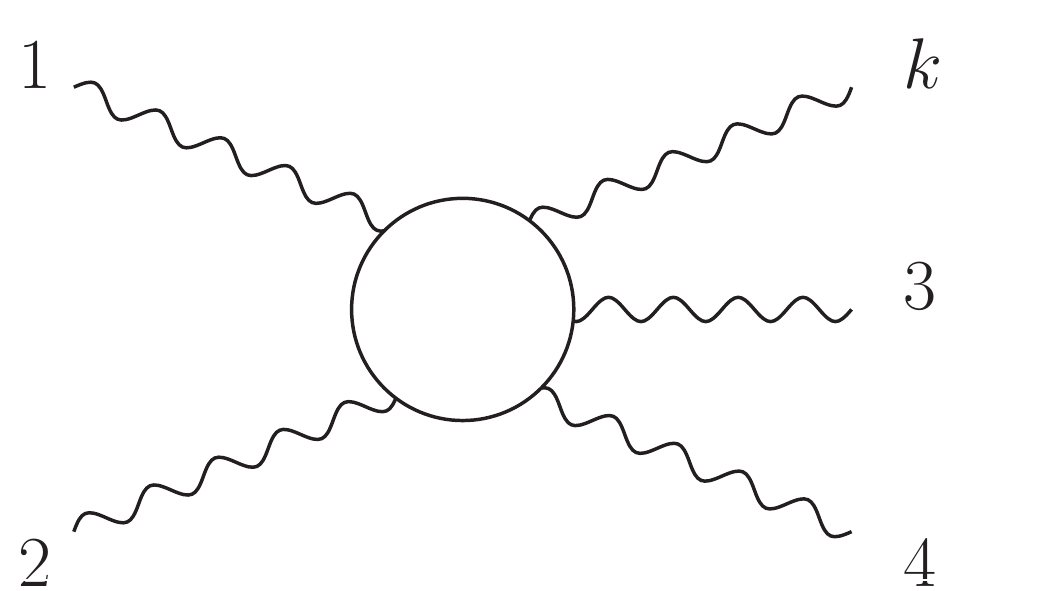} 
 } \quad
\subfloat[]{ 
\includegraphics[width=5.5cm]{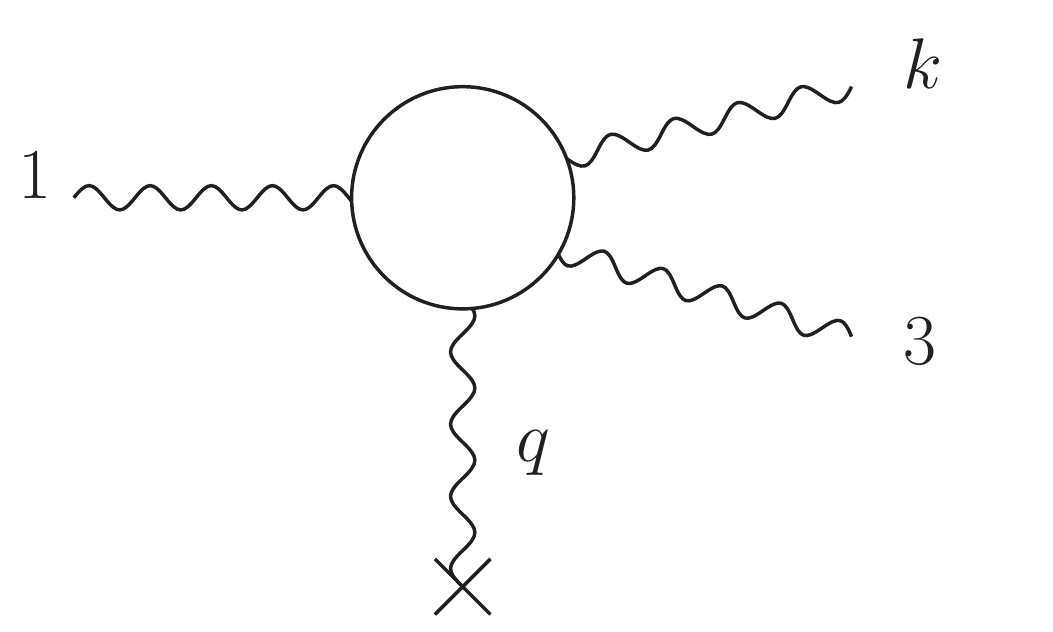}  }
 \caption{(a): Diagrammatic representation of the 2 to 3 matrix element. (b): The leading contribution to the 2 to 3 process corresponding to small momentum exchange $q\ll \omega_i$ ($i=1,2,3,4,k$). In this limit, gluons 1, 3 and $k$ are quasi-collinear. The cross at the end of the vertical gluon line illustrates the fact that the collinear gluons 2 and 4 are integrated over so that their respective momenta do not affect the (upper) branching process. } \label{inel-leading}
\end{figure}

Hence, in 2 to 3 processes, depicted in Figure~\ref{inel-leading} (a), one of the out going gluons tends to be collinear to one of the incoming ones, for instance gluons 2 and 4. Conservation of the total 3-momentum implies that the three remaining gluons, 1,3 and $k$, are also nearly collinear. So effectively, keeping only the singular part this process can be regarded as a quasi-collinear 1 to 2 splitting (of gluon 1 into 3 and $k$) induced by the small angle scatterings off gluons (2) whose momentum is integrated out (a diagram representing this process is shown in Figure~\ref{inel-leading} (b)). 

This quasi-collinear splitting was largely discussed in the context of in-medium radiative energy loss  (see Refs.~\cite{Mehtar-Tani:2013pia,Blaizot:2015lma} for a recent review and references therein) developed in the mid 90's then applied to the problem of thermalization of the quark-gluon plasma where the effective theory used in this work was proposed \cite{Baier:2000sb,Arnold:2002zm}. 

A standard approach that accounts for the small angle scattering by construction consists in computing the 1 to 2 splitting in the presence of a stochastic background field, $A_0^\mu(q)$ generated by the hard gluons, with $q\sim m$, whose two-point correlation function is given by
\bel{2A-correlator}
\langle A_0^{\mu,a}(Q) A_0^{\ast \nu, b}(Q') \rangle \equiv (2\pi)^4 \delta^{(4)}(Q-Q')\,\frac{\delta^{ab}}{g^2 N_c}  v^\mu v^\nu \, \cC(q_\perp),
\eeq
where for an isotropic plasma
\beq
{\cal C}(q_\perp) \equiv \frac{g^2 N_c T_\ast m_D^2}{q_\perp^2 (q_\perp^2+m_D^2)}, 
\eeq
in the Hard-Loop approximation \cite{Aurenche:2002pd,Arnold:2002zm}. Here, the momentum transfer  $q_\perp$ is transverse to the momenta of the collinear gluons in the 1 to 2 process. The $v^\mu v^\nu$ factor in the r.h.s. of \eqn{2A-correlator}, where $v^\mu \equiv k^\mu/|\k|$ is the gluon velocity, stems from the eikonal coupling (small angle scattering) of high energy gluons, with the soft collective modes of the background field $q \sim m \ll k$. 

Integrating the function $\cal C$ over $q_\perp$ yields the inverse elastic mean-free-path, 
\bel{mfp}
\ell^{-1}_\text{mfp} \sim \int_\qp \cC(q_\perp) \sim g^2 T_\ast, 
\eeq
where we have used the shorthand notation  $\int_{\qp} \equiv \int \rmd^2\q/(2\pi)^2$. The effective temperature $T_\ast$ and the Debye mass $m_D$ are given by \eqn{T-def} and \eqn{therm-mass}, respectively. 

\begin{figure}[ht] 
\begin{center} 

\includegraphics[width=9cm]{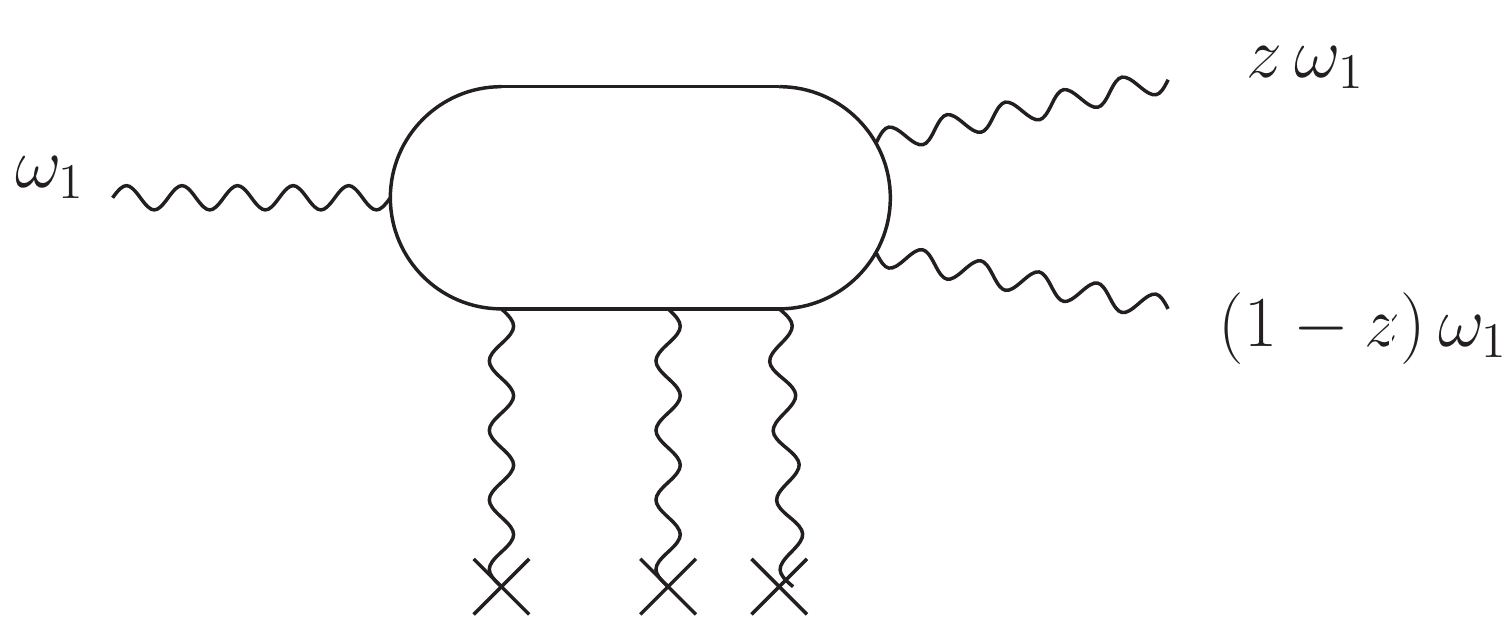} 
 
 \caption{Similar to Figure~\ref{inel-leading}(b) in the regime of multiple small momentum exchanges, that is the LPM regime. Here only three scatterings are depicted. } \label{inel-LPM}
\end{center} 
\end{figure}

From the uncertainty principle, the time it takes for a gluon $\omega$ to split into two gluons sharing fractions $z$ and $1-z$ of its energy reads 
\bel{br-time}
t\sim \frac{z(1-z)\omega}{ k^2_\perp},
\eeq 
where $k_\perp$ is the transverse momentum generated in the branching process. 
When the branching process is triggered by many coherent scatterings the transverse momentum squared accumulated is typically  $k_\perp^2 \sim \hat q \, t $, which together with \eqn{br-time} yields the coherence time
\beq
t = t_\coh \sim\sqrt{\frac{z(1-z)\omega }{\hat q }}.
\eeq
There exist two regimes that are characterized by the largest of either $\ell_\mfp$ or $t_\coh$, and which we summarize in the following:
\begin{enumerate}
\item The Bethe-Heitler (BH) regime: when $t_\coh \ll \ell_\mfp$, that is, $\omega \ll \omega_\BH\sim \hat q\, \ell_\mfp^2$, the collinear branching involves a single scattering. Using  Eqs.~(\ref{qhat-def}), (\ref{T-def}) and (\ref{mfp}), we find that $\omega_\BH$ depends only on the integrals $I_a$ and $I_b$ (cf. \ref{Ia-Ib}),
\bel{ome-BH}
\omega_\BH \sim \frac{I_b^2}{I_a}.
\eeq
In this regime, the matrix element squared can be approximated by the collinear limit of the 2 to 3 matrix element with proper treatment of screening effects (cf. \ref{app:branching-rate}). The rate for the branching of a gluon of frequency  $\omega_1=\omega/z$ into two offsprings with frequencies $\omega $ and $\omega_1=(1-z)\omega/z$ as shown in Figure~\ref{inel-leading}(b), scales as $ \cK_\BH  \sim \bar \alpha/ \ell_\mfp$ (see \eqn{K-BH}). 
\item The Landau-Pomeranchuk-Migdal (LPM) regime: when $t_\coh \gg \ell_\mfp$ or  $\omega \gg\omega_\BH$, multiple scatterings off the medium constituents occur during the branching process depleting the inelastic rate compared to the incoherent BH process (cf. Figure~\ref{inel-LPM}). In this case,  the matrix element includes singular contributions from higher order processes of the form $N+1$ to $N+2$.  In this coherent limit, the branching rate scales as $  \cK_\text{LPM}  \sim \bar\alpha / t_\coh$. In turns out that in this case the branching rate is only logarithmically sensitive to the screening mass. 
\end{enumerate}

The branching rate, first derived by Baier {\it et al} \cite{Baier:1996kr} and Zakharov \cite{Zakharov:1996fv}, can be easily recovered from the literature \footnote{Different notations were adopted in the literature making it difficult to connect directly to the present work which is based on the derivation presented in \cite{Blaizot:2012fh}. The facilitate the task we refer the reader to Ref.~\cite{Arnold:2008iy}  where a useful dictionary relating the various approaches is provided.}. It reads,\footnote{The branching rate $\cK(z)\equiv \cK(z,\omega_1)$ is an implicit function of the energy of the parent gluon $\omega_1$. }
\bel{br-rate}
{\cal K}(z)\equiv \frac{2\alpha_s}{z^2(1-z)^2\omega_1^2} P(z) \text{Re} \int_0^\infty \rmd \tau  \int_{\p_\perp,\q_\perp} (\p_\perp\cdot \q_\perp) S(\p_\perp, \q_\perp,z,\tau),\nn
\eeq
where,
\beq
P(z) = 2 N_c \left[\frac{1-z}{z}+\frac{z}{1-z}+z(1-z)\right], 
\eeq
is the Altarelli-Parisi gluon-gluon splitting function. The Green's function $S$ obeys the two dimensional Schr\"odinger equation 
\beq\label{S-green}
&& \left[i\frac{\del}{\del \tau } -\frac{\p_\perp^2+m_z^2}{2z(1-z)\omega_1 }\right]S(\p_\perp,\q_\perp,z,\tau)=\nn
&&  + \frac{i}{2}\int_\lp\left[\tilde \cC(\lp/(1-z))+\tilde\cC(\lp/z)+\tilde\cC(\lp) \right] S(\p_\perp-\lp,\q_\perp,z,\tau),\nn 
\eeq
with the initial condition $S(\p_\perp,\q_\perp,z,\tau)= (2\pi)^2 \delta^{(2)}(\pp-\qp)$. Here, $m_z^2 =(1-z+z^2) \, m^2$ where $m^2$ is the  gluon thermal mass squared (cf.~\eqn{therm-mass}) and 
\beq
\tilde \cC(\lp) \equiv \cC(\lp)  -(2\pi)^2\delta^{(2)}(\lp) \int_{\l'_\perp} \cC(\lp') .
\eeq
In \eqn{br-rate}, $\tau$ stands for the time between the branchings in the amplitude and the complex conjugate amplitude where only the transverse momentum of gluon $z$ is measured and $q_\perp$ and $p_\perp$ are respectively its initial and final transverse momenta relative to the parent gluon. The Green's function (\ref{S-green})  describes the in-medium propagation, characterized by the differential scattering rate $\cC$ of the system made of the two offspring gluons (carrying the energy fractions $z$ and $1-z$ of the parent gluon) in the amplitude and the parent gluon in the complex conjugate amplitude (assuming that the branching in the amplitude occurs first).

Analytic solutions of \eqn{S-green} can only be obtained in the limiting cases aforementioned. This is worked out in \ref{app:branching-rate}. In the Bethe-Heitler regime the branching kernel reads
\bel{K-BH}
{\cal K}(z) \simeq \frac{\pi \bar\alpha^2 \, C\,T_\ast }{N_c} P(z),
\eeq
where $C\simeq 1.83$ and in  the LPM regime it reads
\bel{K-LPM}
\cK(z)= \bar\alpha P(z)\sqrt{\frac{\hat q (1-z+z^2)}{z (1-z)\omega_1} }. 
\eeq

Let us turn now to the effective kinetic equation corresponding to the quasi-collinear 1 to 2 process. In general, the 3-wave collision integral takes the form 
\beq\label{c-inel-2}
I_\inel[n] &=&\frac{1}{\omega^2}\int \left ( R_{12\omega} +R_{1\omega 2} - R_{\omega 12}\right)\, \rmd\omega_1 \rmd \omega_2\, 
\eeq
where
\beq
R_{12\omega}\equiv V_{12\omega}\, F_{12\omega} \,  \delta(\omega_1-\omega_2-\omega)
\eeq
with
\beq
F_{12\omega} &=& n_1 (n_2+1) (n_\omega+1) - (n_1+1) n_2 n_\omega\nn
&\simeq & n_1 n_2 +(n_1 - n_2 )n_\omega,
\eeq
in the overoccupied case, i.e., $n\gg 1$.\footnote{The underoccupied case was discussed in Refs.~\cite{Blaizot:2013hx} and \cite{Blaizot:2015jea}, in the context of parton cascade evolution in a dense medium. There, it is shown that an inverse energy cascade forms, that is characterized by (local) democratic branchings with KZ exponent $-7/2$ ( or $-1/2$ for the distribution $D(\omega) \equiv \omega \rmd N/\rmd \omega$).}. The prefactor $1/\omega^2$  ensures energy conservation:
\beq
\frac{\del  E}{\del t} \equiv  \frac{1}{2\pi^2}\int_0^\infty \rmd \omega \, \omega^3 I_\inel[n] = 0. 
\eeq
The matrix element $V$ relates to the branching rate $\cK$ as follows,
\beq
V_{12\omega} \equiv \frac{1}{2}\omega_1 \cK \left( \frac{\omega}{\omega_1}\right),
\eeq
Using the symmetry in the exchange $1 \leftrightarrow 2$, we readily see that the first two terms in the r.h.s of \eqn{c-inel-2} are equal, that is, $R_{12\omega}=R_{1\omega 2}$. In the last term, multiplying the integrant by $1\equiv (\omega_1+\omega_2)/\omega $, then replacing the dummy variable $\omega_2$ by $\omega_1$ in the second term (owing to the aforementioned symmetry), amounts to the following replacement 
\beq
 V_{\omega 12}  \,  \to \,  \frac{2 \omega_1}{\omega }\, V_{\omega 12}. 
\eeq
Thereby, \eqn{c-inel-2} can be rewritten as
\beq \label{c-inel-3}
I_\inel[n] &=&\frac{2}{\omega^2}\int  \,  \left[ V_{12\omega}\, F_{12\omega} \,  \delta(\omega_1-\omega_2-\omega)\right. \nn
&&-\left. \frac{\omega_1}{\omega }V_{\omega 12}\, F_{\omega 12} \,  \delta(\omega-\omega_1-\omega_2) \right] \rmd\omega_1 \rmd \omega_2\nn
\eeq
After integrating over $\omega_1$ using the delta functions and defining the kernel:
\beq
K(\omega,\omega_2)\equiv \omega_1 \omega \cK\left( \frac{\omega}{\omega_1}\right) = 2 \omega V_{12\omega}.
\eeq
one can express \eqn{c-inel-3} as follows (with $q = \omega_2$),
\bel{c-inel-4}
I_\inel[n] &=& \frac{1}{\omega^3} \left[ \int_0^\infty \rmd q K(\omega,q) F(\omega+q, \omega)\right. \nn
&&\left. \qquad\qquad -\int_0^\omega \rmd q K(\omega-q,q) F(\omega, \omega-q) \right], \nn
\eeq
Here, it is understood that 
\bel{F-inel-def}
F(\omega+q, \omega) \equiv F_{\omega+q, \omega,q}=n(\omega+q) n(\omega) + \left[n(\omega+q)-n(\omega) \right] n(q).
\eeq 
and similarly for $F(\omega, \omega-q)$.

\begin{figure}[ht] 
\begin{center} 

\includegraphics[width=9cm]{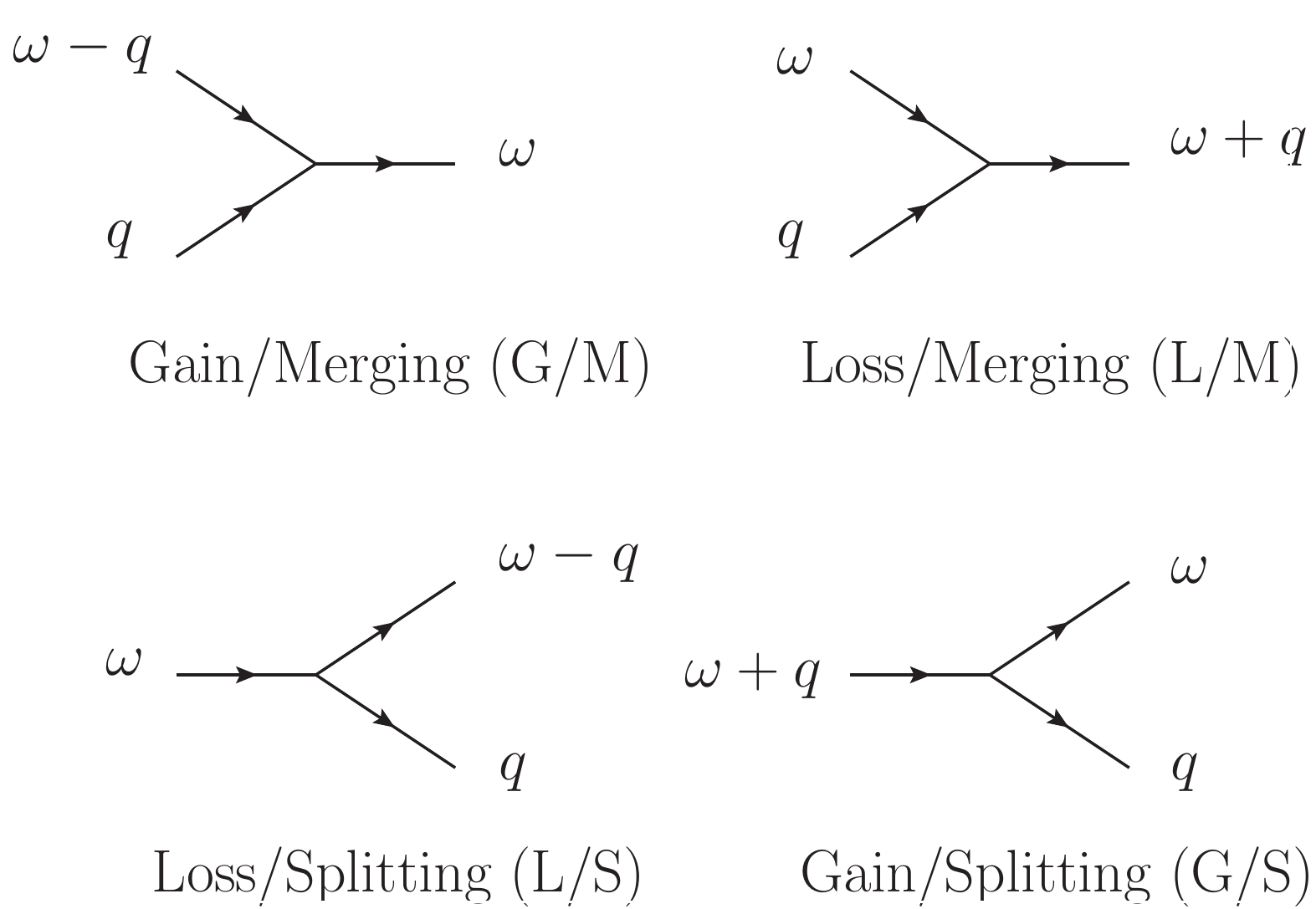} 
 
 \caption{The four contributions to the three wave collision integral.  } \label{gain-loss-1}
\end{center} 
\end{figure}

\begin{figure}[ht] 
\begin{center} 

\includegraphics[width=8cm]{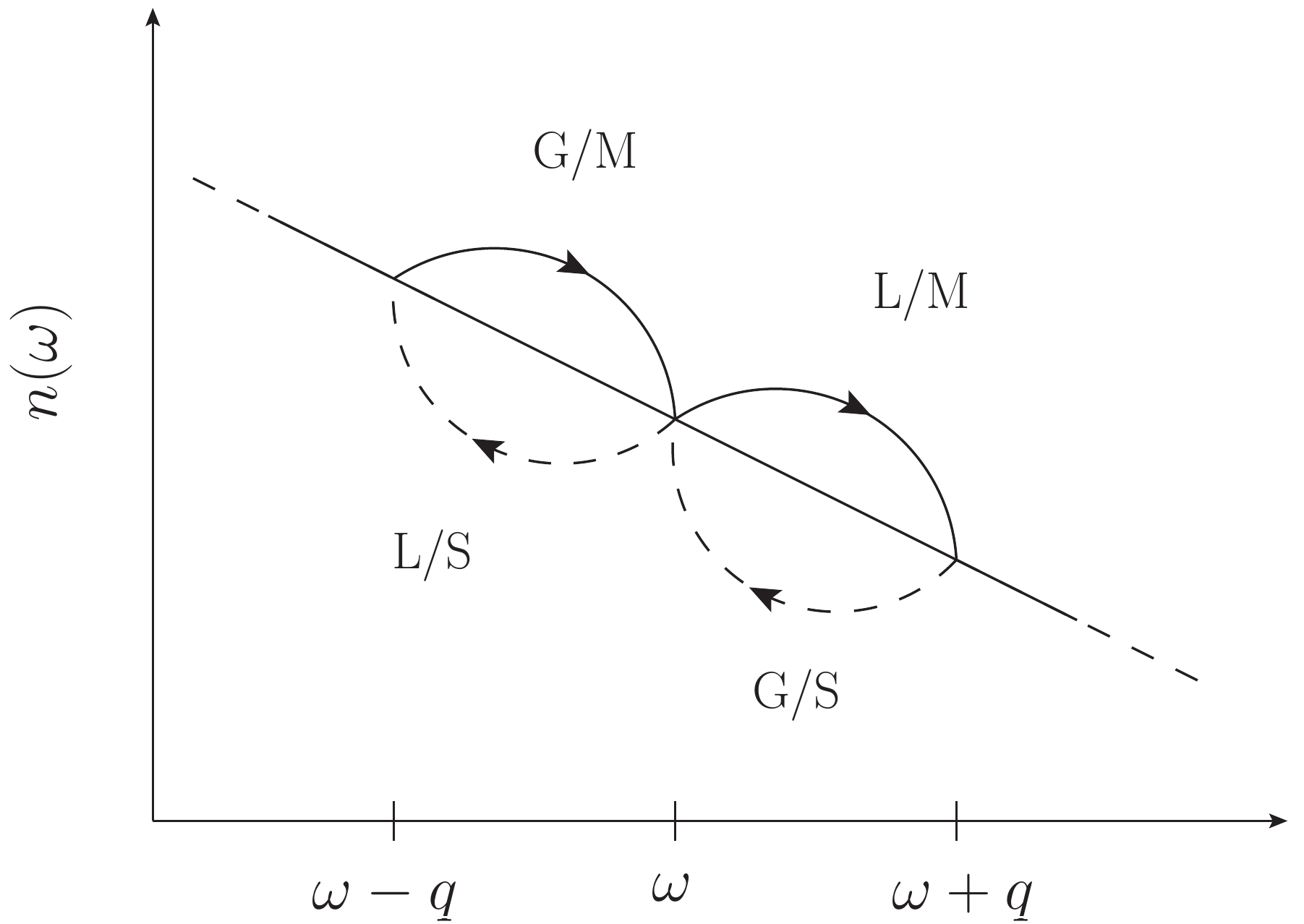} 
 
 \caption{A sketch of the occupation number with an illustration of the four processes (cf. Figure~\ref{gain-loss-1}) involved in its time evolution: gain and loss by merging and splitting are depicted, respectively,  by G/M and L/M (full arrows), and by G/S, L/S (dashed arrows). } \label{gain-loss-2}
\end{center} 
\end{figure}

A few remarks are in order. \eqn{c-inel-4} contains four contributions. The first term corresponds to rate for producing a gluon $\omega$ from a higher excitation $\omega+q$. It involves a gain term (positive part of \eqn{F-inel-def}) that enhances the occupancy at frequency $\omega$ by branching of a gluon $\omega+q$ into two gluons $\omega$ and $q$,  and a loss term that corresponds to the inverse process, that is,  gluon the merging of gluon $\omega$ with a gluon $q$ to from a the harder gluon $\omega+q$. The latter contribution is related to the negative part of \eqn{F-inel-def}. The second term in \eqn{c-inel-4} accounts for interactions between the $\omega$ bin with lower frequencies, it is related to the first term through the transformation $\omega \to \omega+q$. 
A depiction of these four components is shown in Figs.~\ref{gain-loss-1} and \ref{gain-loss-2}. 


\subsection{Nonlocality of KZ spectra for the effective 1 to 2 gluon process }\label{nonlocal-KZ}
As we have discussed above the kernel $K$ is not scale invariant due to the existence of two regimes, the BH and LPM regimes that are separated by the frequency $\omega_\BH$. Nevertheless, assuming that either $\omega\ll \omega_\BH $ or $\omega\gg \omega_\BH $, the Kernel becomes a homogeneous function of the frequencies and one can look for KZ spectra in the respective regimes.

In order to generically study the fixed points of \eqn{c-inel-4} it is useful to parametrize the Kernel such that it includes both cases. In addition, and without loss of generality we shall consider a simplified kernel that captures correctly the limit $z\to 0 $ and $z \to 1$, by making the replacement $P(z) \to 2N_c /z(1-z)$ and $1-z-z^2\to 1$. Hence, we define 
\bel{kernel-beta}
K(\omega,q) \equiv \kappa\,\frac{(\omega+q)^{3+\beta}}{\omega^{\beta}q^{1+\beta}}\,.
\eeq
From Eqs.~(\ref{K-BH}) and (\ref{K-LPM}), we read off $\beta=0$ and $\kappa \equiv 2 \pi \bar\alpha^2 C T_\ast$ in the BH regime, and  $\beta=1/2$ and $C=2 N_c \bar\alpha \sqrt{\hat q }$, in the LPM regime.

Because the quasi-collinear 1 to 2 process under consideration conserves energy only, one expects a single turbulent spectrum associated with a finite flux of energy throughout scales from the source $\omega_f$ to the sink. To derive the corresponding KZ spectrum, it is useful to return to the $z$ variable, which is equivalent to performing a Zakharov conformal transformation, as in \ref{Z-transformation}, for 3-wave interaction collision integral. Thus, let $ z = \omega/(\omega+q)$ in the gain term and $z= (\omega-q)/\omega$ in the loss term. Under this change of variables the collision integral (\ref{c-inel-4}) becomes 
\beq\label{c-inel-z}
\frac{\kappa}{\omega^\beta}\left[ \int_0^1 \rmd z \frac{1}{z^4 (1-z)^{1+\beta}} F(\omega, \omega/z) -  \int_0^1 \rmd z \frac{1}{z^\beta (1-z)^{1+\beta}}F(z\omega,\omega)\right].  \nn
\eeq
Following Zakharov's method, we assume a power spectrum 
\beq
n(\omega) = A  \omega^{x},
\eeq
and look for $x$ that cause the collision integral to vanish or, equivalently, that generates a constant flux of energy,
\beq
P(\omega) =P= \text{const.}
\eeq
Using \eqn{c-inel-z} in \eqn{P-def}, one obtains
\beq \label{flow}
P(\omega) =  \kappa \int_0^1 \rmd z \frac{1}{z^\beta(1-z)^{1+\beta}}  \int_{\omega}^{\omega/z} d\omega_0\,\omega_0^{3-\beta} F(z\omega_0, \omega_0),
\eeq
with 
\beq\label{R-inel}
F(z\omega_0, \omega_0) = A^2  \omega_0^{2x} z^x (1-z)^x \left[  (1-z)^{-x} +z^{-x}-1\right].
\eeq
Note first that $F$ vanishes identically (and hence the flux $P$) at the thermal fixed point $x=-1$.
To obtain the non-thermal fixed point, one requires the energy flux to be independent of $\omega$. Therefore, the $\omega_0$ integral must yield
\beq
\int_{\omega}^{\omega/z} \rmd\omega_0 \,\omega_0^{3-\beta+2x} = \int_{\omega}^{\omega/z} \frac{\rmd\omega_0}{\omega_0} =\ln\frac{1}{z},
\eeq 
which implies  
\beq\label{KZ-exp}
x=\frac{\beta-4}{2}.
\eeq
This corresponds to the KZ solutions
\bel{KZ-spect-inel}
n(\omega) =  \begin{dcases}  \quad \omega^{-2} \qquad \text{for} \quad\beta=0\quad (\text{BH regime})\\
  \quad\omega^{-7/4} \quad \text{for} \quad \beta=1/2\quad  (\text{LPM regime}),
\end{dcases} 
\eeq
which are associated with a constant and negative energy flux $P(\omega) \equiv - P_f <0$, 
\bel{KZ-flux-inel}
P_f =  A^2 \kappa \int_0^1 \rmd z \frac{ (1-z)^{-x} +z^{-x}-1}{z^{\beta-x}(1-z)^{1+\beta-x}} \ln\frac{1}{z},
\eeq
that is, with a direct energy cascade. 

The physical relevance of the KZ solution relies on the convergence of the $z$ integral on the KZ spectra. For $x<-1$, we see that this is only the case when $\beta-x-2<0$, i.e., $\beta <0$ and $x<-2$. This condition is therefore not fulfilled by the KZ spectra (\ref{KZ-spect-inel}). 
Nevertheless, we observe that the energy flux for the KZ spectrum $\omega^{-2}$ in the BH regime diverges double logarithmically, as $P \sim \int_0 \rmd z/z \ln (1/z)$. As a result, it is marginally nonlocal and thus, can be corrected, to yield a constant flux, by a mild logarithmic dependence on the forcing scale, that is, $n(\omega)\sim \omega^{-2} \ln^{-1} (\omega/\omega_f)$ \cite{Kraichnan1971}. 

\begin{figure}[ht] 
\centering

 \includegraphics[width=10cm]{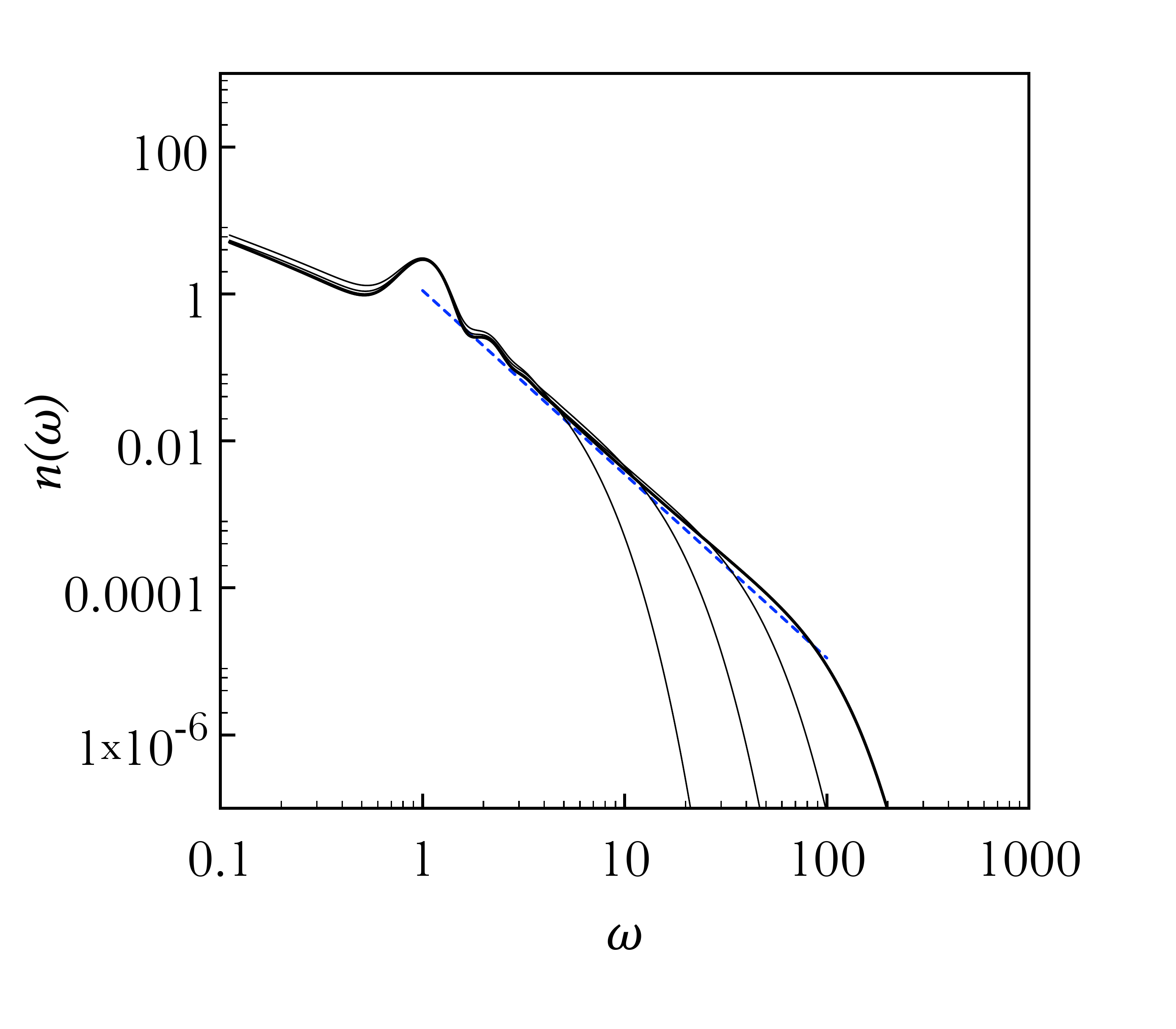} 

\caption{(Color online) Numerical simulation of the 3-wave kinetic equation with $\beta=-1$ (characterized by local interactions). From left to right, the various curves correspond to the times $\tau=10, 41, 163, 651$, where the rescaled time $\tau$ defined in \eqn{tau-def} . The dashed blue line corresponds to the local KZ spectrum $A_\KZ \,P^{1/2}\, \omega^{-5/2}$ with $A_\KZ\simeq 1.11 $ computed using Eq.~(\ref{KZ-flux-inel}) letting $\kappa=1$ and $P_f=1$. } 
\label{local-KZ-inel}
\end{figure}

As a proof of principle, we have computed the KZ constant for the local KZ spectrum $x=-5/2$ associated with $\beta=-1$. Integrating \eqn{KZ-flux-inel} and solving for $A$ we find $A\equiv A_\text{KZ} =1.1$. In Figure~\ref{local-KZ-inel}, we compare the time evolution of the occupation number, computed by solving the 3-wave kinetic equation numerically, the analytic KZ spectrum shown by the blue (dashed) line.  

Let us now briefly comment on the transition between the BH and the LPM regimes. The BH frequency (\ref{ome-BH}) is invariant under a uniform rescaling of the occupation number, therefore, it should be independent of the energy flux, $P_f$. Moreover, when the occupation number is steeper than $\omega^{-2}$, then  $\omega_\BH\sim \omega_f$, parametrically. Hence, for a large inertial range $\omega_f \ll \omega \ll  \omega_\max$, the turbulent spectrum lies in the LPM regime where interactions are strongly nonlocal. Remarkably, we will see that in the limit of strongly asymmetric interactions the 3-wave kinetic equation reduces to a diffusion equation whose stationary solution is $\omega^{-2}$ regardless of the regime in which the interactions occur. This limiting case yields a mild logarithmic dependence of $\omega_\BH$ on the ultraviolet sector (see \eqn{ome-BH-log} and discussion above).
The fact that the flux diverges on the KZ spectrum (\ref{KZ-exp}) indicates that the dominant contributions to the collision term are given by strongly asymmetric merging or branching. 
Therefore, one expects that in the region above the forcing scale, $\omega \gg \omega_f$, where the direct energy cascade forms, the integral receives a major contribution from the region  $q\sim \omega_f \ll \omega $, while, in the opposite limit, $\omega \ll \omega_f$, we expect to find $q\sim \omega_f \gg \omega$.  In what follows, we shall analyze the kinetic equation in these two regimes.

\subsection{Thermalization of the soft sector ($\omega \ll \omega_\for$)}
Consider first the soft sector defined by $\omega \ll \omega_f$. Since energy is carried by a direct cascade from the source, we expect the flux to vanish below the forcing scale at late times. This can only be achieved if the spectrum vanishes in this region or the soft sector is in thermal equilibrium, that is, $n(\omega)\sim T_\ast/\omega$. We will see that the latter is the physical solution. 

Assuming the collision term to be regular in the limit $\omega \to 0$, one can neglect  in \eqn{c-inel-4} the second term, which corresponds to interactions of the mode $\omega$ with a lower mode $\omega-q$ (cf. Figure~\ref{gain-loss-2}). Thus,
\beq \label{c-inel-5}
I_\inel[n]\simeq  \frac{1}{\omega^3} \int_0^\infty \rmd q\, K(\omega+q,q) F(\omega+q,\omega), 
\eeq
In the region $\omega \ll q$, the kernel (\ref{kernel-beta}) reduces to 
\bel{K-soft}
K(\omega+q,q) \simeq  \kappa \frac{q^{2}}{\omega^\beta}\,.
\eeq
and the expansion of $F(\omega,\omega+q)$ for large $q$ yields 
\bel{F-soft}
F(\omega+q,\omega) &\simeq&  n(\omega)[n(\omega+q)-n(q)]+n^2(q).\nn
&\simeq&  \omega n(\omega) \frac{\del}{\del q}n(q) +n^2(q).
\eeq
Note that although the first term is proportional to $\omega $ it is enhanced by the potentially large occupation number, $n(\omega) \gg n(q)$, in the soft sector. 
Inserting Eqs.~(\ref{K-soft}) and (\ref{F-soft}) in \eqn{c-inel-5} yields
\beq\label{evol-eq-soft}
\frac{\del n(\omega)}{\del t} = \frac{2 \pi^2 \kappa I_b  }{\omega^{2+\beta}}\left[ \frac{T_\ast }{\omega}- n(\omega) \right],
\eeq
where we have integrated by part in the first term as follows (cf. \eqn{Ia-Ib})
\beq
\int_0^\infty \rmd q \,q^2\, \frac{\del}{\del q }\,n(q) =-2  \int_0^\infty \rmd q \,q \,n(q)= - 2\pi^2 I_b,
\eeq
by neglecting the boundary terms and assuming that $q^2 n(q) \to 0$ when $q \to 0$ and $q \to \infty$. Interestingly, we encounter again the effective temperature $T_\ast=I_a/I_b$ (cf. \eqn{T-def}) that appears in the Fokker-Planck equation for elastic scatterings (Eqs.~(\ref{I-el-2}) and (\ref{flux-Q-FP})). 

As argued above, the Bethe-Heitler scale is of the order of the forcing scale (or slightly higher).
Therefore, to the left of the forcing scale the spectrum is in the BH regime and \eqn{evol-eq-soft} is to be evaluated for $\beta=0$. 

Now, \eqn{evol-eq-soft} takes the form a Boltzmann equation in the relaxation time approximation
\bel{evol-eq-soft-relax}
\frac{\del n(\omega)}{\del t} = \frac{n_\eq(\omega)-  n(\omega) }{\tau_\rel(\omega)},
\eeq
where the equilibrium distribution is given by the Rayleigh-Jeans distribution (\ref{RJ-dist-0}), with vanishing chemical potential $\mu=0$,
and the relaxation time, given by 
\beq
\tau_\rel(\omega) = \frac{\omega^2}{4 \pi^3 \bar\alpha^2 C I_a}  \sim \frac{\omega^2}{\hat q },
\eeq
is decreasing with $\omega$ and depends on time through $\hat q\equiv \hat q(t)$. 
It is of the order of the characteristic time scale for elastic processes described by the FP equation (\ref{I-el-1}). This is a formal demonstration of the fact that elastic and inelastic processes are in effect comparable. This will be completed by a similar observation in the next section, which is devoted to the UV sector. 

Recalling that $\tau_\el(\omega)\equiv \tau_\el(\omega,t)$ and $T_\ast\equiv T_\ast(t)$ are functions of time, the solution to Eq.~(\ref{evol-eq-soft-relax}) can be rewritten formally as,
\beq\label{evol-eq-soft-2}
n(\omega) = \int_0^t \rmd t' \exp\left[-\int_{t'}^t\frac{\rmd t''}{\tau_\rel(\omega,t'')}\right] \,  \frac{n_\eq(\omega,t')}{\tau_\el(\omega,t')},
\eeq
where we have used the initial condition $n(\omega,t=0) = 0$. 

\begin{figure}[ht] 
\centering

\includegraphics[width=10cm]{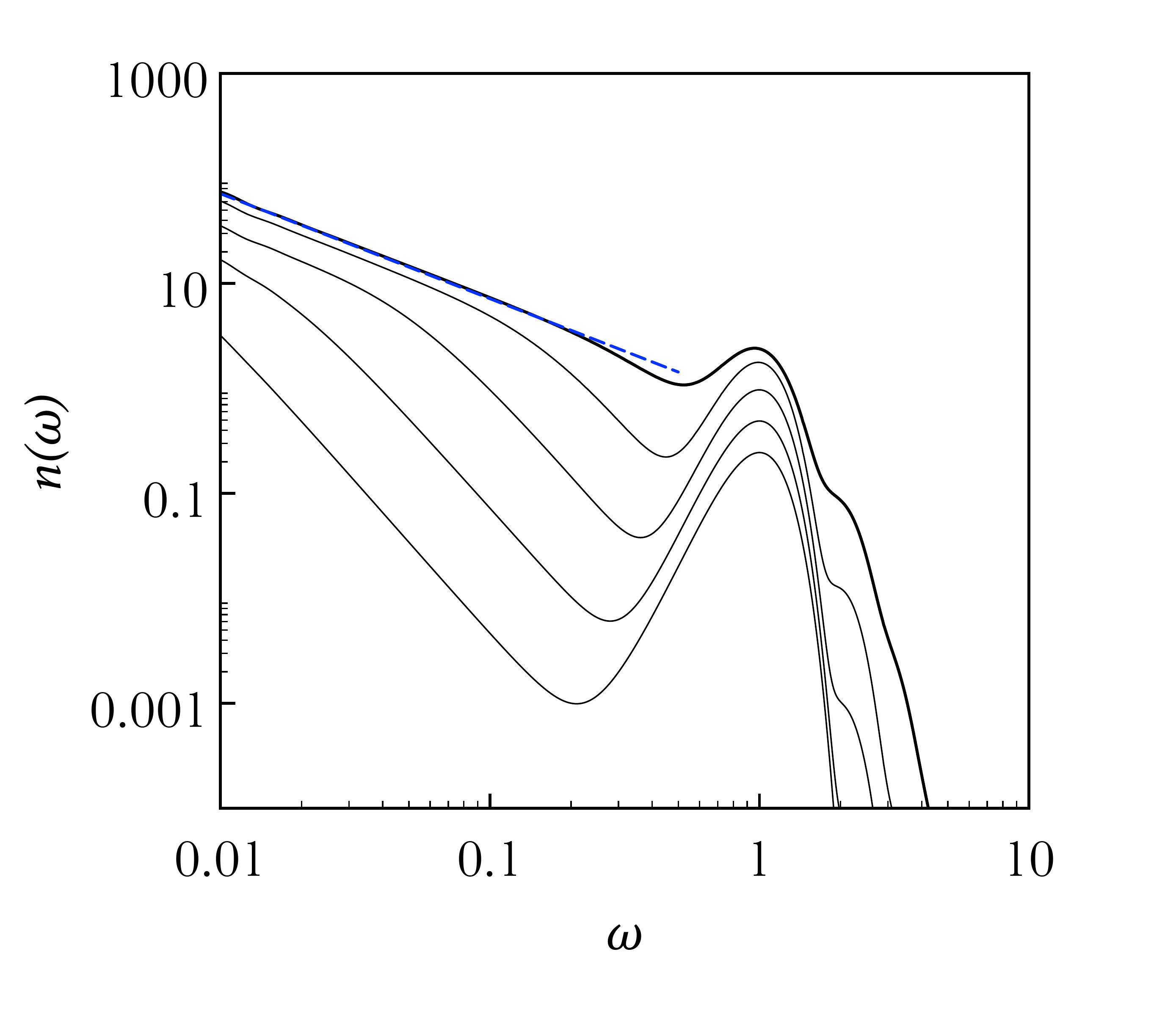} 

\caption{(Color online) Numerical solution of the kinetic equation with the inelastic collision integral given by \eqn{c-inel-4}, in the BH regime (for $\beta =0$), at early times. The occupation number is plotted at the times: $\tau=$ 0.08, 0.16, 0.32, 0.64 and 1.28, from bottom to top, with $\tau$ given by \eqn{tau-def}. The tangent (blue) dashed line corresponds to the thermal distribution $T_\ast/\omega$ with the effective $T_\ast\simeq 0.716$ evaluated at $\tau=1.28$.  } 
\label{turb-inel-early}
\end{figure}

This solution exhibits two regimes: a radiation dominated regime at early times and a thermal regime at late times (or small $\omega$). In the former, the source deposits gluons at $\omega_f$ via radiation. Those gluons then rarely scatter or radiate so long as the occupation number is low. During this phase, the occupation number at the source grows linearly with time until nonlinear effects become important enough to saturate the distribution, this happens parametrically when $t \sim \tau_\el(\omega_f)$. Hence, in this transient regime, $t\ll \tau_\el(\omega_f)$, $n(\omega_f) \sim t$ and as a result, $\hat q \sim t^2$ and $T_\ast \sim t$. The rare scatterings induce a radiation spectrum:
\bel{rad-dist}
n(\omega)  \simeq \frac{n_\eq(\omega)}{\tau_\rel(\omega)} t \sim  \frac{ \omega_\ast ^2(t)}{ \omega ^2 }, 
\eeq
so long as  $t\ll \tau_\rel(\omega)$, which implies $\omega_f > \omega \gg  \omega_\ast(t) \equiv \sqrt{\hat q t} $. The distribution below $\omega_\ast(t)$ has already relaxed to the thermal distribution 
\bel{therm-dist}
n_\eq(\omega)\equiv \frac{ T_\ast}{\omega},
\eeq
as a result of an exact balance between the soft gluon radiation and absorption rates. To see this more clearly let us assume that $\hat q $ and $T_\ast$ are constants. Hence, \eqn{evol-eq-soft-2} can be integrated to yield the simple form: 
\beq
n(\omega) = \left[1- \exp\left(-\frac{t}{\tau_\rel(\omega)}\right)\right] n_\eq(\omega),
\eeq
which clearly exhibits the two limiting cases (\ref{rad-dist}) when $\omega\gg \omega_\ast(t)$ and (\ref{therm-dist}) when $\omega\ll \omega_\ast(t)$. 
Note that the thermal distribution $n_\eq(\omega)$ is instantaneously generated in the soft sector and propagates toward the ultraviolet with a wave front characterized by the (diffusion-like) scale $\omega_\ast(t)= \sqrt{\hat q t}$, that we have already encountered in the elastic case, confirmed by the numerical simulation of the kinetic equation (see \ref{numerics} for details on the numerical procedure) with the collision integral  (\ref{c-inel-4}), shown in Figure.~\ref{turb-inel-early}. In this regime, a thermal bath forms in the region $\omega < \omega_\ast(t)$ due nonlocal interactions of the soft modes with the source. The thermalization of the soft modes have been recently thoroughly discussed in a related problem, that is, the thermalization of an isotropic plasma \cite{Blaizot:2016iir}. In this case, the hard modes present initially act effectively as a source for the soft gluons. 

When $t\sim \tau_\rel(\omega_f)$, the front reaches $\omega_f$ and the entire soft sector is in thermal equilibrium. At this point, energy starts flowing to the right of the forcing scale as a result of the excitation of higher modes. As a result, the effective temperature $T_\ast$ and the diffusion coefficient $\hat q$ vary slowly in time as they are mostly sensitive to the occupancy at the forcing scale, which saturates after the initial phase.

The thermalization of the soft sector in the free turbulence scenario proceeds in a similar fashion, where the initial spectrum acts as a reservoir of energy for the soft modes. The presence of elastic processes does not change this conclusion, however, the interplay between elastic and inelastic terms in the kinetic equation uniquely determines the first power correction to the thermal spectrum \cite{Blaizot:2016iir}.  

\subsubsection{Nonlocal turbulent spectrum for $\omega \gg \omega_f$ }

In the previous section we have seen that after a time $t\sim \tau_\el(\omega_f)$ a thermal bath forms to the left of the forcing scale. We turn now  to the late time dynamics of the system, i.e., $t \gg \tau_\el(\omega_f)$, in the region above $\omega_f$ and investigate stationary state solutions of the kinetic equation, associated with the steady transport of energy toward the ultraviolet. 

As we have shown in Section \ref{nonlocal-KZ} interactions on the KZ spectra are nonlocal and thus cannot be achieved. Therefore, to gain more insight on the nonlocal steady state we shall analyze the kinetic equation in the regime of strongly asymmetric gluon branching (merging).  This amounts to assuming either $\omega \gg q $, or  $\omega \ll q $ in the first term and $\omega \simeq q $ in the second term of (\ref{c-inel-4}). 

Let us start by expanding Eq.~(\ref{F-inel-def}) near $q=0$.  To do so, we first multiply Eq.~(\ref{c-inel-4}) by a test function $\omega^3 g(\omega)$ and integrate over $\omega$:
\beq
&& \int_0^\infty \rmd \omega\, \omega ^3 \,g(\omega) \, I_\inel[n] =\nn
&& \qquad-\int_0^\infty \rmd \omega \int_0^\infty \rmd q K(\omega,q) F(\omega,\omega+q) \left[g(\omega+q)-g(\omega) \right],\nn
\eeq
where, we have made the substitution: $\omega \to \omega+q$, in the first term. This allows to make the following approximation,
\beq
g(\omega+q)-g(\omega)\simeq q \frac{\del }{\del \omega} g(\omega),
\eeq
Integrating by part and noting that $g(\omega)$ is an arbitrary function of $\omega$, we obtain for contribution to the collision integral from the region $q \ll \omega$ (denoted by $I^<$), 
\beq\label{c-inel-hard}
I^<_\inel[n] \simeq \frac{1}{\omega^3} \frac{\del }{\del \omega} \int^{q_\text{max}}_0 \rmd q\, q K(\omega,q) \,F(\omega+q,q),
\eeq
where by consistency the $q$ integration must be cutoff at $q_\text{max} < \omega $. 

Assuming that the distribution $n(\omega)$ is steeper than $1/\omega$ one can neglect the first term in $F(\omega,\omega+q)$ (cf. Eq.~(\ref{F-inel-def})), and thus, one can write
\beq\label{R-overoc-2}
F(\omega,\omega+q) \simeq [n(\omega+q)- n(\omega)]n(q) \simeq q \, n(q) \frac{\del}{\del \omega }\, n(\omega).
\eeq
In this approximation, the kernel (\ref{kernel-beta}) reduces to 
\beq\label{kernel-hard}
K(\omega,q) \simeq \kappa \frac{\omega^3}{q^{\beta+1}}.
\eeq
Inserting Eqs.~(\ref{R-overoc-2}) and (\ref{kernel-hard}) in the collision term (\ref{c-inel-hard}), 
we obtain a the following diffusion equation 
\beq\label{inel-FP}
I^<_\inel[n]= \frac{1}{4\omega^3} \,\frac{\del }{\del \omega}\,  \omega^3 \hat q_\inel\, \frac{\del }{\del \omega} \, n(\omega),
\eeq
where the diffusion coefficient reads $\hat q_\inel \equiv 4 \pi^2 \kappa \,I_c$,
with
\beq\label{Ic}
 I_c \equiv \frac{1}{\pi^2 }\int_0^{q_\text{max}} \rmd q \, q^{1-\beta} \, n(q),
\eeq
that reduces to $I_b$ in the BH regime, i.e. when $\beta=0$. 

Note that \eqn{inel-FP} has two fixed points $n(\omega) = \text{const.}$ and $n(\omega) = \omega^{-2}$, which are independent of $\beta$. The second solution is the one related to constant flux of energy $P$ carried by a direct energy cascade. For a power spectrum that is steeper than $q^{-2+\beta}$, $I_c$ is dominated by the lower bound, thus, in the LPM regime, where $\beta=1/2$ one can safely send $q_\text{max}$ to infinity. On the other hand, in the BH regime, $I_c=I_b$ is only logarithmically dependent on $q_\text{max}$ which should not affect any power like behavior and, similarly to the Coulomb logarithm in the elastic collision integral, may be ignored. 
More generally, we can write 
\bel{qinel-general}
\hat q_\inel = \frac{4 \kappa}{\omega^3}  \int_0^\infty \rmd q \, q^2 \, K(\omega,q) \, n(q).
\eeq
Interestingly, \eqn{inel-FP} describes a diffusion process in four dimensions as a consequence of energy conservation. This is in contrast with three dimensional diffusion whose integral of motion is the total particle number. 

Likewise, the contribution from $q \gg \omega $ in the first term of Eq.~(\ref{F-inel-def}) yield, 
\beq
&& \frac{1}{\omega^3} \int_{q_\text{min}}^\infty \rmd q\, q K(\omega,q) \,F(\omega+q,q)\simeq \frac{\kappa}{\omega^{2+\beta}} \int_{q_\text{min}}^\infty \rmd q q^2 \frac{\del n(q) }{\del q} n(\omega),\nn
\eeq
where we have neglected the term $n^2(q)$ in $F(\omega+q,\omega)$ assuming that the spectrum is steeper than  $\omega^{-1}$.  Here, we recover the integral $I_b$ (cf. Eq.~\ref{Ia-Ib}), which is dominated by the UV when the spectra exponent of the distribution $n(q)$ is larger than $-2$ the integral $I_b$, in which case, one can set the lower bound to 0. 

We consider now the region $q\simeq \omega$ in the second term and make the change of variables $q\to \omega-q$. We obtain
\beq
\frac{1}{\omega^3} \int_{0}^\omega \rmd q\, q K(\omega-q,q) \,F(\omega,q)&=& \frac{1}{\omega^3} \int_{0}^\omega \rmd q\, q K(q,\omega-q) \,F(\omega,\omega-q)\nn
&\simeq &  \frac{1}{\omega}  \int_{0}^{q_\text{max}} \rmd q q^{1-\beta} n(q) \frac{\del n(\omega)}{\del \omega}\nn
&\simeq &  \frac{\hat q_\inel}{4\omega} \frac{\del n(\omega)}{\del \omega},
\eeq
where $\hat q_\inel$ is related to the integral $I_c$ (cf. Eq.~(\ref{Ic}) and above) that we already encountered when analyzing the region $q \ll \omega$.

Putting all the pieces together and assuming $\hat q_\inel $ to be independent of $ \omega$, the inelastic collision integral reads 
\beq\label{inel-PDE}
\frac{\del n(\omega)}{\del t} \simeq \frac{\hat q_\inel}{4 \omega^3 } \frac{\del }{\del \omega } \omega^3 \frac{\del }{\del \omega } n(\omega)  -  \frac{\hat q_\inel}{4\omega} \frac{\del n(\omega)}{\del \omega}+\frac{2\pi^2 \kappa}{\omega^{2+\beta}} I_b n(\omega).\nn
\eeq 
It is straightforward to verify that the first diffusive term and the last two terms conserve energy separately. 
Close to the ultraviolet cutoff the last term is suppressed since it accounts for interactions of the mode $\omega$ with very hard modes. In this case the first two terms combine to yield the diffusion equation 
\beq\label{inel-PDE-2}
\frac{\del n(\omega)}{\del t} \simeq  \frac{\hat q_\inel}{4 \omega^2 } \frac{\del }{\del \omega } \omega^2 \frac{\del }{\del \omega } n(\omega) \nn
\eeq
It follows that a diffusive process is responsible for the energy transport toward to UV. Hence, we expect the UV cutoff to scale as $t^{1/2}$. This will be checked soon in numerical simulations of the complete inelastic collision term.

Because the collision term Eq.~(\ref{inel-PDE}) is inhomogeneous in $\omega$ no steady power law solution is expected. Of course, close to the UV cutoff, one can neglect the third term, in which case, the power spectrum $\omega^{-1}$ becomes a steady state solution Eq.~(\ref{inel-PDE}). But since this solution is related to the conservation of particle number through the diffusion term given by Eq.~(\ref{inel-PDE-2}) it cannot be associated with the direct energy cascade. 

\begin{figure}[ht] 
\centering

\includegraphics[width=10cm]{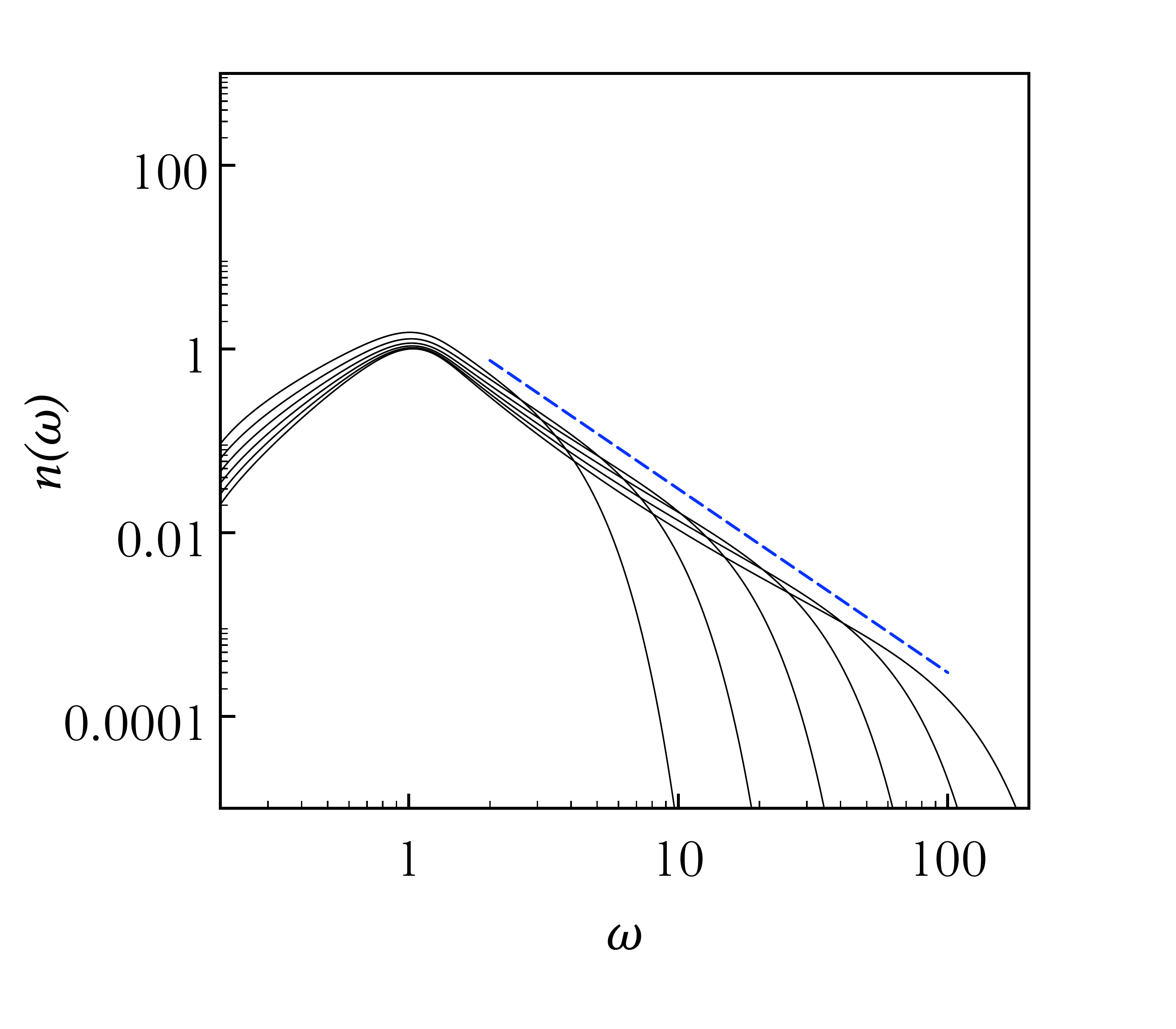} 
\caption{(Color online) Numerical solution of Eq.~(\ref{inel-PDE}) for $\kappa=1$ at $t=$ 1, 4, 16, 64, 256 and 1000 (from left to right) in the LPM regimes, i.e. $\beta=1/2$.  The (blue) dashed line corresponds to the power law $\omega^{-2}$ and is given as reference.   } 
\label{inel-spectrum-diffusion}
\end{figure}

In Figure~\ref{inel-spectrum-diffusion}, the numerical solution of Eq.~(\ref{inel-PDE}) is plotted for various times.
First, we note that the power exponent is close to $-2$. However, the spectrum is not steady in the wave of the right moving front due to the presence of the last two terms in the r.h.s. of Eq.~(\ref{inel-PDE}) that are responsible for the slow decrease of the spectrum as a function of time. The red(dotted) curve corresponds to the analytic solution of Eq.~(\ref{inel-PDE}) where only the first diffusive term is kept (cf. Eq.~(\ref{sol-diff-inel-1})). 


To gain more insight on the scaling of the turbulent spectrum in the inertial range let us consider a simplified equation. Let us only keep the first energy conserving term in Eq.~(\ref{inel-PDE}). Away from the source, namely when $\omega \gg \omega_f$, the solution reads (see details in \ref{sol-inel-FP}) 
\beq\label{sol-diff-inel-1}
n(\omega)\simeq   \frac{4 \pi^2 P_f}{ \hat q_\inel\, \omega^2}\, \exp\left(- \frac{\omega^2}{\hat q_\inel\, t }\right),
\eeq
which describes a quasi-steady spectrum $\omega^{-2}$ at late times, provided $\hat q_\inel$ is a slowly varying function of time, which will be shortly verified. This spectrum forms in the wake of a diffusive front characterized by the scale $ \omega_\ast(t)=\sqrt{\hat q_\inel \, t} > \omega_f $. It is associated with a direct energy cascade that carries a constant energy flux toward growing $\omega$'s.  

By matching the solutions to the left and to the right of the forcing scale, parametrically,  we have $n(\omega_f)\sim T_\ast / \omega_f \sim P_f/ \hat q_\inel \, \omega_f^2$. This implies that $T_\ast \sim P_f/ \hat q_\inel\,  \omega_f$.
Moreover, using the forms of the spectrum for $\omega< \omega_f$, and $\omega> \omega_f$, namely, $\omega^{-1}$ and $\omega^{-2}$, respectively, we observe that the modes that contribute to the diffusion coefficient are of order $q\sim \omega_f$. Hence, the diffusion coefficient scales as $\hat q \sim \bar\alpha^2 \, \omega_f ^3 n^2(\omega_f)$.

On the other hand, the inelastic diffusion coefficient is also dominated by the forcing scale, which is consistent with the nonlocal nature of the interactions justifying the gradient expansion. In the LPM regime we find
\bel{qhat-inel-1}
\hat q_\inel \simeq  4 \bar\alpha \sqrt{\hat q } \int \rmd q \, q^{1/2} n^2(q)\sim  \bar\alpha^2 \, \omega_f ^3 n^2(\omega_f). 
\eeq
 Using $T_\ast \sim P_f/ \hat q_\inel\,  \omega_f$ together  with \eqn{qhat-inel-1} yield 
 \bel{qhat-T-steady}
\hat q_\inel\sim \hat q  \sim \abar^{2/3} P_f^{2/3}  \omega_f^{-1/3}, \quad \text{and }\quad T_\ast \sim\abar^{-2/3} P_f^{1/3}  \omega_f^{-2/3}.
 \eeq
This parametric estimate was derived in the LPM regime, but it can be easily checked that it holds in the BH regime too. 

Finally, inserting \eqn{qhat-T-steady} in \eqn{sol-diff-inel-1} for $\omega_f\ll \omega\ll  \omega_\ast(t)$, i.e., in the inertial range, we find that the steady state spectrum associated with a constant flux reads
\bel{inel-steady-spec}
n(\omega) \sim \frac{A}{\omega^2},\quad \text{with}\quad A\sim \frac{P_f^{1/3} \omega_f^{1/3}}{\abar^{2/3}}.
\eeq
Note that, as anticipated in the introduction, the steady state spectrum for quasi-collinear (inelastic) 1 to 2 processes,  scales as $P_f^{1/3}$ similarly to 2 to 2 processes (cf. \eqn{factor-A-res}). Moreover, there is an explicit dependence on the forcing scale $\omega_f$ demonstrating the nonlocality of inelastic interactions in a turbulent state. 

\begin{figure}[ht] 
\centering

\subfloat[]{ \includegraphics[width=6.5cm]{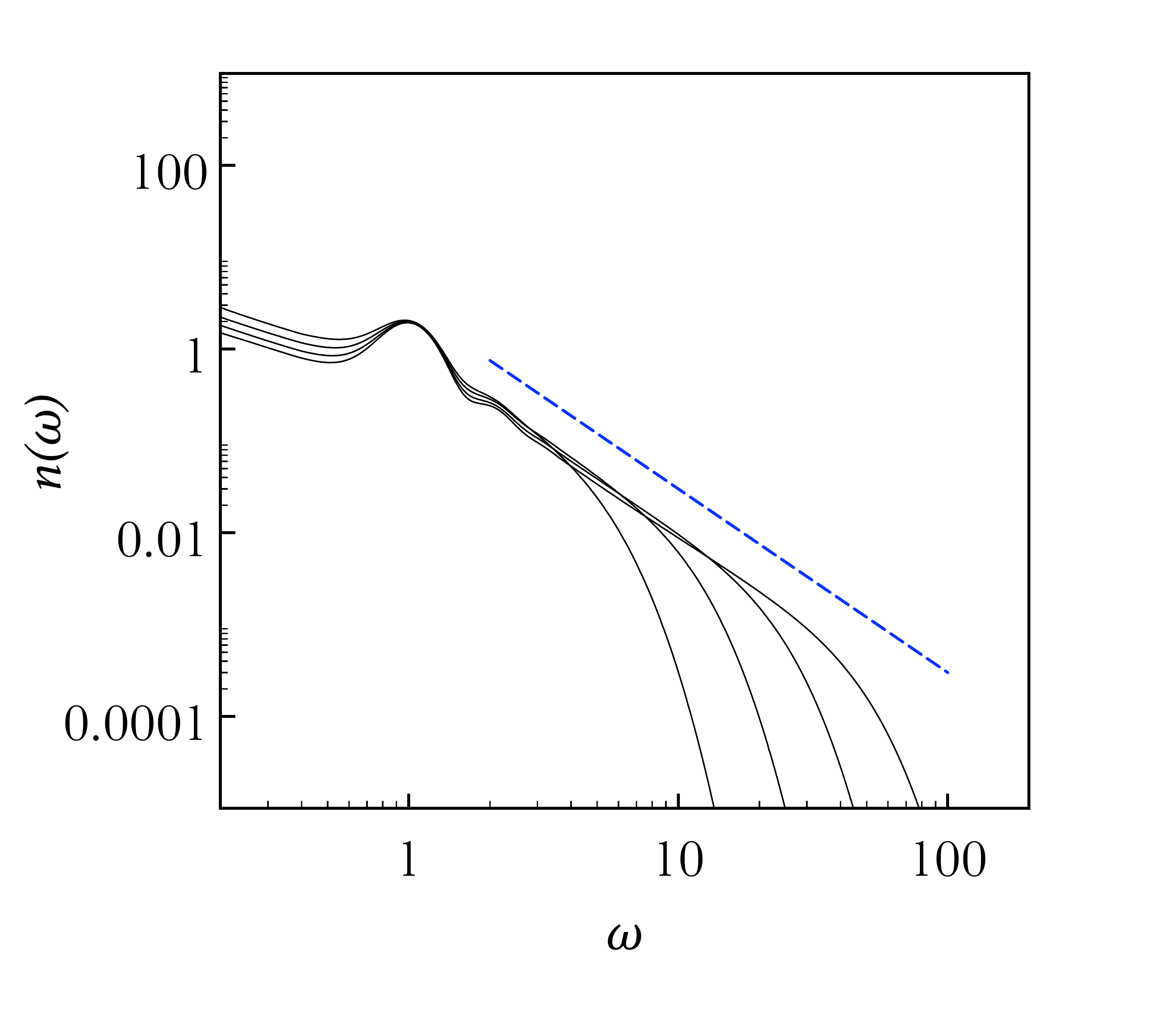} }  \subfloat[]{ \includegraphics[width=6.5cm]{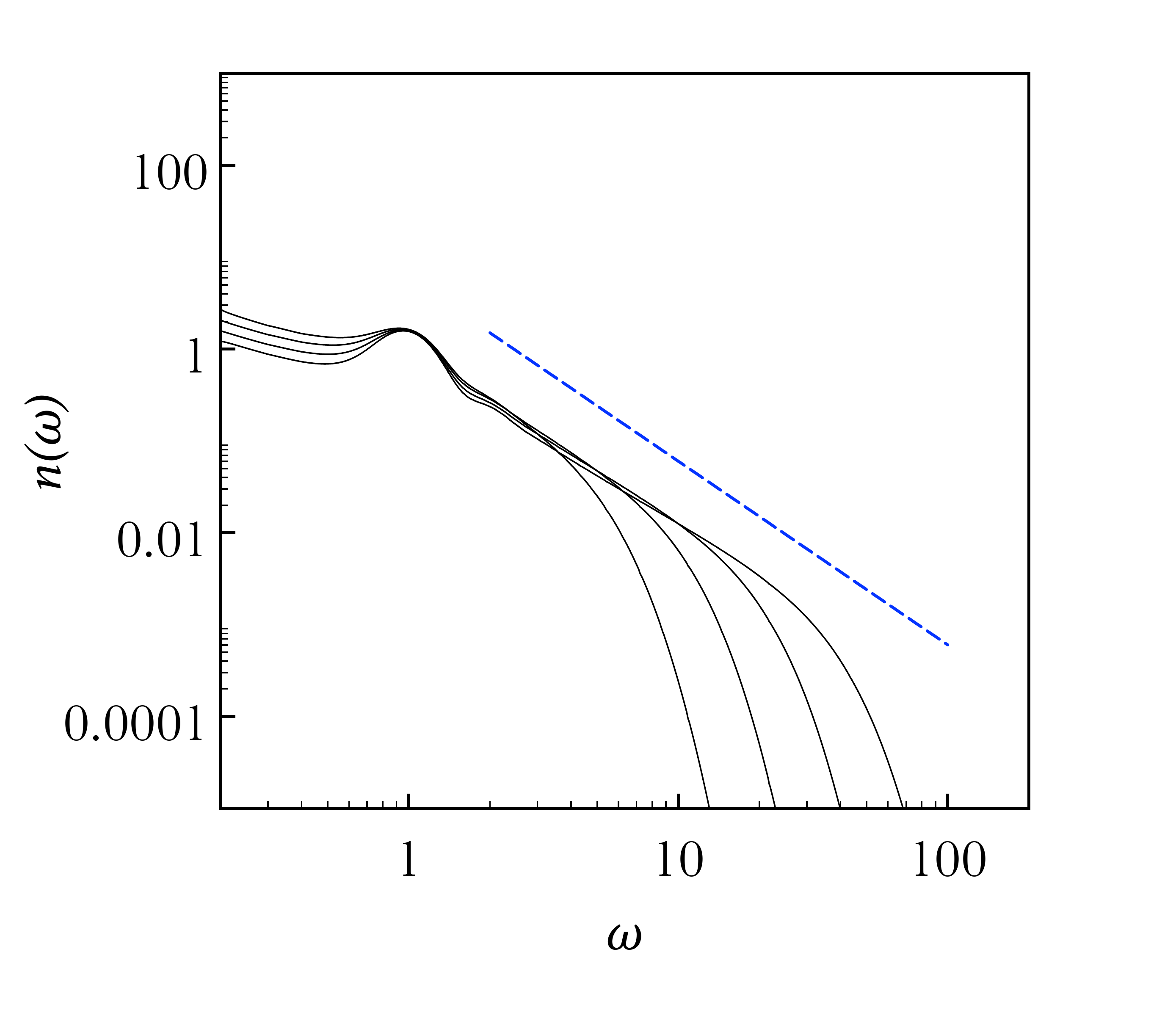} } 

\caption{(Color online) The gluon spectrum in the pure inelastic case for $\kappa=1$ at $t=$ 1, 4, 16 and 64 (from left to right) in the BH (left panel) and the LPM regimes (right panel), respectively.  The (blue) dashed line corresponds to the power law $\omega^{-2}$ and is given for reference.   } 
\label{inel-spectrum}
\end{figure}

In the above analysis, we have neglected the contributions of democratic splittings to the collision integral and concluded that energy is transport by diffusion toward the UV.  
To assess the validity of this approximation we have computed numerically the occupation number at late times in the BH and LPM regimes. This is shown in Figure~\ref{inel-spectrum}. In the BH regime a clear stationary spectrum with spectral exponent $-2$ forms in the wake of the right moving front.


\begin{figure}[ht] 
\centering

\subfloat[]{ \includegraphics[width=6.5cm]{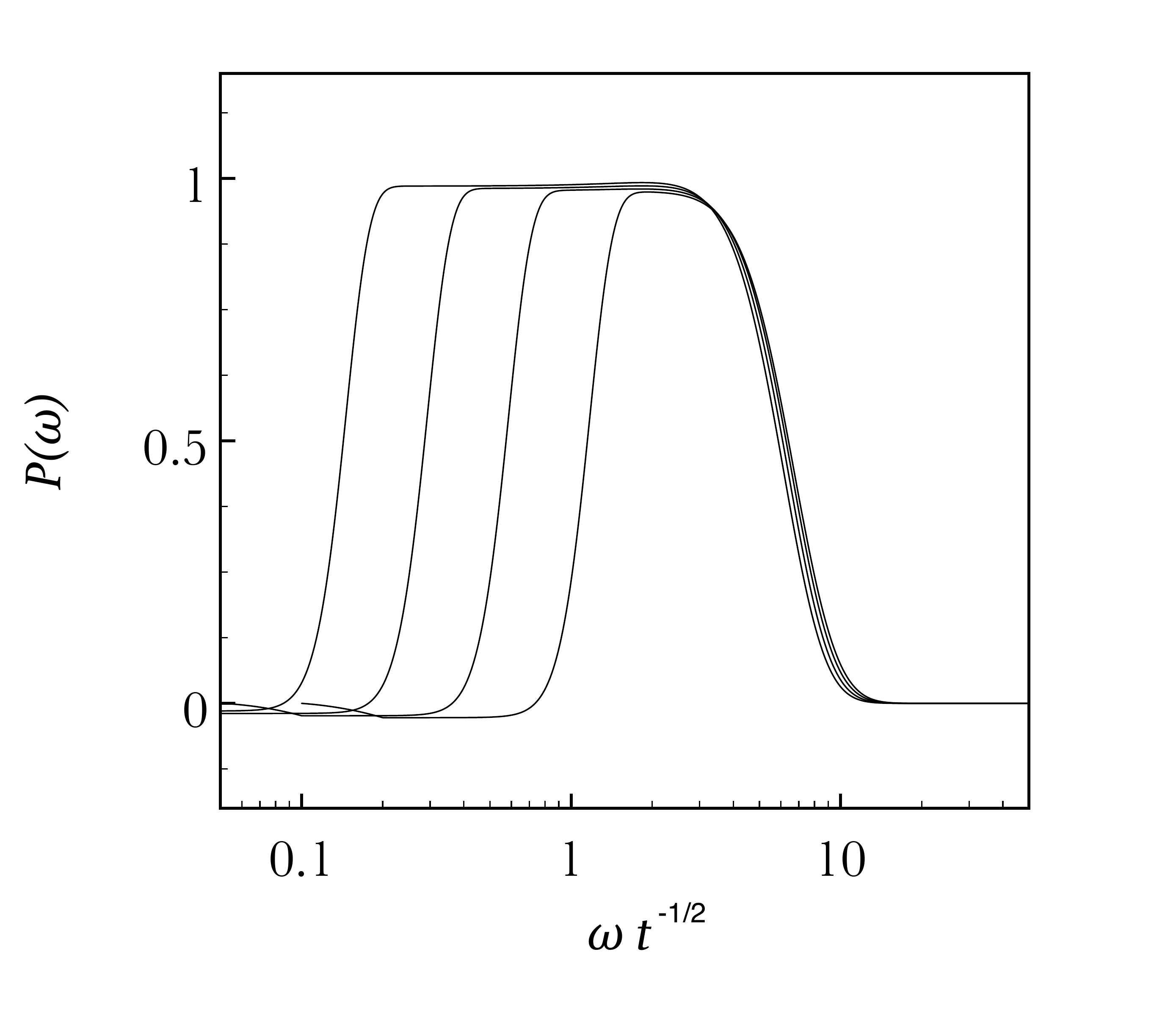} }  \subfloat[]{ \includegraphics[width=6.5cm]{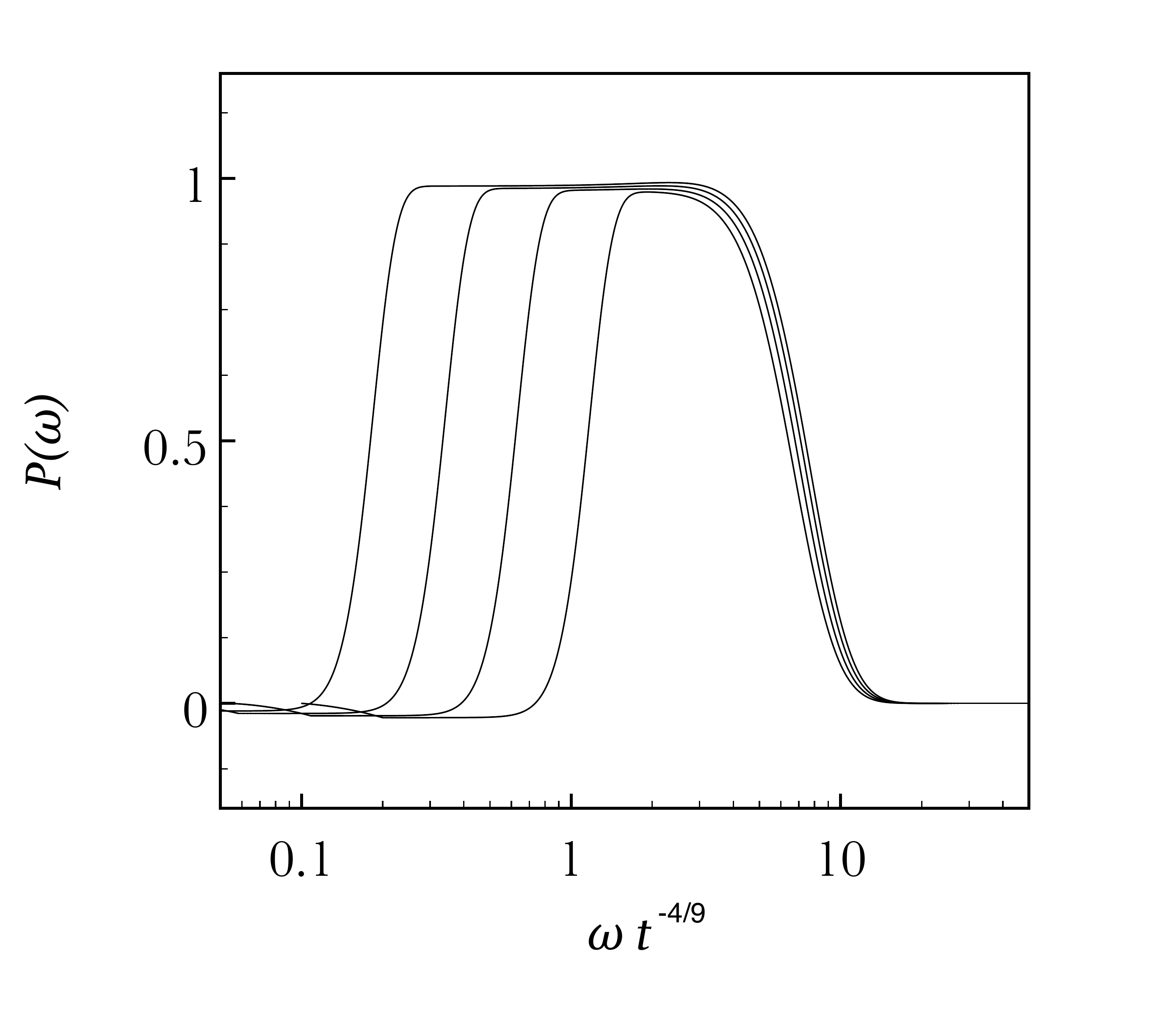} }

\caption{(Color online) The energy flux density as a function of the rescaled momentum $\omega \,t^{-1/2}$ (left panel) and $\omega \, t^{-4/9}$ (right panel)  in the LPM regime. From left to right: $t=$ 1, 4, 16 and 64. } 
\label{inel-flux-LPM}
\end{figure}

 
Incidentally, the exponent $-2$ is also that of the KZ spectrum established in Section \ref{nonlocal-KZ}, which we argued was only marginally nonlocal and hence, could be realized with a mild logarithmic corrections. 
However, no evidence of this logarithmic dependence of the spectrum was found is the numerical study. 

In Figure~\ref{inel-spectrum}(b), we have plotted the spectrum at various times in the LPM regime. We observe that as the front moves to the right the spectrum slowly decreases in its wake, in particular near the source. This is the result of nonlocal interactions that flatten the spectrum as the time goes on. 

If the LPM regime were dominated by local interactions one should expect the front to scale as $t^{9/4}$, as shown in the next section. However, Figure~\ref{inel-flux-LPM}, where the energy flux is plotted, shows that the wave front scales as $t^{1/2}$, rather than the unphysical local scaling $t^{9/4}$. Indeed, the former scaling implies a faster wave propagation and therefore, nonlocal interactions are more efficient in transporting energy to the UV.  Similarly, the wave front in the BH regime is also found to scale as $t^{1/2}$. 

The above analysis is valid as long as nonlocal interactions are dominated by exchanges with the source, that is, when the integral $I_c$ is dominated by the IR. This assumption must be relaxed when the spectrum exponent is larger that $-2+\beta=-3/2$. In this case, the interactions will be dominated by exchanges with the UV end of the spectrum and source would playing a marginal role. I this case, according to the conclusion in the previous section, the spectrum must be in thermal equilibrium with the UV that plays now the role of the source.

For completeness,  we shall evaluate the BH scale in order to estimate the relevance of either of the regimes in the inertial range and in the soft sector. The integral $I_a$ (cf.  \eqn{Ia-Ib}) is dominated by momenta of the order of the forcing scale, $I_a \sim \int \rmd q \, q^2 \, n^2(q) \sim A^2 \omega_f$, whereas,  the integral $I_b$ evaluated for the turbulent spectrum  \eqn{sol-diff-inel-1} yields a logarithmic dependence on the ultraviolet cutoff $\omega_\ast(t) \gg \omega_f$,
$I_b \sim A \, \ln \omega_\ast(t)/\omega_f $. 
Using these estimates in \eqn{ome-BH}, we find that
\bel{ome-BH-log}
\omega_\BH \sim  \omega_f \ln^2\frac{\omega_\ast(t)}{\omega_f}. 
\eeq
Apart from the mild logarithmically increase, $\omega_\BH$ is of the order of the forcing scale $\omega_f$. Since hard modes in the inertial range interact mainly with modes near $\omega_f$, one may argue that the BH branching rate is a better approximation of the full kernel, which obeys \eqn{br-rate}. 

Similarly, using \eqn{inel-steady-spec} in \eqn{therm-mass}, we find that the Debye mass scales parametrically as 
\beq
m^2_D \sim \abar^{1/3} \,P_f^{1/3} \, \omega^{1/3}_f \, \ln\frac{\omega_\ast(t)}{\omega_f},
\eeq
which must be much smaller than $\omega_f$ for this analysis, based on neglecting the gluon mass dependence in the dispersion relation and the small angle approximation, to be valid. 

These estimates must be reevaluated when the spectrum flattens toward the thermal one at late times. In this situation, the integrals $I_a$ and $I_b$ will be dominated by the UV cutoff, and we obtain $\omega_\BH \simeq \omega_\ast(t)$, while the Debye mass reads $m_D^2 \sim \abar T_\ast \omega_\ast(t)$ for $n\sim T_\ast/\omega$, and remains suppressed at small coupling. 

\subsubsection{Analytic form of the front: $\omega \gg \omega_\ast(t)$}
The merging of gluons of comparable energies, which is not taken into account in the diffusion approximation, is the dominant process in the region of the wave front, namely when $\omega \gg \omega_\ast(t)$ and yields an exponential suppression rather than Gaussian, as shown in \eqn{sol-diff-inel-1}. To see this, note that this regime corresponds to the edge of the phase space, where a gluon $\omega$ is dominantly produced by interactions with lower bins (merging). This amounts in neglecting the first term in Eq.~(\ref{c-inel-2}). Moreover, due to the stiffness of the spectrum, $ n(\omega) \ll n(q) \sim n(\omega-q)$, and one can neglect terms promotional to $n(\omega)$. It follows that $F(\omega-q,\omega)  \approx -n(\omega-q)n(q)$ and  Eq.~(\ref{c-inel-2}) reduces to 
\beq\label{eq-front-forcing}
\frac{\del }{\del t}n(\omega) \approx  \frac{1}{\omega^3} \int_0^\omega \rmd q\, K(\omega,q) \, n(\omega-q)\,n(q).
\eeq
Before going on solving \eqn{eq-front-forcing}, let us first determine the speed of the front by assuming a self-similar evolution of the occupation number, 
\beq
n(\omega)\equiv t^a f(\omega t^b),
\eeq
where $a$ and $b$ will be fixed uniquely by energy conservation and the homogeneity of the kinetic equation.
Then, using the kernel form (\ref{kernel-beta}) where $\kappa$ is roughly constant as a function of $\hat q $,  \eqn{eq-front-forcing} yields 
\bel{front-similar-eq}
t^{a-1}\left[ a f(x) +b x f'(x)\right]&=&\kappa\, t^{2a+\beta b} x^{-\beta}\nn
&\times& \int_0^1 \rmd z z ^{-1-\beta}(1-z)^{-\beta} f(zx) f((1-z)x),\nn
\eeq
where $x=\omega t^b$. The homogeneity condition implies that
\bel{front-exp-1}
a+\beta b +1=0,
\eeq
Since most of the energy is carried by the hard sector a second equation can be obtained requiring energy conservation
\beq
E=P_f t \sim  \int \rmd \omega \omega^3 n(\omega) =   t^{a-4 b}\int \rmd x x^3 f(x), 
\eeq
from which we must have
\bel{front-exp-2}
a-4 b -1=0.
\eeq
Solving Eqs.~(\ref{front-exp-1}) and (\ref{front-exp-2})  for $a$ and $b$, one finds
\bel{exp-similar}
a =  -\frac{4-\beta}{4+\beta}, \quad  b= - \frac{2}{4+\beta},
\eeq
In the BH regime, $\beta=0$,  we have
\bel{exp-similar-BH}
a =  -1, \quad  b= - \frac{1}{2}.
\eeq
This result is indistinguishable with that of the previous section, where we have found that in the diffusion approximation the position of the front propagates as $t^{1/2}$. However, in the LPM  regime, where $\beta=1/2$, one obtains a slower propagation of the front, 
\bel{exp-similar-LPM}
a =  -\frac{7}{9}\approx -0.78 \, \quad  b= - \frac{4}{9}\approx - 0.44.
\eeq
One may argue that the mismatch between the time evolution of characteristic scales in the diffusion regime, $t^{1/2}$, and in the wave front, $t^{9/4}$,  would generate an accumulation of particles in the wake of the wave front, which may cause the slope to be smaller than the predicted power $-2$, as observed in the numerical simulation given in Figure~\ref{inel-spectrum}.

Owing to the form of the r.h.s. of \eqn{front-similar-eq}, we assume the following ansatz for the wave front scaling function  
\beq\label{ansatz-front}
f(x) = A \, x^\gamma \, \rme^{-Bx}.
\eeq
Injecting (\ref{ansatz-front}) in Eq.~(\ref{front-similar-eq}) we obtain 
\beq\label{eq-front-forcing-2}
(a+b \gamma)x^\gamma  - B b x^{\gamma+1}=\kappa \, A x^{2\gamma-\beta}\, \int_0^1 \rmd z \, z ^{\gamma-1-\beta}(1-z)^{\gamma-\beta}.
\eeq
For the $z$ integral to converge one must have $\gamma>\beta$, and because $\beta>0$, when $x\gg 1$, one can neglect the first term in the l.h.s. of \eqn{eq-front-forcing-2}. By matching the exponents of $x$ in the second term with the r.h.s. we find that $\gamma= 1+\beta$. As a result, the $z$ integral is independent of $\beta$ and yields
$\int_0^1 \rmd z (1-z) =1/2$. Finally, solving  for $A$, one gets $A= 2B / [(4+\beta) \kappa]$, and the solution for the wave front takes the form
\beq\label{solution-front}
n(\omega) \simeq \frac{2B\, t^{-\frac{4-\beta}{4+\beta}}}{(4+\beta) \kappa} \left(\frac{\omega}{ \omega_\beta(t)}\right)^{1+\beta} \exp\left[-B\left(\frac{\omega}{ \omega_\beta(t)}\right)\right].
\eeq
where $\omega_\beta(t) \sim t^{\frac{2}{4+\beta}}$.

As shown in the previous section, numerical simulations of the kinetic equation in the LPM regime, where $\beta=1/2$, show that the wave front propagates faster than $t^{4/9}$ and is more compatible with $\omega_\ast(t)\sim t^{1/2}$, which characterizes the nonlocal solution (\ref{sol-diff-inel-1}).

\section{Turbulent spectrum in the presence of elastic and inelastic processes}\label{sec:el-inel}
Up to now, we have discussed elastic and inelastic processes separately and shown that they result in different power spectra, however, in both cases the energy cascade is found to be governed by a diffusion like transport toward the ultraviolet.

\begin{figure}[ht]
\centering 
\subfloat[]{ \includegraphics[width=6.5cm]{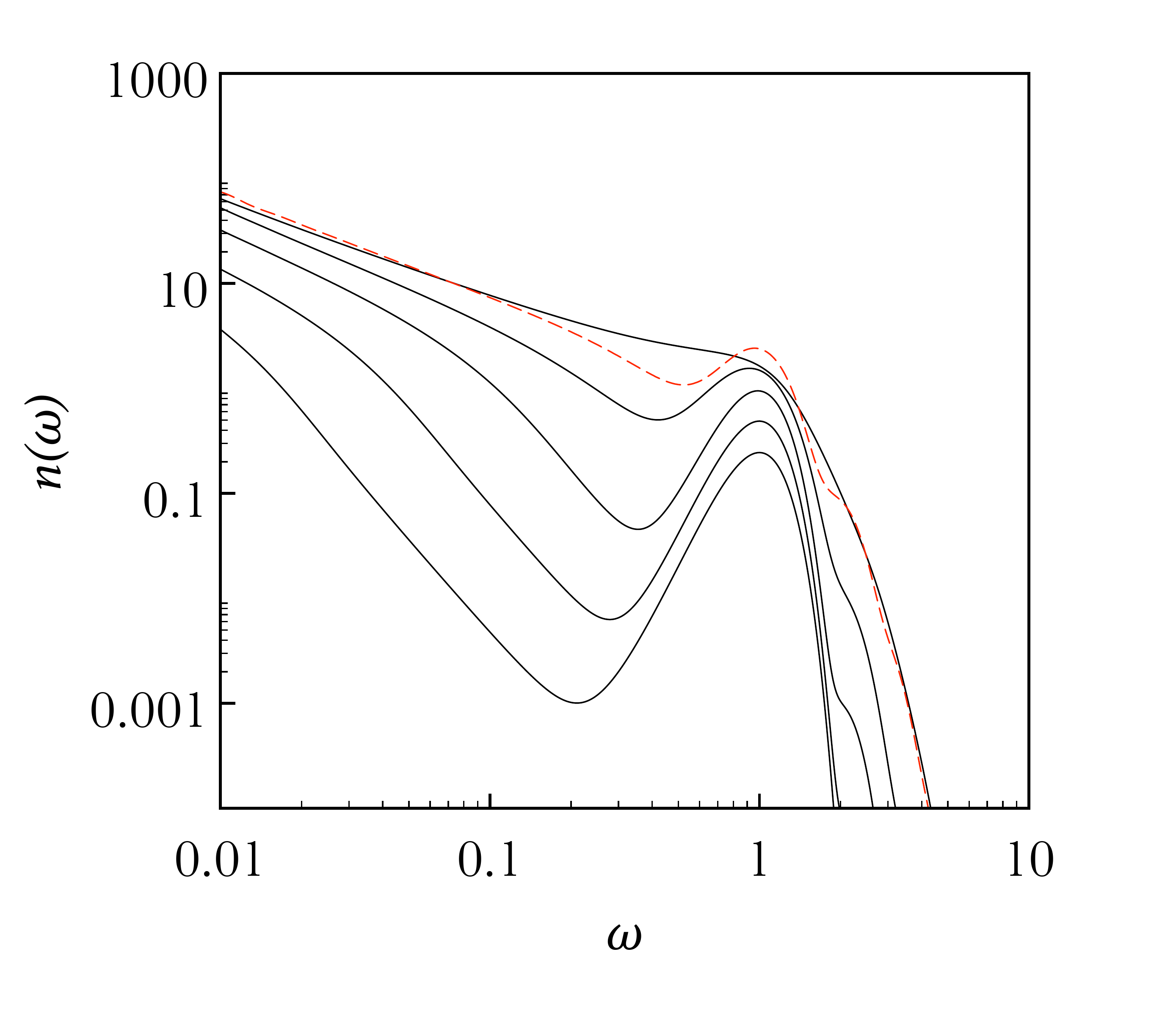} }  \subfloat[]{ \includegraphics[width=6.5cm]{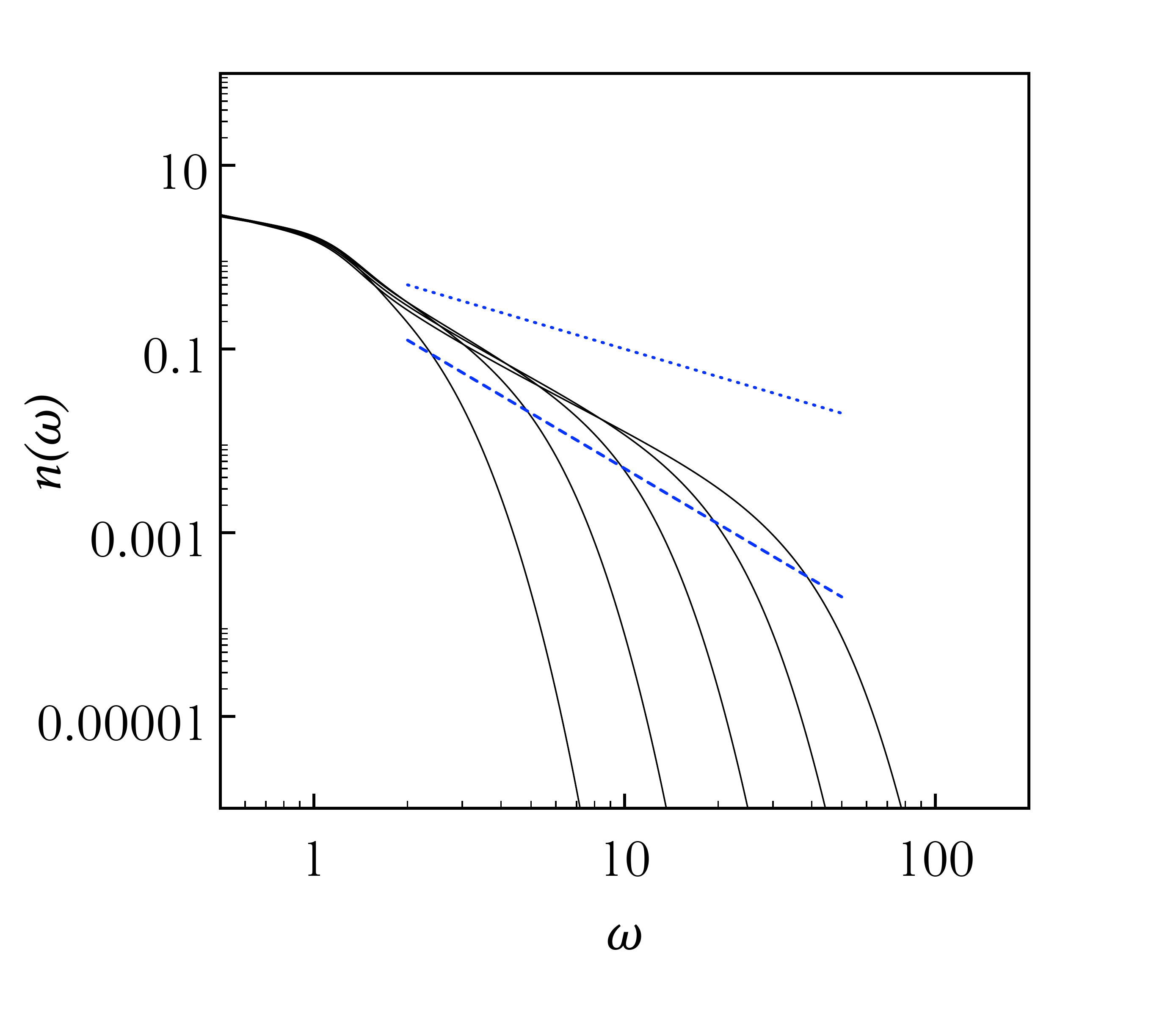} } 
\caption{(Color online) The numerical solution of the full kinetic equation, with elastic and inelastic processes taken into account. Left:  black curves are for $\tau=0.08, 0.16, 0.32, 0.64$ and 1.28, from  bottom to top. The dashed (red) curve shows the result of inelastic processes alone in the BH regime plotted at $\tau=1.28$ (to be compared to the corresponding black curve). Right:  Same as in the left panel  for $\tau=2, 8, 32,  128, 512$ (left to right). The dashed and dotted (blue) curves stand for the power spectra $\omega^{-2}$ and $\omega^{-1}$ and are shown as reference.    } 
 \label{fig:inel-el-steady}
\end{figure}

\begin{figure}[ht]
\centering 
\includegraphics[width=8cm]{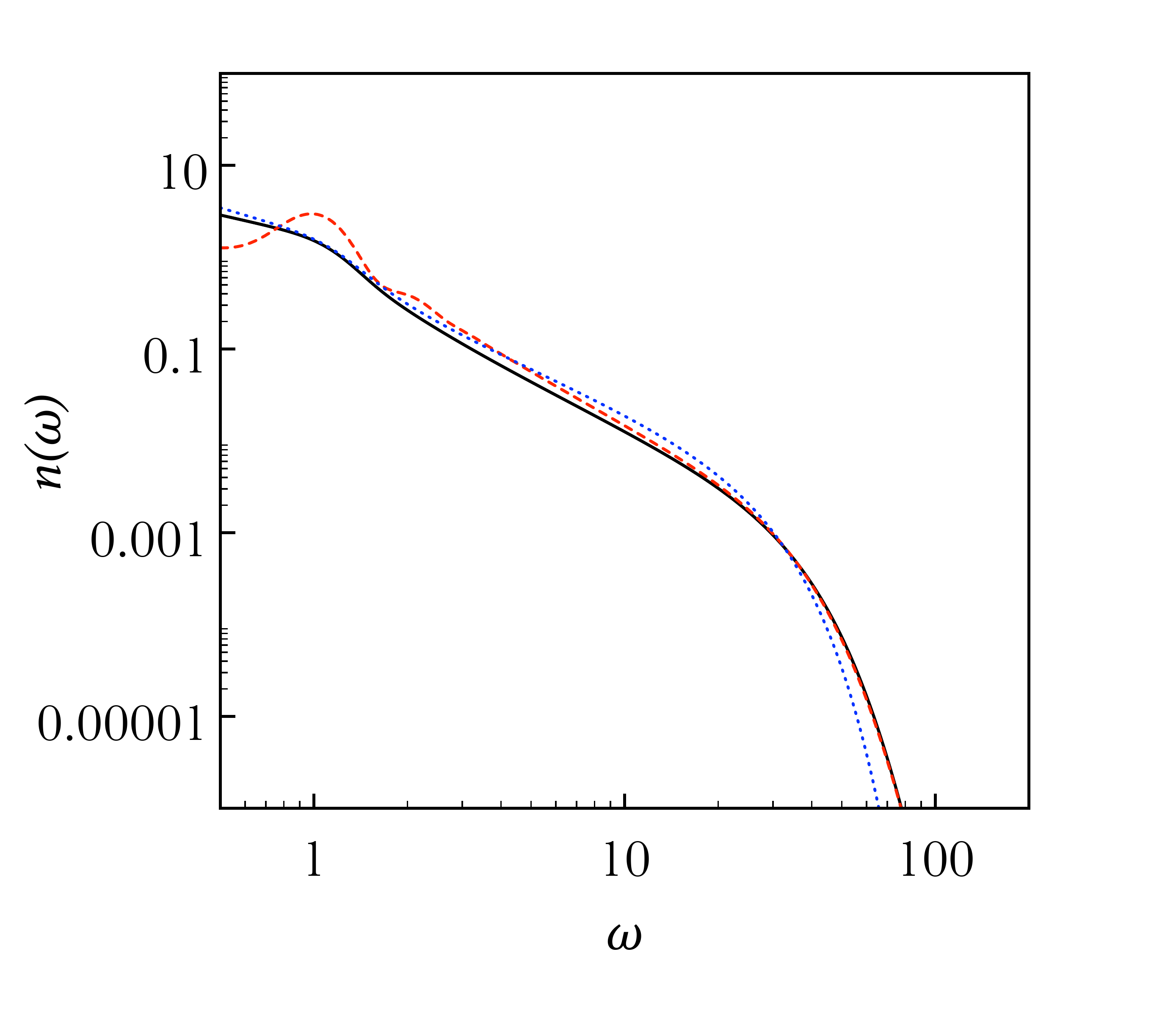} 
\caption{(Color online) The full spectrum at $\tau=512$  (full line) compared to the those in the purely elastic and inelastic cases,  which are depicted by dotted (blue) and dashed (red) curves, respectively. } 
 \label{fig:inel-el-steady-2}
\end{figure}

Before we turn to the direct energy cascade, let us first analyze the soft sector (to the left of the forcing scale). While in the presence of  elastic processes alone we have seen that the steady state corresponds to a non-thermal distribution $A/\omega$, with $A> T_\ast $, associated with a constant flux of particles, we expect the inelastic term to prevent such a spectrum due to the singular behavior of the radiation rate in the infrared that is tamed only when the soft sector is in thermal equilibrium, regardless of the presence of the elastic term. In Figure~\ref{fig:inel-el-steady} (a), we see that at early times gluons are injected into the soft sector via radiation then a thermal spectrum develops from the left to the right with temperature $T_\ast$. The similarity with the inelastic alone is highlighted by comparing the largest time $\tau=1.28$ (full thick curve) with the result obtained with only the inelastic term taken into account (dashed red curve) and we can see that both are in semi-quantitative agreement. Hence, we conclude that inelastic processes dominate over elastic ones. 

A more detailed analysis of the interplay of elastic and inelastic processes in given in Ref.~\cite{Blaizot:2016iir}, where a universal power correction to the thermal spectrum is derived, 
\beq
n(\omega)\simeq \frac{T_\ast}{\omega}+A \, \omega^\alpha, \quad \text{with}\quad \alpha= \frac{-3+\sqrt{1+4C}}{2},
\eeq
where for $C\simeq 1.83$ (cf. \eqn{K-BH}) we have $\alpha\simeq 0.4$, which qualitatively accounts for the deviation of the complete kinetic equation from the pure inelastic case as shown in Figure~\ref{fig:inel-el-steady} (a). 

This feature is not obvious since one could expect, if inelastic processes were negligible, particle number to be approximately conserved causing a finite flux of particle to develop in the infrared and thus, the formation of a transient Bose-Einstein condensate. 

Let us turn now to the ultraviolet sector where the direct energy cascade develops. Due to nonlocal interactions there is no steady power spectrum. However, since the thermal-like spectrum $\omega^{-1}$ is a fixed point of the full kinetic equation, one may expect at very late times, the system to be driven toward the thermal fixed point  with a small correction that would account for finite energy flux as argued in Ref.~\cite{Mueller:2006up}. This is known as ``warm turbulence" in the literature. When the spectral exponent becomes larger than $-3/2$, nonlocal interactions that were initially dominated by interactions with the source are now dominated by interactions with modes located at the UV cutoff $\omega_\ast(t)$. This is reflected by the fact that the global integrals $I_a$, $I_b$ and $I_c$ cease to be dominated by the source scale, $\omega_f$. In this case, the UV modes play the role of the source with which softer modes must be in thermal equilibrium.

The result of the numerical simulation for relatively late times is shown in Figure~\ref{fig:inel-el-steady} (b) where the slope of the spectrum appears to be more compatible with the power $-2$ than the thermal one. The simulations for the purely elastic and inelastic cases exhibits a similar behavior as shown in Figure~\ref{fig:inel-el-steady-2}.  However, in the region of the front wave the result for inelastic scatterings alone is closer to the full result than the purely elastic case. This may indicate that inelastic processes are dominant in transporting energy toward the UV.

\section{Summary}
In this paper, we have investigated analytically and numerically, in the framework of kinetic theory,  the spectrum of gluons that forms at late times in the presence of a steady and spectrally narrow source of energy.  Both elastic and inelastic processes are taken into account in the small angle approximation allowing for a semi-analytical treatment. Owing to the different nature of the two contributions, we have first discussed them separately. Elastic processes are described by a Fokker-Planck equation that conserves energy and particle number, for which non-thermal stationary solutions are derived. Below the forcing scale a inverse particle cascade characterized by the spectrum $\omega^{-1}$ develops, no finite steady spectrum is associated with the direct energy cascade. This results strongly differs from scaler $\phi^4$ theory, where KZ turbulent spectra are expected to form, because of the strong nonlocal nature of interactions in momentum space in Yang-Mills theory as argued in \ref{Mueller:2006up}. Deep in the infrared regime, below the forcing scale, the dynamics is governed by scattering of the soft modes off gluons at the forcing scale.

We have shown that the inelastic processes are also strongly nonlocal, hence, the corresponding KZ spectra, are not physically relevant except in the BH regime where the KZ spectra $\omega^{-2}$ yields a logarithmic divergent flux and therefore, is marginally nonlocal. Below the forcing scale at late times, radiation and absorption of soft gluons by gluons at the forcing scales are exactly balanced. As a result the soft sector is in thermal equilibrium at late times. To the right (above) the forcing scale, assuming strongly nonlocal (asymmetric) interactions the kinetic equation describing the effective 1 to 2 scattering processes reduces to a diffusion-like equation that conserves energy. As a result, energy is transported to the UV by diffusion and therefore, the spectrum spreads as $t^{1/2}$. 

Numerical simulations of the inelastic collision integral in the LPM regime agree qualitatively with the results obtained in the nonlocal approximation. In particular, the wave front is shown to scale as $t^{1/2}$ which is consistent with the aforementioned diffusion-like energy transport that characterizes nonlocal inelastic processes. In the inertial range, there is no steady power law spectrum due to the sensitivity of the spectrum to the UV cutoff. 

Combining elastic and inelastic processes, which are qualitatively similar, one  expects the spectrum to evolve toward the thermal fixed point. However, in the numerical simulations of the kinetic equation the power spectrum appears to be close to  $\omega^{-2}$ which is most likely a transient power spectrum. This result, based on kinetic theory,  agrees with the turbulent spectrum  observed in the context of Chromo-Weibel Instabilities \cite{Arnold:2005ef}.

Finally, the above analysis breaks down when quantum effects can no longer be neglected, namely, when $n(\omega_\ast)\sim 1$, forcing the spectrum to evolve toward the Bose-Einstein distribution. 

In summary, nonlocal interactions are dominant in non-Abelian gauge plasma. Thus, the resulting turbulent spectrum is not universal and depends strongly on the physics at the forcing and dissipation scales. We have shown that the spectrum in the presence of elastic and inelastic processes flattens when the time goes on and hence is not steady. However, further investigation are necessary in order to better understand the late time dynamics of the turbulent spectrum in non-Abelian theories.

\noindent{\bf Acknowledgements}

We would like the thank J.-P. Blaizot, D. B. Kaplan, J. Liao, A. Kurkela, A. H. Mueller, M. Strickland and K. Tywoniuk for many discussions, and in particular J.-P. Blaizot and K. Tywoniuk  for the careful reading of the manuscript. We would also like to thank J. Liao and J.-P. Blaizot for fruitful discussions and collaboration on issues related to this work. 
This research is supported by the U.S. Department of Energy under Contract No. DE-FG02-00ER41132.

 \appendix

\section{Zakharov conformal transformations}\label{Z-transformation}
In this appendix we derive the Zakharov transformation that make the power-like non-thermal  fixed points of the collision integral explicit. After integrating over $\omega_1$ with the help of the delta function in Eq.~(\ref{coll-int-omega}), the remaining double integration over $\omega_2$ and $\omega_3$ is to be performed over the region $\Delta$ defined by $\omega_2>0$, $\omega_3>0$ and $\omega_2+\omega_3>\omega$, with $\omega_1+\omega=\omega_2+\omega_3$ from energy conservation. It is useful to divide $\Delta$ into four subregions of integration as shown in Figure~(\ref{KZ-int-regions}). Each integration region corresponds to the various possible orderings of $\omega$ and the other three frequencies. In Region 1, defined by 
\beq
\Delta_1\equiv \Big\{\omega>\omega_2>0, \omega>\omega_3>\omega-\omega_2  \Big\},
\eeq 
where the largest frequency of the quartet  is $\omega $. Likewise, for region 2, 3 and 4, we have respectively
\beq
\Delta_2&\equiv& \Big\{\omega>\omega_2>0, \omega_3>\omega  \Big\},  \quad\text{with }  \quad\text{max}(\omega,\omega_1,\omega_2,\omega_3)=\omega_3, \\
\Delta_3&\equiv& \Big\{\omega_2>\omega, \omega_3>\omega \Big\}, \quad \quad\text{with }  \quad\text{max}(\omega,\omega_1,\omega_2,\omega_3)=\omega_1,\\
\Delta_4&\equiv& \Big\{\omega>\omega_2>0, \omega_3>\omega \Big\},  \quad\text{with }  \quad\text{max}(\omega,\omega_1,\omega_2,\omega_3)=\omega_2. 
\eeq

Now, we look for a conformal transformation for each region that will transform the largest frequency of the quartet into $\omega$. Consider for instance Region 2. We seek a transformation $(\omega, \omega_1,\omega_2,\omega_3) \to (\omega', \omega'_1,\omega'_2,\omega'_3)$, that preserves the constraint $\omega+\omega_1-\omega_2-\omega_3=\omega+\omega'_1-\omega_2'-\omega_3'=0$, such that
\beq\label{KZ-ineq}
\omega_2< \omega (\omega_1)< \omega_3\quad  \to \quad \omega'_1< \omega_3' (\omega'_2)< \omega. 
\eeq
To do so, we multiply the quartet by a scale factor $\lambda$ such that \footnote{There exists two possible transformations owing to the symmetry in the exchange of $\omega_2$ and $\omega_3$.} 
\beq \label{KZ-trans-lambda}
 (\lambda \omega, \lambda\omega_1,\lambda\omega_2,\lambda\omega_3) = (\omega'_3, \omega'_2,\omega'_1, \omega').
\eeq

Of course, since $\omega$ is the only non-dummy variable we require $\omega'=\omega$. Eqs.~(\ref{KZ-ineq}) and (\ref{KZ-trans-lambda}) imply that $\lambda=\omega/\omega_3=\omega'_3/\omega$, thereby, we can readily identify
\beq\label{trans-region-2}
\omega'_1 = \frac{\omega \omega_2}{\omega_3}, \quad \omega'_2 = \frac{\omega\omega_1}{\omega_3}, \quad \omega'_3 = \frac{\omega^2}{\omega_3}. 
\eeq
Let us now perform the transformation (\ref{trans-region-2}) on the corresponding term in the collision integral. First, the integration measure transforms as follows
\beq
\rmd \omega_3\rmd  \omega_2 = \rmd \omega'_3\rmd  \omega'_2 \left(\frac{\omega'}{\omega'_3}\right)^3
\eeq
The homogeneous function $U$ (cf. \eqn{U-def}), yields
\bel{U-homo-deg}
U(\omega,\omega_1,\omega_2,\omega_3)=\lambda^{-\nu}  U(\lambda\omega,\lambda\omega_1,\lambda\omega_2,\lambda\omega_3)=\left(\frac{\omega}{\omega_3'}\right)^{\nu}U(\omega_3',\omega'_2,\omega'_1,\omega)\nn
\eeq
Similarly, for the factor $F[n]\equiv F(\omega,\omega_1,\omega_2,\omega_3)$, containing the occupation numbers, we obtain
\beq
F(\omega,\omega_1,\omega_2,\omega_3)&=&\left(\frac{\omega'_3}{\omega}\right)^{-3x} F(\omega_3',\omega'_2,\omega'_1,\omega)\nn
&=& - \left(\frac{\omega'_3}{\omega}\right)^{-3x} F(\omega,\omega'_1,\omega_2',\omega'_3),\nn
\eeq
where in last equality we have used the symmetry properties of $F[n]$, namely,
\beq
F(\omega,\omega_1,\omega_2,\omega_3) &=&F(\omega_1,\omega,\omega_2,\omega_3)=F(\omega,\omega_1,\omega_3,\omega_2)\nn
&=&-F(\omega,\omega_1,\omega_2,\omega_3).\nn
\eeq
Finally, the result of Zakharov transformation on Region 2 reads 
\beq\label{coll-int-region-2}
- \frac{g^2 }{2(4\pi)^3}  \int_{\Delta_1} \rmd \omega_2\rmd \omega_3 \,\frac{\omega_1}{\omega^2}\,\left(\frac{\omega_3}{\omega}\right)^{-3x-4}\,F(\omega,\omega_1,\omega_2,\omega_3).
\eeq
where we have relabelled the dummy variables $\omega_2'$ and $\omega_3'$ as $\omega_2$ and $\omega_3$.

Following the same reasoning regions 3 and 4 are mapped into region 1, using the conformal transformations, 
\beq\label{trans-region-3}
\omega'_1 = \frac{\omega^2}{\omega_1}, \quad \omega'_2 = \frac{\omega\omega_2}{\omega_1}, \quad \omega'_3 = \frac{\omega\omega_3}{\omega_1}, 
\eeq
for Region 3 and 
\beq\label{trans-region-4}
\omega'_1 = \frac{\omega\omega_3}{\omega_2}, \quad \omega'_3 = \frac{\omega\omega_1}{\omega_2}, \quad \omega'_2 = \frac{\omega^2}{\omega_2}.
\eeq
for Region 4.

\section{Angular integration in Eq.~(\ref{BN-eq})}\label{angular-integration}
In this appendix, we perform the angular integrations in \eqn{BN-eq} that yield \eqn{U-def}. We have
\beq
U(\omega,\omega_1,\omega_2,\omega_3)&& \equiv \frac{g^4  }{32} \omega\omega_1\omega_2\omega_3  \nn
&& \times \int \frac{ \rmd \Omega_1}{(2\pi)^3}\int \frac{ \rmd \Omega_2}{(2\pi)^3}\int \frac{ \rmd \Omega_3}{(2\pi)^3}(2\pi)^3\delta(\k+\k_1-\k_2-\k_3).\nn
\eeq
Recall that $U$ shares the same symmetries as the matrix element as shown in \eqn{sym-U-22} and that it is a homogeneous function of gluon frequencies whose degree of homogeneity is $\nu=-1$ (cf. \eqn{U-homo-deg}).
Using the following integral representation of the delta distribution 
\beq
(2\pi)^3\delta(\k+\k_1-\k_2-\k_3) = \rme^{i \x\cdot (\k+\k_1-\k_2-\k_3)},
\eeq
one can integration over the solid angles separately, 
\beq
\int \frac{ \rmd \Omega_1}{(2\pi)^3}  \rme^{i \x\cdot \k_1} &=& \frac{1}{4\pi^2} \int_{-1}^{1} \rmd \cos \theta  \, \rme^{i |\x| \omega_1\cos\theta }  =\frac{1}{2\pi^2} \frac{1}{ |\x| \omega_1} \sin ( |\x| \omega_1),\nn
\eeq
and one obtains
\beq
&&U(\omega,\omega_1,\omega_2,\omega_3)\nn 
&&\qquad\qquad=\frac{\lambda^2  }{2(2\pi)^4}  \frac{1}{\omega^2 } \int_0^\infty \frac{\rmd |\x|}{|\x|^2 }\sin ( |\x| \omega)\sin ( |\x| \omega_1)\sin ( |\x| \omega_2)\sin ( |\x| \omega_2).\nn
 \eeq
Then using the following result
\beq
\int_0^\infty \frac{\rmd |\x|}{|\x|^2 }\sin ( |\x| \omega)\sin ( |\x| \omega_1)\sin ( |\x| \omega_2)\sin ( |\x| \omega_3)=\frac{\pi}{4} \text{min}(\omega, \omega_1,\omega_2,\omega_3),\nn
\eeq
we readily find 
\beq
U(\omega,\omega_1,\omega_2,\omega_3)= \frac{g^4 }{2(4\pi)^3} \text{min}(\omega, \omega_1,\omega_2,\omega_3).
\eeq

\section{Boltzmann-Nordheim equation in the small angle approximation }\label{gradient-exp}
Due to the Coulomb singularity, the 2 to 2 gluon scattering matrix element (\ref{matrix-22-g}) diverges when $t\to 0 $ (and $s\simeq -u$) in the third term and $u \to 0$ (and $s \simeq t$) in the fourth.  In this limit, we have
\beq\label{matrix-22-approx}
|M_{23\to1k}|^2 \equiv 8 N_c^2 g^4 \left(-\frac{us}{t^2}-\frac{st}{u^2}\right) \simeq 16 N_c^2 g^4 \frac{s^2}{t^2} \equiv 16 N_c^2 g^4 \frac{(k-k_1)^4}{(k-k_2)^4} .
\eeq
Introducing the momentum transfer  $\q = \k-\k_2$ and by using the conservation of momenta, we can rewrite the incoming momenta $\k_2$  and $\k_3$ as a function of the outgoing momenta $\k$  and $\k_1$, and the new variable $\q$,
\beq
\k_2 = \k+\q, \quad \k_3 = \k_1-\q, 
\eeq
Now, integrating over $\k_2$, the collision integral (\ref{BN-eq}) reads
\beq\label{coll-int-22-2}
I[n]\equiv  \int_{1\q}  w(\k1\q)  F[n], 
\eeq
with $F[n]$ given by \eqn{F-occup} and
\beq\label{w-def}
w(\k\k_1\q) \equiv W(kk_1k_2k_3)  \, (2\pi) \delta(\omega+\omega_1-\omega_2-\omega_3) \Big|_{\k_3=\k+\k_1-\k_2},
\eeq
where
\beq
W(kk_1k_2k_3) =\frac{1}{32\,\omega \omega_1 \omega_2 \omega_3 } |M_{23\to1k}|^2,
\eeq
which is symmetric under any permutation of the four momenta. This symmetry translates in $w(\k\k_1\q)$ by
\beq\label{w-sym}
w(\k\k_1\q) = w(\k_1\k\q) = w(\k_1\k, - \q) =w(\k_1-\q, \k+\q,\q). 
\eeq
In the limit where $\q \ll \k_i$ one can perform a gradient expansion of the collision integral. To do so, it is useful to multiply the Eq.~(\ref{coll-int-22-2})  by a test function $g_\k$ and integrate over $\k$,
\beq
\int_\k g_\k \, I[n] & &\equiv \int_{\k\k_1\q} g_\k \,w(\k \k_1\q) , F[n]\nn
 & &\equiv \frac{1}{2}\int_{\k\k_1\q}  (g_\k -g_{\k+\q})\, w(\k\k_1\q)\, (n_{\k_1}+1)(n_\k+1)n_{\k+\q}n_{\k_1-\q} \nn
  & & + \frac{1}{2}\int_{\k\k_1\q}  (g_{\k-\q} -g_\k ) \,w(\k\k_1\q)  \, n_{\k_1}n_\k (n_{\k-\q}+1)(n_{\k_1+\q}+1), \nn
\eeq
where we have made an extensive use of Eq.~(\ref{w-sym}). The first non-vanishing term in the expansion in powers of $\q$ is quadratic: the zeroth order vanishes identically, and owing to the fact that
\beq
\int_\q  \, q^i\, w(\k \k_1\q) =0,
\eeq
the linear term vanishes too. Thus, we obtain
\beq
&& \int_\k g_\k \, I[n] \simeq  \frac{1}{2} \int_{\k \k_1\q}  q^i q^j (\nabla^i_\k g_\k) w(\k\k_1\q)\left[(n_{\k_1}+1)(n_\k+1) \right. \nn
&&\left.  \times (n_{\k}\nabla^i_1 n_{\k_1}-n_{\k_1}\nabla^i_\k n_{\k})+n_{\k_1}n_{\k}((n_{\k_1}+1)\nabla^i_\k n_{\k} -(n_{\k}+1)\nabla^i_1 n_{\k_1} ) \right]\nn
&& = \frac{1}{2} \int_{\k\k_1\q}  q^i q^j (\nabla^i_\k g_\k) w(\k\k_1\q)  \left[n_{\k} (n_\k+1)\nabla^i_1 n_{\k_1} - n_{\k_1}(n_{\k_1}+1)\nabla^i_\k n_{\k}\right] \nn
&& = \frac{1}{2} \int_{\k} g_\k  \nabla^i_\k \int_{\k_1\q}  q^i q^j w(\k\k_1\q)  \left[n_{\k} (n_\k+1)\nabla^i_1 n_{\k_1} - n_{\k_1}(n_{\k_1}+1)\nabla^i_\k n_{\k}\right]. \nn
\eeq
In the last step we have performed an integration by part where the surface term at infinity is assumed to vanish. This allows us, since the test function is arbitrary, to identify the integrant in the l.h.s to the one in the r.h.s., and as a result we obtain the Landau equation
\beq
I[n] \equiv \nabla^i_\k \int_{\k_1}  B^{ij}(\k\k_1)  \left[n_{\k_1}(n_{\k_1}+1)\nabla^i_\k n_{\k}-n_{\k} (n_\k+1)\nabla^i_1 n_{\k_1} \right],
\eeq 
where 
\beq
B^{ij}(\k\k_1) = \frac{1}{2}\int_{\q}  q^i q^j w(\k\k_1\q) . 
\eeq
In a relativistic on-shell particle of momentum $k$, we have $k^2=0$, which yields the dispersion relation  $\omega(\k)=|
\k|$. If follows that
\beq
(k+k_1)^4 &=& (2 k_1\cdot k)^2 = 4 \omega^2 \omega_1^2 \left(  1 -\v\cdot\v_1\right)^2, 
\eeq
where $\v \equiv \k/|\k|$ and $\v_ 1\equiv \k_1/|\k_1|$ stand for the velocity vectors.  Similarly, using $\omega_2=|\k_2|\simeq |\k|+\q\cdot\v$ when $\q \ll \k$, we have
\beq
(k-k_2)^4 &=& \left[(\omega-\omega_2)^2 -\q^2\right]^2 \simeq    \left[(\v\cdot\q)^2 -\q^2\right]^2. 
\eeq
Finally,  in this approximation  the argument of the delta function in Eq.~(\ref{w-def}) reads
\beq
\omega(\k)+\omega(\k_1)-\omega(\k_2)-\omega(\k_3)\simeq  \v\cdot\q-\v_1\cdot\q.
\eeq
As a result, we find in the small angle approximation, i.e. $\q \ll \k,\k_1$,
\beq
w(\k\k_1\q) \simeq 4\pi N_c^2 g^4 \frac{(1-\v\cdot\v_1)^2}{ \left[(\v\cdot\q)^2 -\q^2\right]^2} \delta(\v\cdot\q-\v_1\cdot\q). 
\eeq
Hence, (dropping for the moment the constant factor $4\pi N_c^2 g^4$)
\beq\label{B-matrix}
B^{ij}(\k\k_1) = \frac{1}{2}\int_{\q}  q^i q^j  \frac{(1-\v\cdot\v_1)^2}{ \left[(\v\cdot\q)^2 -\q^2\right]^2} \delta(\v\cdot\q-\v_1\cdot\q). 
\eeq
Now, it is useful to make the following change of variables:
\beq
\w\equiv \v-\v_1 \quad \text{and }\quad \u\equiv \v+\v_1.
\eeq
Using the fact that $\v^2=\v_1^2=0$  and that the vectors $\u$ and $\w$ are orthogonal $\w\cdot\u=0$
Eq.~\ref{B-matrix} can be decomposed as follows
\beq\label{B-matrix-2}
B^{ij}(\k\k_1) = B_1 \left( \delta^{ij} -\frac{w^iw^j}{\w^2}\right) +B_2 \frac{u^i u^j}{\u^2}, 
\eeq
where 
\beq\label{B1B2}
\u \cdot \B_1\u = ( B_1 + B_2) \u^2,  \quad \text{and}\quad \text{Tr} \B = 2 B_1+ B_2.
\eeq
Inverting Eq.~(\ref{B1B2}) yields
\beq\label{B1B2-2}
B_1 = \text{Tr} \B -\frac{\u\cdot \B \u}{\u^2}, \quad \text{and}\quad  B_2 = -\text{Tr} \B +2\frac{\u\cdot \B \cdot \u}{\u^2}.
\eeq
Note also the following identities
\bel{vv1-id}
1-\v\cdot\v_1= \frac{1}{2} \w^2,  \quad 1+\v\cdot\v_1= \frac{1}{2} \u^2, \quad \text{and}\quad \w^2+\u^2 = 4
\eeq
Thereby, we have
\beq
\text{Tr} \B  \equiv  \int_{\q}  \frac{ \q^2  \w^2}{ \left[(\u\cdot\q)^2 -4\q^2\right]^2} \delta(\w\cdot\q).
\eeq
and 
\beq
\frac{\u\cdot \B \cdot \u}{\u^2}.  \equiv  \int_{\q}  \frac{ (\u\cdot\q)^2 }{\u^2  \left[(\u\cdot\q)^2 -4\q^2\right]^2} \delta(\w\cdot\q).
\eeq
In order to perform the $\q$ integration we chose the reference frame where $\w$ is along the $z$ axis and thus  $\u$ is in the $xy$ plan. 
In spherical coordinates, we have $\w\cdot \q = |\w| |\q| \cos\theta$ and $\u\cdot \q = |\w| |\q| \sin\theta \sin \phi$.
Hence, 
\beq\label{B1B2-3}
B_1 &=&\frac{2}{(2\pi)^3} \int \rmd |\q| |\q|^2 \int_0^{2\pi} \rmd \phi\int_0^{2\pi}\rmd \cos\phi \nn
&\times & \frac{ (\u\cdot\q)^2 }{\u^2  \left[(\u\cdot\q)^2 -4\q^2\right]^2} \delta(|\w||\q| \cos\phi)\nn
&=& \frac{2|\w|^3}{(2\pi)^3} \int \frac{\rmd |\q|}{|\q|} \int_0^{2\pi} \rmd \phi \frac{1-\sin^2\phi}{\u^2 \sin^2\phi-4}.
\eeq
The integration over $\q$ is logarithmically divergent and yields the Coulomb logarithm $\lambda $. The remaining integral over $\phi$ yields
\beq
 \int_0^{2\pi} \rmd \phi \frac{1-\sin^2\phi}{\u^2 \sin^2\phi-4} = \frac{\pi}{8 \sqrt{4-\u^2}} = \frac{\pi}{8 |\w|}.
\eeq
We finally obtain,
\beq
B_1  = \frac{\lambda}{8 (2\pi)^2} |\w|^2, \quad B_2 = \frac{\lambda}{8 (2\pi)^2}, 
\eeq
where, for $B_2$, we have used
\beq
 \int_0^{2\pi} \rmd \phi \frac{2\sin^2\phi-1}{\u^2 \sin^2\phi-4} = \frac{\pi \u^2 }{8 (4-\u^2)^{{3/2}}} = \frac{\pi \u^2 }{8 |\w|^{3/2}}.
\eeq
Finally, $B^{ij}$ reads
\beq\label{B-def}
B^{ij}=\frac{\lambda}{8(2\pi)^2} \left[ \left(\delta^{ij}-\frac{w^iw^j}{|\w|^2}\right)|\w|^2+u^iu^j\right]. 
\eeq
which implies that $w^i B^{ij} =0$. 
For an isotropic medium, the Fokker-Planck equation simplifies further. We first rewrite the collision term as
\beq
I[n]\equiv \nabla_\k^i \left[ D_\k^{ij}\nabla_\k^j n_\k - C^i_\k n_\k(n_\k+1) \right],
\eeq
with 
\beq\label{D-C-factors}
D_\k^{ij} \equiv \int_1 B^{ij}(\k \k_1) n_{\k_1} (n_{\k_1} +1),   \quad \text{and}\quad C^i_\k\equiv \int_1 B^{ij}(\k \k_1) \nabla_1^j n_{\k_1}
\eeq
Now, in the isotropic case, $C^i_\k\propto v^i$ and using the fact that $w^i B^{ij} =0$, we find 
\beq
C^i_\k &\equiv&  v^i  \int_1 \v\cdot \B(\k \k_1)\cdot \bnab_1 n_{\k_1} =\frac{1}{2}v^i  \int_1 \u\cdot \B(\k \k_1)\cdot \bnab_1 n_{\k_1} \nn 
\eeq
Using \eqn{B-def} and recalling that $\u= \v_1-\v$, the term proportional to $\v$ is a total divergence and therefore vanishes assuming that $n_\k$ vanishes at $\k\to \infty$. Hence, integrating by part using $\bnab_1 \v_1 = 1/|\k_1|$, we obtain
\beq
C^i_\k &=&    \frac{\lambda}{8(2\pi)^2}v^i  \int_1\frac{ n_{\k_1}}{|\k_1|}\equiv   \frac{\lambda}{16 (2\pi)^2}v^i I_b.
\eeq
where $I_b$ is given in \eqn{Ia-Ib}.
Let us turn now to $D^{ij}$.  It takes to form $D^{ij} = D_1 \delta^{ij} + D_2 v^i v^j$ where, 
\beq\label{D1D2-1}
D_1 \equiv \frac{1}{2}\left( \tr \D -\v\cdot\D\cdot \v\right) \quad \text{and} \quad D_2 \equiv -\frac{1}{3} \tr \D + \v\cdot\D\cdot \v. 
\eeq
From \eqn{B-def} and using \eqn{vv1-id}, it follows that
\beq\label{trace-D}
\tr \D & = &  \frac{\lambda}{8(2\pi)^2}  \int_1 (6+2\v_1\cdot\v)  n_1(n_1+1)\nn
& = & \frac{\lambda}{8(2\pi)^2}   6 \int_1 n_1(n_1+1) \nn
& = & \frac{\lambda}{8(2\pi)^2}   6 I_a,
\eeq
where the second term in the r.h.s. of the first vanishes due to the symmetry under the parity transformation $\v_1 \to - \v_1$. The integral $I_a$ is introduced  in \eqn{Ia-Ib}. Similarly, we obtain
\beq\label{vDv}
\v\cdot  \D  \cdot \v & = &\frac{\lambda}{8(2\pi)^2}    2 I_a. 
\eeq
Now, using Eqs.~(\ref{trace-D}) and (\ref{vDv}) in (\ref{D1D2-1}), we find
\beq\label{D1D2-2}
D_1 \equiv  \frac{\lambda}{4(2\pi)^2}  I_a \quad \text{and} \quad D_2 =0, 
\eeq
or 
\beq\label{D-factor}
D_\k^{ij} \equiv \frac{\lambda}{4(2\pi)^2}   I_a \delta^{ij}. 
\eeq

The Fokker-Planck equation can be then rewritten in the form (putting back the prefactor $4\pi N_c^2 g^4$)
\beq
I[n]\equiv   4\pi N_c^2 g^4 \frac{\lambda}{16 (2\pi)^2} \bnab_\k\cdot \left[ I_a \bnab_\k n_\k + I_b \frac{\k}{|\k|} n_\k (n_\k+1)\right] 
\eeq
\section{Steady state solution of the Fokker-Planck equation}\label{solution-FP}
In this appendix we will  derive the stationary solution to  the Fokker-Planck equation (\ref{Landau-source}), in the presence of a steady source that inject particles at a forcing scale $\omega_f$ with a constant rate $Q_f$. 
We recall that
\bel{Ricatti-Q}
Q\equiv  \frac{1 }{4}\hat q \omega^2 \left[ \, \frac{\del }{\del \omega}  n(\omega)\, + \frac{ \,n^2(\omega)}{T_\ast}\right]+Q_f \theta(\omega-\omega_f) =\text{const}. 
\eeq
Eq.~(\ref{Ricatti-Q}) is a first order  non-linear differential equation
where $Q$ is a constant to be fixed by boundary conditions. It is a Ricatti type equation that can be reduced to a first order linear ordinary differential equation. Let us  look for a solution of the form
\beq
n(\omega) = T_\ast \left[\frac{C}{\omega}+\frac{1}{y(\omega)} \right],
\eeq
$C$ being a constant. In terms of the new function $y(\omega)$, Eq.~(\ref{Ricatti}) reads
\beq
-\frac{C}{\omega^2}-\frac{y'}{y^2}+\frac{C^2}{\omega^2}+\frac{2C}{\omega y }+\frac{1}{\omega ^2}-\frac{D}{ \omega^2}=0, 
\eeq
with 
\beq 
D\equiv D(\omega)= \frac{4(Q-Q_f ) }{\hat q  T_\ast}\theta(\omega-\omega_f) 
\eeq
Now, $C$ can be chosen to cancel the terms that are proportional to $\omega^{-2}$. As a result, we are left with a quadratic equation to solve for $C$,
\beq\label{B-eq}
C^2 - C -D =0,
\eeq
and a first order differential equation  for $y$, 
\beq\label{y-eq}
- y' + \frac{2 C}{\omega} y+1 =0.
\eeq
Eq.~(\ref{B-eq}) admits two solutions, 
\beq
C_{\pm}= \frac{1}{2} \left( 1 \pm \sqrt{1+ 4 D }\right), \quad \text{for} \quad D > -\frac{1}{4},
\eeq
and Eq.~(\ref{y-eq}) admits the particular solution 
\beq
y_p=  \frac{\omega}{1-2 C}=\mp \frac{\omega}{\sqrt{1+ 4 D}},
\eeq
which is to be added to the general solution of the homogenous equation 
\beq\label{y-eq-2}
-y' + \frac{2 C}{\omega} y= 0,
\eeq
that is
\beq
y_h= A\, \omega^{2 C}. 
\eeq
Finally the general solution of Eq.~(\ref{Ricatti-Q}), that is symmetric in the exchange of $C_+$ and $C_-$, reads 
\beq\label{g-sol}
n_\text{st}(\omega) = \frac{T_\ast}{2\omega}\left(1+\beta+\frac{2\beta}{ A k^{\beta} - 1} \right),
\eeq

where $\beta(\omega) =\sqrt{1+ 4 D(\omega) }$. 

Let us now specify the boundary conditions. We require the presence of sinks to remove particles below the infrared scale $\omega_\min \ll \omega_f$ and above the UV scale $\omega_\max \gg \omega_f $. The sinks are accounted for by imposing the following Dirichlet boundary conditions: 
\beq\label{dirichlet}
n(\omega_\min) = n(\omega_\max) =0.
\eeq
These boundary conditions (\ref{dirichlet}) applied to the general solution (\ref{g-sol}), 
\beq
1+\beta+\frac{2\beta}{ A\omega_\min^{\beta} - 1}  = 0,
\eeq
to the left of the forcing scale, allow us to determine the constant $C$, 
\beq
A= \frac{1-\beta}{1+\beta}\, \omega_\min^{-\beta}.
\eeq
Thereby, for $\omega < \omega_f$,  the steady state solution reads
\beq\label{g-sol-min}
n_\st(\omega) = \frac{T_\ast (1+\beta)}{2\omega}\left(1+\frac{2\beta}{(1-\beta) (\omega/\omega_\min)^{\beta} - 1-\beta} \right), \eeq
with $ \beta= \sqrt{1+ \frac{16 Q}{\hat q T_\ast}}$, 
and similarly, to right of the forcing scale,  $\omega > \omega_f$, we obtain
\beq\label{g-sol-max}
n_\st(\omega) = \frac{T_\ast (1+\beta)}{2\omega}\left(1+\frac{2\beta}{(1-\beta) (\omega/\omega_\max)^{\beta} - 1-\beta} \right),
\eeq
with  $\beta= \sqrt{1+ \frac{16(Q-Q_f)}{\hat q T_\ast}}$. It is instructive to analyze the solution away from the sinks. For $\omega_\min \ll \omega < \omega_f$, we have
\beq\label{g-sol-min-approx}
n_\text{st}(\omega) \simeq \left(1+\sqrt{1+ \frac{16 Q}{\hat q T_\ast}}\right) \frac{T_\ast }{2\omega},\nn
\eeq
and for $\omega_f < \omega \ll  \omega_\max$, we have
\beq\label{g-sol-max-approx}
n_\text{st}(\omega) \simeq \left(1-\sqrt{1+ \frac{16(Q-Q_f)}{\hat q T_\ast}}\right)\frac{T_\ast }{2\omega} ,\nn
\eeq
We note a discontinuity of the solution at the forcing scale which is confirmed by explicit numerical solutions. In the case where the sinks and sources are well separated we expect $Q\simeq Q_f$, which implies that $n_\st(\omega)=0$ to the right of the forcing scale. 

\section{Collinear divergence in 2 to 3 processes}\label{app:inelatis-2to3}
\begin{figure}[ht] 
\centering
\subfloat[]{ 
\includegraphics[width=6cm]{turb-inel-2.pdf} 
 }\quad
\subfloat[]{ 
\includegraphics[width=6cm]{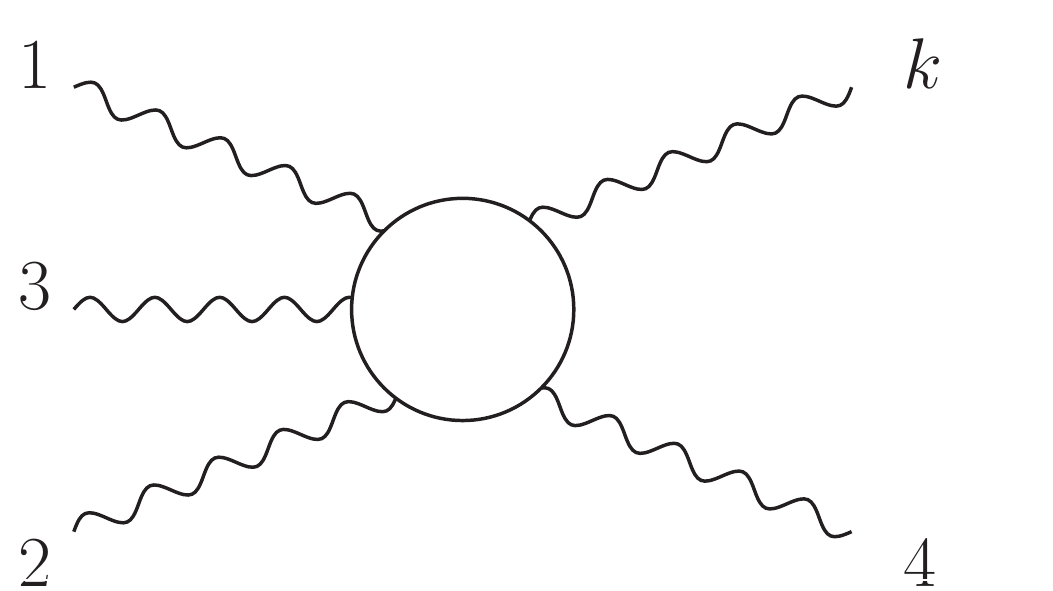} }
\caption{Depiction of the 2 to 3 (a) and 3 to 2 (b) inelastic processes. Only the gain contributions to the collision integral are shown (with the gluon $k$ is measured in the final state).  } 
\label{2to3-gain}
\end{figure}

In this appendix we analyze the next to leading order contribution to the collision integral that corresponds to inelastic 2 to 3 (and 3 to 2) processes depicted in Figure.~\ref{2to3-gain} \cite{Berends:1981rb}. In particular, we show that the corresponding matrix element is singular in the collinear limit and see how this singularity is cured, in the Hard-Loop approximation, by accounting for collective effects in the gluon self energy (See also Ref.~\cite{Huang:2013lia} for a similar discussion). 

Consider the 2 to 3 scattering process, $k_1+k_2 \to k_3+k_4+k$, where $k$ and $k_i$, with ($i=1,2,3,4$), stand for the measured gluon in the final state and the remaining unmeasured ones, respectively.  The collision integral reads 
\beq\label{col-int-23}
I_{2\to3}[n]&\equiv &\frac{1}{2!2! (2\omega)}\,\int_{1,2,3,4} |M_{12\to34k}|^2 \, \nn
&\times &F_{2\to 3}[n]\, (2\pi)^4\delta^{(4)}(k_1+k_2-k_3-k_4-k),
\eeq
where 
\beq\label{F-23}
F_{2\to 3}[n]\equiv n_1n_2 (n_3+1)(n_4+1)(n_k+1)-(n_1+1)(n_2+1) n_3n_4n_k,
\eeq
Similarly, for the 3 to 2 process, $k_1+k_2 +k_3\to k_4+k$, we have
\beq\label{col-int-32}
I_{3\to2}[n]&\equiv& \frac{1}{3! (2\omega)}\,\int_{1,2,3,4} |M_{123\to4k}|^2 \nn
&\times&\, F_{3\to 2}[n]\, (2\pi)^4\delta^{(4)}(k_1+k_2+k_3-k_4-k),
\eeq
where
\beq\label{F-23-2}
F_{3\to 2}[n]\equiv n_1n_2 n_3 (n_4+1)(n_k+1)-(n_1+1)(n_2+1)(n_3+1) n_4n_k. 
\eeq
The factors $1/2!2!$ and $1/3!$ are symmetry factors. 

Here, the invariant matrix element squared is given by \cite{Berends:1981rb}
\beq\label{matrix-23} 
 && |M_{12\to345}|^2 = \frac{1}{2}g^6 \frac{N_c^3}{N_c^2-1} \, \frac{N}{D}\,  [ (12345)+(1254)+ (12435) + (12453) \nn &&+ (12534) + (12543) + (13245) + (13254) + (13425) + (13524) \nn 
 && + (14235) + (14325)]\nn
 \eeq
 with
 \beq
 N&=& (12)^4+(13)^4+(14)^4+(15)^4+(23)^4\nn
 &+&(24)^4+(25)^4+(34)^4+(35)^4+(45)^4 
 \eeq
 and 
 \beq
 D=(12)(13)(14)(15)(23)(24)(25)(34)(35)(45)
 \eeq
 where $(ij)\equiv k_i\cdot k_j$ and $ (ijklm)\equiv (ij)(jk)(kl)(lm)(mj)$. 
 
If the perturbation expansion were well behaved, this contribution should be sub-leading. However, as we shall see in this appendix, the 2 to 3 matrix element possesses a collinear divergence that invalidate the naive perturbative expansion.
 
\begin{figure}[ht] 
\centering
\subfloat[]{ 
\includegraphics[width=6cm]{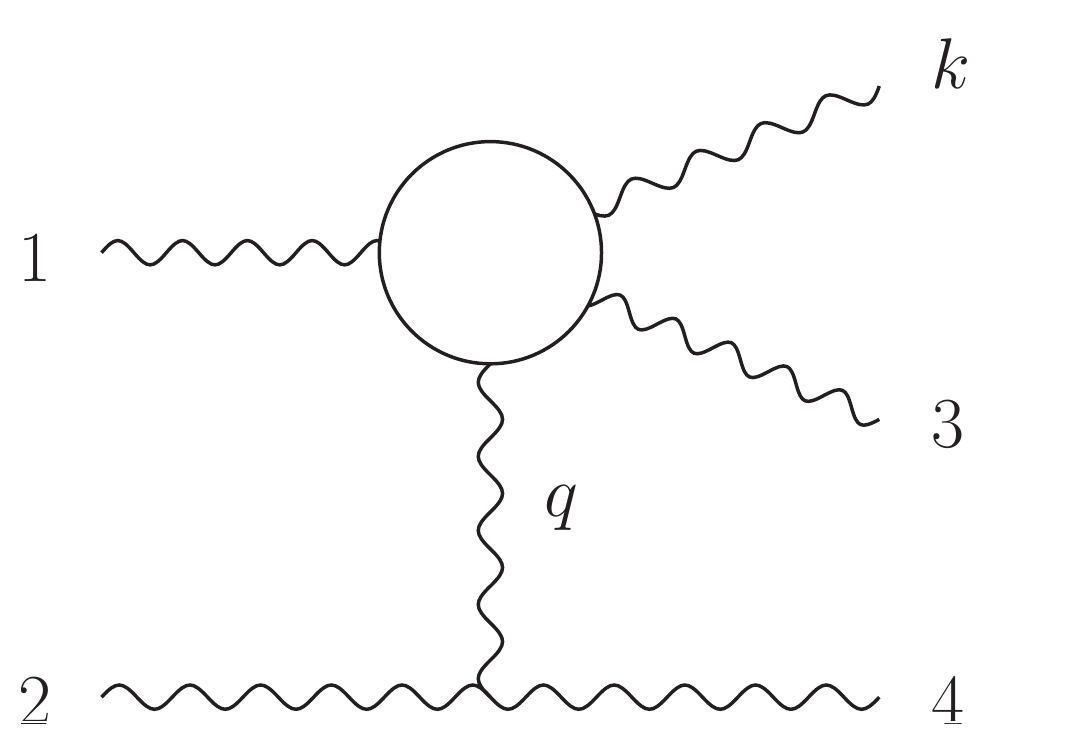} 
 }\quad
\subfloat[]{ 
\includegraphics[width=6cm]{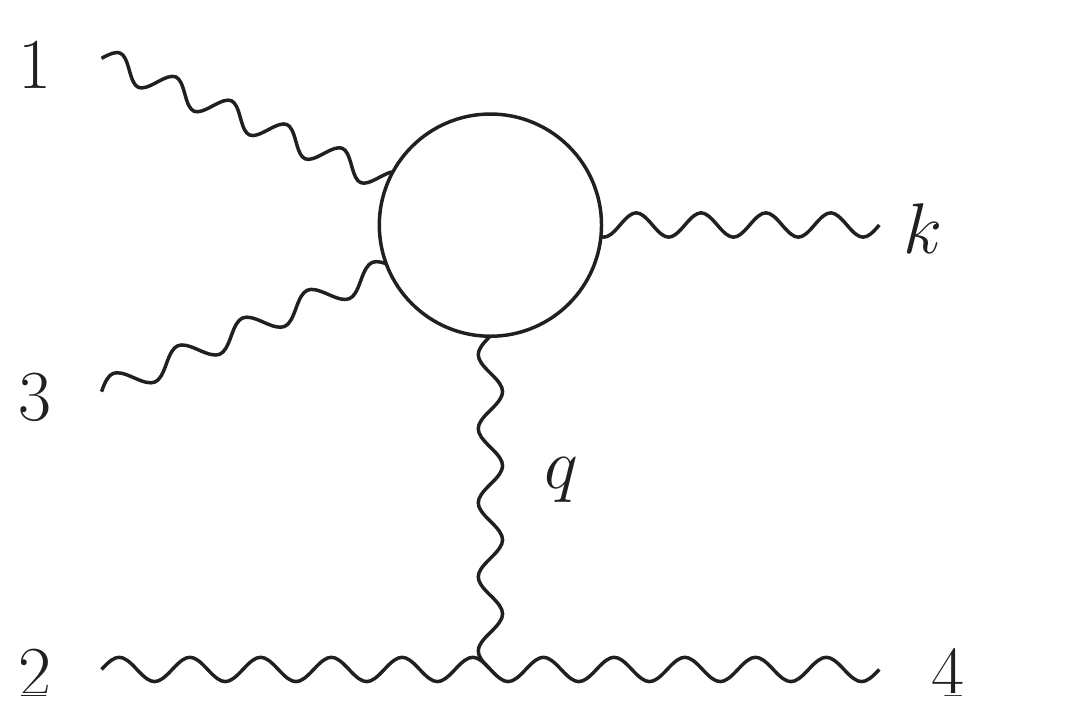}  }
 \caption{Diagrammatic representation of the dominant contribution to the 2 to 3 (a) and 3 to 2 (b) processes shown in Figure~\ref{2to3-gain}, that corresponds to the quasi-collinear 1 to 2 branching (merging) induced by a small momentum transfer $q$.} 
\label{2to3-gain-dom}
\end{figure}

As alluded to above, this matrix element possesses a collinear divergence that invalidate the naive perturbative expansion. To see this, let us consider \eqn{matrix-23} with $k_5\equiv k$, in the limit where gluons 2 and 4 are collinear. The result will then be multiplied by $2!2!$ to account for the remaining symmetric contributions. In this case, we have $t'\equiv q^2 = (k_4-k_2)^2=-2 k_2\cdot k_4 \to 0$ and as a consequence of momentum conservation,  it follows that the outgoing gluons 3 and $k$ are collinear to gluon 1, that is, $t\equiv (k_3-k_1)^2=-2 k_1\cdot k_3 \to 0$ and $t\equiv (k-k_1)^2=-2 k_1\cdot k \approx \omega \omega_1 \theta_{1k}^2\to 0$, that effectively corresponds to a small angle scattering (or small momentum transfer $q$) of gluon 2 into 4. This process is depicted in Figure~\ref{2to3-gain-dom}.

For reasons that will later be clear, it is convenient to choose a frame where $\k_1$ is along the $z$ axis and adopt the light-cone variables, i.e., $k_1\equiv (k^+_1, 0^-,0_\perp)$. Keeping only the leading contributions in the collinear limit, $q\to 0$, and defining the momentum fraction $z = k^+/k_1^+ \approx \omega /\omega_1$, we find that $2(k_2\cdot k_4) \equiv q^2 \simeq -\q_\perp^2$ and 
\bel{ij-factor}
&& (k_1\cdot k_2 ) \, \simeq\,  (k_1\cdot k_4)\, \simeq \frac{s}{2}, \qquad\qquad (k_2\cdot k) \simeq (k_4 \cdot k) \simeq  \frac{s}{2}z, \nn
&& (k_2\cdot k_3) \simeq (k_3\cdot k_4) \simeq  \frac{s}{2} (1-z), \qquad\qquad (k_1\cdot k)  \simeq \frac{\k^2_\perp}{2 z},\nn &&    (k_1\cdot k_3) \simeq \frac{(\q-\k)^2_\perp}{2(1-z)}, \qquad\qquad\qquad (k_3\cdot k ) \simeq \frac{(\k-z\q)_\perp^2}{2z(1-z)}.\nn 
\eeq
In the numerator $N$ (cf. \eqn{matrix-23}) yields
\bel{numerator-2}
 && (12)^4+(13)^4+(14)^4+(15)^4+(23)^4+(24)^4+(25)^4\nn
 && +(34)^4+(35)^4+(45)^4 \nn
 && \simeq   (12)^4+(14)^4+(15)^4+(23)^4+(25)^4+(34)^4+(45)^4 \nn
 && \simeq   \frac{s^4 }{8 } [1+z^4 + (1-z)^4] =  \frac{s^4 z(1-z) }{8 N_c} P(z),
\eeq
Similarly, the sum of the terms $(ijklm)$ in \eqn{matrix-23} reads
\bel{numerator-1}
 \frac{s^4}{2^4}\left[ (1-z)^2 \kp^2 + z^2 (\k-\q)_\perp^2 + (\k-z\q)_\perp^2\right], 
\eeq
and the denominator $D$ yields
\bel{denominator}
&&(12)(13)(14)(15)(23)(24)(25)(34)(35)(45) \nn
&&\qquad\qquad\qquad \simeq \frac{s^6}{2^{10}} \, \q^2_\perp \k^2_\perp (\k-\q)_\perp^2 (\k-z\q)_\perp^2. \nn 
\eeq
Inserting Eqs.~(\ref{numerator-2}), (\ref{numerator-1}) and (\ref{denominator}) in \eqn{matrix-23}, we finally obtain 
\bel{matrix-23-coll}
 && |M_{12\to34k}|^2 \simeq  \frac{4 g^6 N_c^2 s^2 }{N_c^2-1}  z(1-z)P(z) \nn
 && \times \frac{ (1-z)^2 \kp^2 + z^2 (\k-\q)_\perp^2 + (\k-z\q)_\perp^2 }{ \q^2_\perp \k^2_\perp (\k-\q)_\perp^2 (\k-z\q)_\perp^2}. \nn
\eeq
Note that in the limit $\q_\perp \to 0$, the matrix element, \eqn{matrix-23-coll}, diverges as $\q^{-4} $. Likewise, when $\k_\perp \to 0$,  the matrix element diverges as $ \k^{-4} $.
In the next section we shall see that this divergence is cut-off by screening effects, but for the time being, we assume that \eqn{matrix-23-coll} is a well defined quantity to derive the general form the collision integral in the small angle approximation.

\begin{figure}[ht] 
\centering 
 \subfloat[]{ 
\includegraphics[width=6cm]{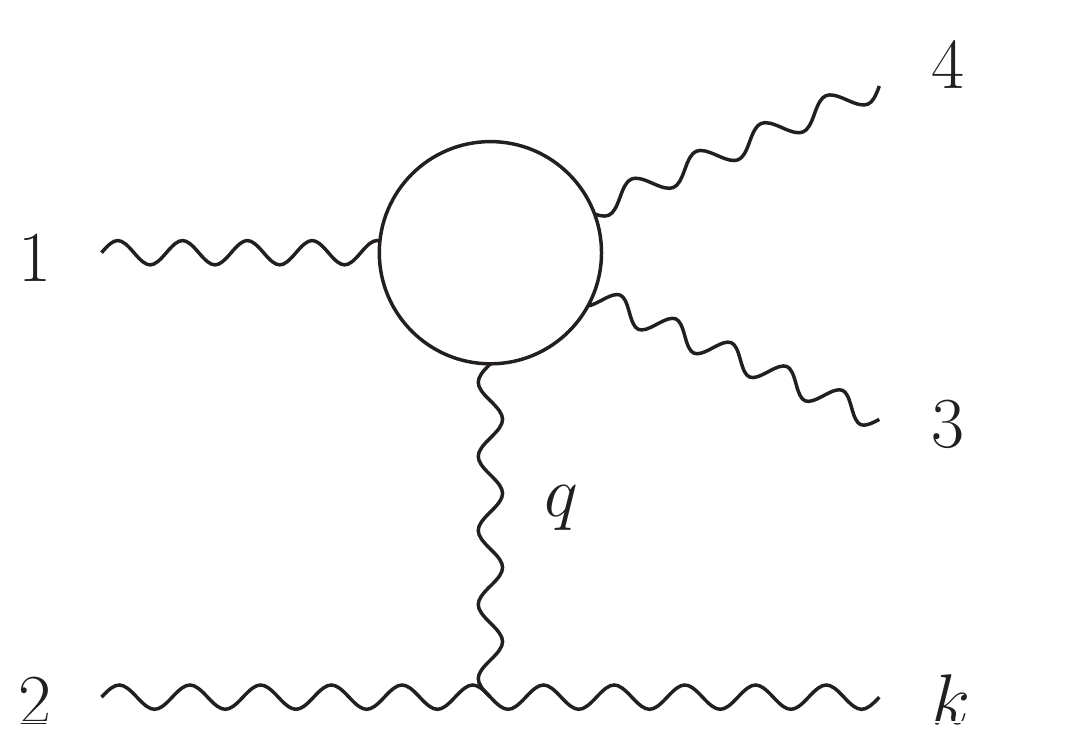} }\quad
\subfloat[]{ 
\includegraphics[width=6cm]{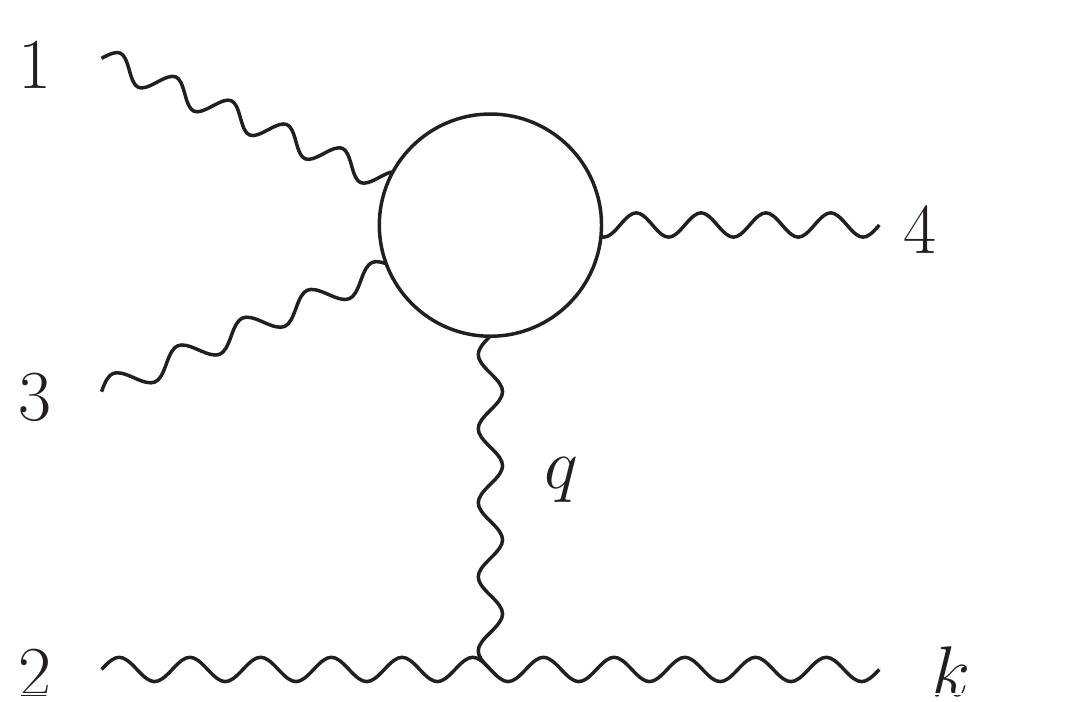} }
 \caption{Diagrammatic representation of the subdominant contribution to the 2 to 3 (a) and 3 to 2 (b) processes shown in Figure~\ref{2to3-gain}, that corresponds effectively to the NLO correction to the elastic rate for small momentum transfer $q$.}\label{2to3-gain-sub}
\end{figure}

Now, the integration over $k_3$ (for instance) is trivial with the help of the 3-momentum delta function, $\k_3= \k_1-\k+\q$. In terms of the new variables,  the integration measure transforms to 
\bel{int-mes}
\frac{1}{4\omega_3\omega}  \int_{1,2,4} &&\equiv\frac{1}{4\omega_3\omega} \int \frac{\rmd^3k_1}{(2\pi)^3 2\omega_1 } \int \frac{\rmd^3k_2}{(2\pi)^3 2\omega_2 } \int \frac{\rmd^3k_4}{(2\pi)^3 2\omega_4 } \nn
&&= \frac{1}{8(2\pi)} \int \frac{\rmd^3k_2}{(2\pi)^3} \int \frac{\rmd z}{z^4(1-z)} \int \frac{\rmd^2\k_\perp}{(2\pi)^2}\int \frac{\rmd q_z \rmd^2\q_\perp}{(2\pi)^3}  \frac{1}{\omega_1^2 \omega^2_2},\nn
\eeq
where $\omega_1=\omega/z$ and $\omega_3= \omega(1-z)/z$, and we have used that $\omega_4\simeq \omega_2$. 
The remaining (energy conserving) delta function yields 
\bel{delta-5}
&& \delta(\omega_1+\omega_2 - \omega_3-\omega_4-\omega) \nn
&&\qquad\qquad= \delta(|\k_1| + |\k_2|-|\k_1-\k+\q|-|\k_2+\q|-|\k|)\nn
&&\qquad\qquad\simeq  \delta\left(|\k_1| -|\k_1-\k|-|\k|-\frac{\q\cdot \k_2}{|\k_2|}-\frac{\q\cdot (\k_1-\k)}{|\k_1-\k|}\right), \nn
\eeq
where, in terms of $q_z$ and $q_\perp$, 
\beq
\frac{\q\cdot \k_2}{|\k_2|}-\frac{ \q\cdot (\k_1-\k)}{|\k_1-\k|} \simeq q_z  \left( \frac{k_{2z}}{|\k_2|}-1 \right)-\frac{\q_\perp\cdot \k_{\perp2}}{|\k_2|}-\frac{\q_\perp\cdot (\k_1-\k)_\perp}{|\k_1-\k|}.
\eeq
This allows us to integrate over $q_z$ after turning the integration over $k_4$ into an integration over $q$ (cf. \eqn{int-mes}). 
As $(|\k_2| -k_{2z})/|\k_2|  \simeq 1-\cos\theta_{12}$, the integral over $q_z$ yields a factor $1/(1-\cos\theta_{12})$. 
Using $s=2\omega_1\omega_2 (1-\cos\theta_{21})^2$ and assuming an isotropic medium the integration over $\theta_{12}$  yields
\beq
 \int_{-1}^1\rmd \cos \theta_{12} (1-\cos\theta_{12})= 2. 
\eeq
Finally, one can express the collision integral in the small angle approximation as an effective quasi-collinear 1 to 2 process
\bel{coll-int-12-approx}
I_{2\to3}[n] \simeq I_{``1\to2"}[n] \equiv \int\frac{ \rmd z}{z^3} \, \cK (z)\, F_{1\to2 } [n],
\eeq
where 
\bel{F-12}
&&F_{1\to 2 } [n]  \equiv n_1 (n_k+1)(n_3+1) - (n_1+1) n_k n_k\nn && = n\left(\frac{\omega}{z}\right)[n(\omega)+1] [ n((1-z)\omega/z)+1] \nn
&& - [ n(\omega/z)+1] n(\omega)n((1-z)\omega/z),
\eeq
and 
\bel{kernel-12-z}
\cK (z)&& \equiv \frac{4\alpha_s g^4N_c^2 I_a}{N_c^2-1}  P(z) \int \frac{\rmd^2\k_\perp}{(2\pi)^2}\int \frac{\rmd^2\q_\perp}{(2\pi)^2}   \nn
&& \times \frac{ (1-z)^2 \kp^2 + z^2 (\k-\q)_\perp^2 + (\k-z\q)_\perp^2 }{ \q^2_\perp \k^2_\perp (\k-\q)_\perp^2 (\k-z\q)_\perp^2}.\nn
&&  =  \frac{4\alpha_s g^4N_c^2 I_a}{N_c^2-1}  P(z) \int \frac{\rmd^2\k_\perp}{(2\pi)^2}\int \frac{\rmd^2\q_\perp}{(2\pi)^2}  \frac{ 1+ z^2 +(1-z)^2}{ \q^2_\perp \k^2_\perp (\k-\q)_\perp^2}.\nn
\eeq
with (cf. \eqn{Ia-Ib})
\beq
I_a \equiv \int \frac{\rmd^3k_2}{(2\pi)^3}  n_2(n_2+1).
\eeq
In the situation where the gluon 2 scattering elastically into gluon $k$ as shown in Figure~\ref{2to3-gain-sub}, with $(k\cdot k_4)\to 0$, the measured gluon is not involved in the collinear $1\to 2 $ branching which can be integrated fully. Is is straightforward to verify that this contribution yields a mild logarithmic divergence and hence can be neglected. 

\section{Effective 1 to 2 process}\label{app:branching-rate}
The probability for a quasi-collinear branching of gluon with energy $\omega$ into two, with energies $z\omega$ and $(1-z)\omega$ respectively, has been calculated in the context of in-medium radiative parton energy loss. This result in the branching rate (\ref{br-rate}). In this appendix, using \eqn{br-rate}, we will derive explicit expressions for the branching rate in the BH and LPM regimes, respectively, that we use in this paper.

\subsection{The Bethe-Heitler regime}
In this regime the branching time is smaller that the typical elastic mean free path such that at most one scattering participates to the radiating process. This regime can be viewed as the reduction of the inelastic 2 to 3 process to an effective 1 to 2 process involving a single small angle scattering. The single scattering approximation amounts to expanding Eq.~(\ref{S-green}) to first order in $\cC(\qp)$. The zeroth order reads
\beq
S_0(\p_\perp,\p'_\perp,z,\tau) = \exp\left[ -i\frac{\p^2_\perp+m_z^2}{2z(1-z)\omega } \tau\right] \theta(\tau)\delta(\pp-\pp'), 
\eeq
but do not contribute to the branching kernel as it yields an imaginary result. For the first order we have
\beq
 && S_1(\p_\perp,\p'_\perp,z,\tau) =-\frac{i}{2} \int_0^\tau \rmd \tau'\, \exp\left[ -i\frac{\p^2_\perp+m_z^2}{2z(1-z)\omega }(\tau-\tau')\right] \nn &&\left[ \tilde\cC((\p-\p')_\perp/(1-z))+\tilde\cC((\p-\p')_\perp/z)+\tilde\cC((\p-\p')_\perp)\right] \nn
 &&\qquad\qquad \times\exp\left[ -i\frac{\p'^2_\perp+m_z^2}{2z(1-z)\omega }\tau'\right].\nn
& &= z(1-z)\omega \frac{\rme^{-i\frac{\pp^2+m_z^2}{2z(1-z)\omega} \tau} -\rme^{-i\frac{\pp'^2+m_z^2}{2z(1-z)\omega} \tau}  }{\pp^2-\pp'^2}\nn
&& \times\left[ \tilde\cC((\p-\p')_\perp/(1-z))+\tilde\cC((\p-\p')_\perp/z)+\tilde\cC((\p-\p')_\perp)\right].\nn
\eeq
Integrating over $\tau$ from zero to $\infty$, we find
\beq
&& \int_0^\infty \rmd \tau S_1(\p_\perp,\p'_\perp,z,\tau) = \frac{2 z^2(1-z)^2 \omega^2}{(\pp^2+m_z^2)(\pp'^2+m_z^2)}\nn
&& \times \left[ \tilde\cC((\p-\p')_\perp/(1-z))+\tilde\cC((\p-\p')_\perp/z)+\tilde\cC((\p-\p')_\perp)\right] .
\eeq
Thus, 
\beq
&& {\cal K}(z)=4\alpha_s  P(z) \int_{\p_\perp,\p'_\perp} \frac{\p_\perp\cdot \p'_\perp}{(\pp^2+m_z^2)(\pp'^2+m_z^2)}\nn
&& \times \left[ \tilde\cC((\p-\p')_\perp/(1-z))+\tilde\cC((\p-\p')_\perp/z)+\tilde\cC((\p-\p')_\perp)\right] .
\eeq
This result is in fact incomplete because it includes only interference diagrams. Including the missing contributions that correspond to splitting before and after the scattering amouts to making  the substitution $\p_\perp,\p'_\perp \to (\p-\p')_\perp^2/2$, and together with the change of variables $\qp \equiv \pp-\pp' $ and  $\kp\equiv \pp$,  we obtain 
\beq
&& {\cal K}(z)=2\alpha_s  P(z) \int_{\k_\perp,\q_\perp} \frac{\q_\perp^2}{(\kp^2+m_z^2)((\k-\q)_\perp^2+m_z^2)}\nn
&& \times \left[ \cC(\q_\perp/(1-z))+\cC(\q_\perp/z)+\cC(\q_\perp)\right] .
\eeq
Note that the delta function in $\cC$ does not contribute to the splitting kernel.  In the limit of vanishing thermal masses we recover the divergent result obtained in the previous section \eqn{kernel-12-z}.  

Let us evaluate the integral in the limit $ z \to 0 $ ot ($z \to 1$). Due to nonlocality of interactions these regions give the dominant contribution to the collision integral. For $z \sim 1$, the splitting kernel yields
\beq
\cK(z) \simeq 4\alpha_s  P(z) \int_{\k_\perp,\q_\perp} \frac{1}{(\kp^2+m^2)((\k-\q)_\perp^2+m^2)(\q_\perp^2+2m^2)}, 
\eeq
where we have used $m_D^2=2m^2$. 
The two angular integrations can be performed as follows: one is trivial and yields a factor $2\pi$, after the change of variables $\lp\equiv (\k-\q)_\perp$, we then have
\beq
&& \int_{\k,\l }\frac{1}{(\k^2+m^2)(\l^2+m^2)\left[(\k-\l)^2+2m^2\right]} \nn&&= \int \frac{2 \pi |\k|  \rmd|\k| |\l|  \rmd|\l|}{(2\pi)^4 (\k^2+m^2)(\l^2+m^2)} \int_0^{2\pi} \frac{\rmd \theta }{\k^2+\l^2+2m^2-2|\k||\l|\cos\theta}.\nn
\eeq
Now the integration over $\theta$ yields
\beq
\int_0^{2\pi} \frac{\rmd \theta }{\k^2+\l^2-2m^2-2|\k||\l|\cos\theta}= \frac{2\pi}{ \sqrt{(\k^2+\l^2+2m^2)^2-4\k^2\l^2}},
\eeq
and the remaining double integration over $x\equiv \l^2/m^2$ and $y\equiv \k^2/m^2$, yields the constant
\beq
C=\int_0^\infty \frac{\rmd x\rmd y}{(x+1)(y+1) \sqrt{(x+y+2)^2-4xy}} \simeq 1.83.
\eeq
We finally obtain   
\beq
{\cal K}(z) = \frac{4\alpha_s N_c^2 g^4 I_a C}{(4\pi)^2 m^2} P(z) = \frac{\pi \bar\alpha^2 \, C\,T_\ast }{N_c} P(z).
\eeq
\subsection{The LPM regime}
Let us turn now to the LPM regime. In this regime multiple soft scatterings contribute to the branching process.  Hence, the typical transverse momentum is larger that the thermal mass and the result will be less sensitive to medium scales. Hence, typical momentum transfer is relatively large, $\q_\perp \gg m_D $, allowing us to expand $\tilde\cC(\u_\perp)$ in \eqn{S-green} to quadratic order in $\u_\perp$, we have
\beq
\tilde\cC(\u_\perp) =\int_{\q_\perp} \,\cC(\q_\perp)\, (1- \rme^{i\x_\perp\cdot\q_\perp}) \simeq  \frac{g^4 N^2_c I_a}{4\pi} \u_\perp^2 \ln \frac{1}{m_D |\u_\perp|}\simeq \frac{1}{4} \hat q \u_\perp^2,
\eeq
where we have absorbed the logarithm in the transport coefficient $\hat q $ that we shall treat as a constant. As a result, the integration over $\x_\perp$ and $\y_\perp$ are Gaussian and yield 
\beq\label{3-pt-function-Gauss}
&& S(\p_\perp,\p'_\perp,z,\tau) = \frac{2\pi i }{ z(1-z) \omega \Omega \sinh (\Omega\tau)} \nn&& \times\exp\left[-i\frac{(\p+\p')_\perp^2}{4 z(1-z) \omega \Omega \coth (\Omega\tau/2)}-i\frac{(\p-\p')_\perp^2}{4 z(1-z) \omega \Omega \coth (\Omega\tau/2)}\right].\nn 
\eeq
with 
\beq
\Omega \equiv \frac{1+i}{2}\sqrt{\frac{\hat q [1+z^2+(1-z)^2]}{z(1-z) \omega}}.
\eeq
Using \eqn{3-pt-function-Gauss} in \eqn{br-rate} and integrating over $\tau$ and the transverse momenta $\pp$ and $\pp'$, we obtain the splitting kernel in the LPM regime,
\bel{kernel-LPM}
\cK(z)= \frac{\alpha_sN_c}{\pi} P(z)\sqrt{\frac{\hat q}{\omega} } \frac{(1-z+z^2)^{5/2}}{z^{3/2} (1-z)^{3/2}}. 
\eeq
\section{Solution of \eqn{inel-FP}}\label{sol-inel-FP}
Let $P$ be a four-vector such that $\omega\equiv |P|$. We have then
\beq
 \nabla^2_P n(|P|) = \frac{1}{\omega^3} \,\frac{\del }{\del \omega}\,  \omega^3\, \frac{\del }{\del \omega} \, n(\omega).
\eeq
Using the Green's function method in 4-dimensions we can write the solution formally as
\beq 
n(\omega) = \int_0^\infty \rmd t' \int \frac{\rmd^4 P'}{(2\pi)^4} \, G(|P-P'|,t-t') S(|P'|).
\eeq
where the retarded Green's function $G$ obeys the following equation,
\beq\label{eq-green}
\left[\frac{\del}{\del t} - \frac{\hat q_\inel }{4} \, \nabla^2_P \,\right] G(|P-P'|,t-t') = (2\pi)^4\delta^{(4)}(P-P')\delta(t-t').
\eeq
whose solution reads
\beq
G(|P-P'|,t-t')  =\frac{16 \pi^2 \theta(t-t')}{ [ \hat q_\inel (t-t')]^2 }  \exp\left[ -\frac{(P-P')^2}{\hat q_\inel (t-t')}\right]
\eeq
In the case of interest, $\omega = |P| \gg |P'| = \omega_f$, that is away from the injection scale, one can neglect the dependence of the Green's on the momentum $P'$, it is then straightforward to integrate over $P'$, 
\beq 
n(\omega,t) &\simeq&  \int_0^t \rmd  t'    G(\omega,t-t' ) \int \frac{\rmd^4 P'}{(2\pi)^4} \, S(|P'|).\nn
&=&  \frac{1}{4}P_f \int_0^t \rmd t'   G(\omega,t' ) .\nn
\eeq
Making the change of variable $x= \omega^2/\hat q_\inel (t-t')$, the last integral yields,
\beq
\int_0^t \rmd t'   G(\omega,t' ) &= & \frac{16\pi^2}{\hat q_\inel \omega^2}  \int_{\omega^2/\hat q_\inel t}^\infty  \rmd x \, \rme^{-x}\nn
&=&  \frac{16\pi^2}{\hat q_\inel \omega^2} \exp\left(- \frac{\omega^2}{\hat q_\inel t }\right).
\eeq
Finally, the spectrum reads
\beq\label{sol-diff-inel}
n(\omega)\simeq  \frac{4 \pi^2P_f}{\hat q_\inel \omega^2}\, \exp\left[- \frac{\omega^2}{\hat q_\inel t }\right].
\eeq

\section{Numerical procedure}\label{numerics}
In the numerical calculations, we use a logarithmic grid  in order to have a larger density of points in the infrared where the distribution diverges as a power law, namely, 
\beq
\ln p(i) = - \ln (p_\text{max}/p_\text{min}) \frac{n-i}{n-1} +\ln p_\text{max},
\eeq
where $i = 1,n$, $n=2000$, $p_\text{max} \equiv p(n)=10-1000$ and $p_\text{min}\equiv p(1)=10^{-7}-10^{-2}$. 

The gluon spectrum at $\tau+\Delta \tau$  is computed as follows
\beq
f(\tau+\Delta \tau) = f(\tau) + C[f]\,  \Delta \tau,
\eeq 
with $\Delta \tau \simeq 10^{-2}-10^{-5}$. We use the Runge-Kutta (RK4) method to integrate over $q$ in the inelastic collision integral $C[f]$. As for the elastic part $f_\text{el}(\tau)$ we use to the Backward-Time-Centered-Space Method ``BTCS" in order to stabilize the diffusion term.






\end{document}